\PassOptionsToPackage{usenames,dvipsnames}{xcolor}
\documentclass[%
 reprint,
 showkeys,
superscriptaddress,
nofootinbib,
 amsmath,amssymb,
 amsfonts,
 aps,
]{revtex4-2}
\usepackage{graphicx}
\usepackage{enumitem}
\usepackage{macros}
\usepackage{dcolumn}
\usepackage{bm}
\usepackage{hyperref}
\usepackage{subfig}
\usepackage{graphicx}
\hypersetup{
    colorlinks=true,
    linkcolor=blue,
    filecolor=blue,      
    urlcolor=blue,
    citecolor=blue, 
    pdftitle={AstroEmulator},
    pdfpagemode=FullScreen,
    }
\usepackage{xcolor,colortbl}
\usepackage{appendix}
\usepackage[mathlines]{lineno}
\usepackage{booktabs}
\newcommand{\ra}[1]{\renewcommand{\arraystretch}{#1}}

\usepackage[mathlines]{lineno}
\usepackage[normalem]{ulem}
\usepackage{tabularx}
\newcolumntype{Y}{>{\centering\arraybackslash}X}

\begin{document}

\preprint{APS/123-QED}

\title{Deep Neural Emulation of the Supermassive Black-hole Binary Population}%

\author{Nima Laal}
\affiliation{Department of Physics and Astronomy, Vanderbilt University, 2301 Vanderbilt Place, Nashville, TN 37235, USA}

\author{Stephen R. Taylor}
\affiliation{Department of Physics and Astronomy, Vanderbilt University, 2301 Vanderbilt Place, Nashville, TN 37235, USA}

\author{Luke Zoltan Kelley}
\affiliation{Department of Astronomy, University of California, Berkeley, 501 Campbell Hall \#3411, Berkeley, CA 94720, USA}

\author{Joseph Simon}
\altaffiliation{NSF Astronomy and Astrophysics Postdoctoral Fellow}
\affiliation{Department of Astrophysical and Planetary Sciences, University of Colorado, Boulder, CO 80309, USA}

\author{Kayhan G\"{u}ltekin}
\affiliation{Department of Astronomy and Astrophysics, University of Michigan, Ann Arbor, MI 48109, USA}

\author{David Wright}
\affiliation{Department of Physics, Oregon State University, Corvallis, OR 97331, USA}

\author{Bence B\'{e}csy}
\affiliation{Department of Physics, Oregon State University, Corvallis, OR 97331, USA}
\author{J. Andrew Casey-Clyde}
\affiliation{Department of Physics, University of Connecticut, 196 Auditorium Road, U-3046, Storrs, CT 06269-3046, USA}
\author{Siyuan Chen}
\affiliation{Kavli Institute for Astronomy and Astrophysics, Peking University, Beijing, 100871, China}

\author{Alexander Cingoranelli}
\affiliation{Department of Physics, University of Central Florida, 4000 Central Florida Blvd, Orlando, FL 32816, USA}

\author{Daniel J. D'Orazio}
\affiliation{Niels Bohr International Academy, Niels Bohr Institute, Blegdamsvej 17, DK-2100 Copenhagen, Denmark}
\author{Emiko C. Gardiner}
\affiliation{Department of Astronomy, University of California, Berkeley, 501 Campbell Hall \#3411, Berkeley, CA 94720, USA}

\author{William G. Lamb}
\affiliation{Department of Physics and Astronomy, Vanderbilt University, 2301 Vanderbilt Place, Nashville, TN 37235, USA}
\author{Cayenne Matt}
\affiliation{Department of Astronomy and Astrophysics, University of Michigan, Ann Arbor, MI 48109, USA}

\author{Magdalena S. Siwek}
\affiliation{Department of Physics and Astronomy, Columbia University, 538 West 120th Street, Pupin Hall, NY 10027, USA}
\author{Jeremy M. Wachter}
\affiliation{School of Sciences and Humanities, Wentworth Institute of Technology, 550 Huntington Avenue, Boston, MA 02115, USA}

\date{\today}
\begin{abstract}
While supermassive black-hole (SMBH)-binaries are not the only viable source for the low-frequency gravitational wave background (GWB) signal evidenced by the most recent pulsar timing array (PTA) data sets, they are expected to be the most likely. Thus, connecting the measured PTA GWB spectrum and the underlying physics governing the demographics and dynamics of SMBH-binaries is extremely important. Previously, Gaussian processes (GPs) and dense neural networks have been used to make such a connection by being built as conditional emulators; their input is some selected evolution or environmental SMBH-binary parameters and their output is the emulated mean and standard deviation of the GWB strain ensemble distribution over many Universes. In this paper, we use a normalizing flow (NF) emulator that is trained on the entirety of the GWB strain ensemble distribution, rather than only mean and standard deviation. As a result, we can predict strain distributions that mirror underlying simulations very closely while also capturing frequency covariances in the strain distributions as well as statistical complexities such as tails, non-Gaussianities, and multimodalities that are otherwise not learnable by existing techniques. In particular, we feature various comparisons between the NF-based emulator and the GP approach used extensively in past efforts. Our analyses conclude that the NF-based emulator not only outperforms GPs in the ease and computational cost of training but also outperforms in the fidelity of the emulated GWB strain ensemble distributions.
\end{abstract}

\maketitle

\section{\label{sec:Introduction}Introduction}

Supermassive black-hole (SMBH) binaries are prime candidates for the source of the gravitational wave background (GWB) signal \citep{SMBHB_prime_1,SMBHB_prime_2,SMBHB_prime_3,SMBHB_prime_4,SMBHB_prime_5}. Strong observational evidence for the existence of SMBHs at the center of galaxies reinforces this notion even further setting the expectation that any detected GWB signal will most likely carry imprints of the underlying physics of SMBH-binary systems and their evolution. In particular, motivated by the fact that the timescale for the gravitational-wave-driven inspiral of a SMBH-binary is generally longer than the Hubble time, the astrophysical environment of the binary must play a crucial role in shortening the time it takes for the system to reach its coalescence. Hence, analyzing the spectral features of a GWB signal can shine light on the significance of the environment of a SMBH-binary on guiding the evolution of the system.

Pulsar timing arrays (PTAs)~\citep{Shazin,Detw,FosterBecker,stevebook} provide the most sensitive data set to search for a GWB signal in the nanohertz frequency region. PTAs are galactic-scale experiments that involve the timing of ultra-precise millisecond pulsars using large radio telescopes. The longevity in observing such stable pulsars allows for the recorded time of arrival (TOA) of pulses to build a deterministic model for the expected behavior of photons as they arrive at Earth. Hence, subtracting this deterministic model from the TOAs enables one to find \emph{timing residuals} in which the impact of elusive and potentially stochastic astrophysical or cosmological phenomena affecting the path of pulses can be studied. One particular source of the stochasticity of the timing residuals is expected to be the nanohertz GWB. Searches for a GWB have been the primary subject of the work of the International Pulsar Timing Array (IPTA) community consisting of the European Pulsar Timing Array (EPTA), the Indian Pulsar Timing Array (InPTA), Parkes Pulsar Timing Array (PPTA), and North American Nanohertz Observatory for Gravitational Waves (NANOGrav) for the past decade.    

The many years of observing, timing, and further searching for a GWB culminated in the works of \citep{15yr, d2, d3} in which the individual constituents of the IPTA found, independently of each other, evidence for a low-frequency noise process possessing inter-pulsar correlations that can be attributed to a GWB. Although the degree of the statistical significance of the measured correlations varies among the IPTA constituents, NANOGrav's latest data set and analyses measured the highest significance for the existence of the Hellings and Downs correlations \citep{hd83}, general relativity's prediction for the functional form of the correlations, between pairs of pulsars. NANOGrav's results show the existence of a GWB with a power-law spectrum that is favored over a model with only independent pulsar noises with a Bayes factor above $10^{14}$. 

Yet, the spectral characterization of GWB remains an active area of research \citep{SMBHB-NANO-1, SMBHB-NANO-2, SMBHB-NANO-3, SMBHB-NANO-4, SMBHB-NANO-5}. In particular, \citet{15yrastro} and \citet{15yrnewphysics} analyze NANOGrav's latest data set in search of astrophysical or cosmological models that can explain the origin of the GWB signal measured by NANOGrav. Even though the findings are consistent with astrophysical expectations and a selected few cosmological models, definite statements about the origin of the GWB cannot be made as the current PTA data sets lack the sensitivity required to make such statements.

First used on PTA data sets by \citet{tss17}, Gaussian processes (GP) have been used to emulate the strain spectrum of a GWB by training on population-synthesis simulations of PTAs. Though GPs solved the computational limitations of the prior simulation-based inference techniques, they suffer from fundamental limitations that make them less desirable especially given that more efficient and capable machine-learning techniques exist that can replace GPs entirely \citep{GP-limit}. In particular, GPs can only be trained on the median and the variance of an ensemble distribution (over many Universes) of GWB characteristic-strain. Indeed, reducing a distribution to just two single numbers limits the degree of complexities of the ensemble distribution that can be captured. For instance, any non-trivial kurtosis and skewness of the distributions as well as covariances across frequency bins are inaccessible to GPs for this very reason. Additionally, the efficiency of training and later using GPs as emulators of ensemble distributions of GWB characteristic-strain scale poorly with the increase in the dimensionality of the parameter space (e.g., more GWB frequency-bins). Such limitations hinder the efforts to establish a connection between SMBH-binaries and the GWB spectrum inferred by PTAs.

In this regard, we adopt a cutting-edge machine learning technique based on `normalizing flows' to emulate an ensemble distribution of GWB characteristic-strain detectable by PTAs. Inspired by \citet{whyQ} and \citet{whyQ2} and following the works of \citet{EarlierNF} and \citet{NN}, we adopt the `Autoregressive Coupling Rational Quadratic Spline' (ACRQS) \citep{Delbourgo1982RationalQS,durkan2019neuralsplineflows} normalizing flow technique to learn the connection between the binary evolution parameters of SMBH-binaries and their predicted GWB spectrum using simulations from \texttt{holodeck}\footnote{\href{https://github.com/nanograv/holodeck}{github.com/nanograv/holodeck}} \citep{15yrastro}. The connection is learned through a conditional relationship in which the binary evolution parameters act as \emph{context} parameters for the GWB spectrum whose distribution is to be learned by ACRQS\footnote{In other words, the binary evolution parameters are treated as parameters that the GWB characteristic-strain distributions are conditioned on.}. Furthermore, we use a trained ACRQS within a Markov Chain Monte Carlo (MCMC) simulation to perform Bayesian inference about the binary evolution parameters given a measured PTA spectrum.

Additionally, we highlight the advantages of using ACRQS over GPs throughout the paper. In addition to being significantly faster and easier to train, ACRQS is shown to be more accurate in learning the complexities of the underlying \texttt{holodeck}'s simulations making it a more capable emulator. ACRQS's ability to learn the entirety of a given distribution, rather than only its median and variance as is in the case of GPs, makes it a more reliable tool for simulation-based inferences in the context of PTA experiments. It is worth noting that GPs become more and more intractable with the increase in the dimensionality of the parameters space through the inclusion of more frequency bins for the GWB and/or incorporating more binary evolution parameters in the simulations. ACRQS can handle such an increase in dimensionality much more gracefully. 

The structure of the paper is as follows. In \S\ref{sec:library}, we discuss the astrophysical considerations behind \texttt{holodeck}'s phenomenological binary evolution simulation library. In \S\ref{sec:gp}, we briefly review GPs as emulators for GWB characteristic-strain ensemble distribution. In \S\ref{sec:nf}, we first introduce the basics of normalizing flows as a machine-learning technique and further explain, in detail, the training procedure we adopted to build our NF-based emulator. In \S\ref{sec:Results}, we provide an in-depth comparison between the GP used in \citet{15yrastro} and our ACRQS NF as it pertains to building an emulator based on \texttt{holodeck}'s phenomenological binary evolution simulation library. Finally, in \S\ref{sec: Conclusion}, we present our concluding remarks. 

\section{\label{sec:library}holodeck’s phenomenological binary evolution library}

We generate populations of SMBH-binaries using the \texttt{holodeck} \citep{15yrastro} package.  We summarize the key aspects of the binary populations here, but a more complete description can be found in \citet{15yrastro}.  The number of simulated SMBH-binaries, $N$, is calculated from the number-density of galaxy mergers, $\eta$, the binary evolution timescale as a function of GW frequency, $t(f)$, and the redshift and comoving volume evolution of the universe, $z(t)$, $V_c(z)$:
\begin{equation}
    \label{eq:pop-synth}
    \frac{\partial^4 N}{\partial M \, \partial q \, \partial z \, \partial \ln f} = 
        \frac{ \partial^3 \eta}{\partial M_\star \, \partial q_\star \, \partial z} \frac{\partial t}{\partial \ln f} \frac{ \partial z}{\partial z} \frac{\partial V_c}{\partial z} \frac{\partial M_\star}{\partial M} \frac{\partial q_\star}{\partial q}.
\end{equation}
The population is constructed over a 4D grid of binary total mass $M=m_1 + m_2$, mass ratio $q = m_2 / m_1 \leq 1$, redshift $z$, and GW frequency $f$.  The total stellar mass of the host galaxies is $M_\star$ and the stellar-mass ratio is $q_\star$.

Given a distribution of SMBH-binaries in a simulated universe, the characteristic-strain of the GWB can be calculated as \citep{Phinney-2001}\footnote{For consistency and cross-checks with earlier works, we defer the inclusion of wave interference effects to future work.},
\begin{equation}
    \label{eq:gwb_mc}
    h_c^2(f) = \int dM dq dz \, \frac{\partial^4 N}{\partial M \, \partial q \, \partial z \, \partial \ln f} \, h_s^2(f).
\end{equation}
The populations used in this analysis are all considered to evolve in circular orbits, such that the GW frequency can be related to the rest-frame orbital frequency as $f = 2 f_\mathrm{orb} / (1 + z)$.  The sky- and polarization-averaged GW spectral strain from a single, circular binary $h_s$ is \citep{Finn+Thorne-2000},
\begin{equation}
    \label{eq:strain_lum_circ}
    h_s^2(f) = \frac{32}{5 c^8} \,  \frac{\left(G \mathcal{M} \right)^{10/3}}{d_c^2} \left(2\pi f_\mathrm{orb}\right)^{4/3}.
\end{equation}
Here, $d_c$ is the comoving distance to a source at redshift $z$, and the rest-frame chirp mass $\mathcal{M} = M q^{3/5}/\left(1 + q\right)^{6/5}$.

The right-hand side of Eq.~\ref{eq:pop-synth} requires a number of components.  To calculate the galaxy merger rate, we choose a \texttt{holodeck} parameterization in terms of a single-Schechter galaxy stellar-mass function, power-law galaxy pair fraction, and power-law galaxy merger time \citep{Chen2019}.  To map from galaxies to black holes, we use a power-law stellar-mass~vs.~bulge-mass relation \citep[`M-$\mathrm{M}_\mathrm{bulge}$']{Kormendy+Ho2013}.  We self-consistently evolve SMBH-binaries from galaxy merger until they reach the PTA band using \texttt{holodeck}'s `phenomenological' binary evolution model.  This model utilizes a small number of free parameters to mimic the effects of dynamical friction (at large separations) and stellar scattering (at intermediate separations) before GW emission dominates the binary inspiral at small separations.  Finally, for the cosmology of our simulated universe, we adopt WMAP9 parameters \citep{wmap9}.

These binary models require a large number of free parameters, many of which are degenerate in their effects on resulting GW predictions.  \citet{15yrastro} find that six parameters are sufficient to explore the parameter space: two parameters governing the overall rate and masses of galaxy mergers ($\phi_0$, $m_{\phi,0}$), two parameters governing the normalization and scatter of the M-$\mathrm{M}_\mathrm{bulge}$ relation ($\mu$, $\epsilon_\mu$), and two parameters governing the binary evolution: the total time between galaxy merger and SMBH binary coalescence ($\tau_f$) and the strength of stellar scattering at small separations ($\nu_\text{inner}$). The six parameters are collectively referred to as $\theta_{\text{evo}}$ from hereon in.
 
\section{\label{sec:gp}Gaussian Processes}
Over the last decade, Gaussian processes (GPs) have become widely used across many fields of astronomy. Their mathematical simplicity and relative flexibility make them a powerful tool for many astronomical analyses \citep{GPsReview}. As an interpolation method, GPs parameterize noisy data as if it were drawn from a multivariate Gaussian distribution with a mean vector and a covariance function. The covariance function is typically built from a suite of versatile kernel functions and is fit by initial training data to map the covariance structure of the parameter space. Once fit, the kernel values are used to quickly make predictions about untrained points across parameter space \citep{tss17}. 

Yet, GPs have many limitations. Most notably, they lose efficiency in high-dimensional parameter spaces (e.g., greater than a few dozen). Additionally, while they are good at mapping single values, most commonly the median, GPs are unable to capture complex features of distributions such as keratosis or skewness \citep{MVSK}. While GP interpolation has done a remarkable job at enabling efficient Bayesian inference on a detailed SMBH-binary population from PTA measurements of a GWB, recent analyses have been hampered by the limitations of the GP interpolator \citep{15yrastro}. 

\citet{tss17} were the first to use GPs to interpolate across GWB spectra created by SMBH-binary simulations and subsequent analyses are all based on their work \citep{11yr, 15yrastro}. In this work, we use the GPs trained in \citet{15yrastro}, which were created using the \textsc{George} GP regression library \citep{George}. Separate GPs were trained at each sampling frequency of the GW spectrum ($f_i$). In an attempt to better model the SMBH-binary population, two separate GPs were trained at each frequency: one on the median value of log$_{\rm 10} (h_c(f_i))$, and one on its standard deviation. This allowed predictions on both the typical value and typical spread of the strain while also accounting for the uncertainty in each value's interpolation. 

The training set in \citet{15yrastro} utilized $2000$ samples drawn from across a six-dimensional parameter space as described in \S\ref{sec:library}. These samples were selected using Latin hypercube sampling, which offers an efficient method to generate stratified samples that span the parameter space \citep[e.g.,][and references therein]{LHC}. At each sample point, $2000$ realizations were simulated to capture the spread arising from the discrete nature of the binary population. Inside the GP covariance function, two types of kernels were used: a rational quadratic kernel for the parameters governing binary evolution ($\tau_f$ and $\nu_{\rm inner}$) and a squared exponential kernel for everything else. Refer to \citet{15yrastro} for more details on the construction and the use of GPs. In \S\ref{sec:Results}, we take our best-trained GP and compare its strengths and weaknesses to a normalizing-flow--based emulator trained on the same simulation library. 

\section{\label{sec:nf}Normalizing Flows}
\subsection{\label{sec:nfoverview}Overview}
Normalizing flows (NFs) \citep{nf0, nf1, NFreview} are a machine-learning technique in which a series of reversible and differentiable mappings are tasked with transforming a simple $d$-dimensional `base' probability distribution to an arbitrarily complicated $d$-dimensional `target' probability distribution. The name 
\emph{normalizing flow} refers to the capability of this technique to \emph{flow} from the target to the base distribution; hence, \emph{normalize} (i.e., convert into Gaussian) the target distribution in the process. Despite this naming convention, the base distribution could be any statistical distribution whose probability density function can be evaluated.

In addition to emulating the target distribution's samples, NF techniques are capable of assigning probabilities to a given sample of the target distribution through the use of calculus's change-of-variable formula. For a single invertible and differentiable mapping $g^{-1}$ that transforms the set of random variables $\bm{x}$ drawn from the base distribution $p_{\rm{base}}$ to the set of random variables $\bm{y}=g^{-1}(\bm{x})$ belonging to the target distribution $p_{\rm{target}}$, the formula states
\begin{align}
   {{p}_{\text{target}}}\left( \bm{y} \right) ={{p}_{\text{base}}}\left( g\left(\bm{y}\right) \right){{\left| \frac{\partial g}{\partial \bm{y}} \right|}}.
\end{align}
 In the above, ${\left| \frac{\partial g}{\partial \bm{y}} \right|}$ is the determinant of the Jacobian of $g$ and both $\bm{x}$ and $\bm{y}$ have the exact same dimensionality. Note that $g$ must be sufficiently complicated to successfully transform a base distribution to a target distribution that exhibits many modes and features. Given common choices for the base distribution (e.g., standard Gaussian or uniform), overly simple mappings such as `affine transformations' (i.e., $g(\bm{y}) = \alpha \bm{y} + \beta$ for two parameters $\alpha$ and $\beta$) lack the flexibility needed to perform a successful transformation to the type of target distributions that one encounters in many real-world experiments. Furthermore, practical constraints such as numerical efficiency and accuracy in estimating the determinant of the Jacobian limit viable choices for $g$. 

A loss function $L$ needs to be defined for NFs in order to perform the learning steps through gradient-based optimization techniques such as ADAM optimizer \citep{ADAM}. One way to define the loss function is through the `Kullback–Leibler' (KL) divergence \citep{KL-divergenece}, $D_\text{KL}$, between an NF's estimate of the target probability distribution, denoted by ${{p}_{\text{em}}}$, and the actual target distribution, denoted by ${{p}_{\text{target}}}$:
\begin{align}
\begin{split}
   L&=D_\text{KL}\\
   &=\int{dy\left\{ {{p}_{\text{target}}}\left( y \right)\ln \left( \frac{{{p}_{\text{target}}}\left( y \right)}{{{p}_{\text{em}}}\left( y \right)} \right) \right\}} \\ 
 & =\int{dy\left\{ {{p}_{\text{target}}}\left( y \right)\ln \left( {{p}_{\text{target}}}\left( y \right) \right) \right\}} \\ 
 &-\int{dy\left\{ {{p}_{\text{target}}}\left( y \right)\ln \left( {{p}_{\text{em}}}\left( y \right) \right) \right\}} \\ 
 & =-\int{dy\left\{ {{p}_{\text{target}}}\left( y \right)\ln \left( {{p}_{\text{em}}}\left( y \right) \right) \right\}}+\text{const} \\ 
 & =- \mathbb{E} \left[ \ln \left( {{p}_{\text{em}}}\left( y \right) \right) \right]  +\text{const} \\ 
 & \approx -\frac{1}{m}\sum\limits_{i=1}^{m}{\ln \left( {{p}_{\text{em}}}\left( {{y}_{i}} \right) \right)}+\text{const}.
 \end{split}
 \label{first loss}
\end{align}
In the above, $\mathbb{E}$ represents the expectation value operator and $m$ is the total number of draws made from the target distribution. As shown above, the goal is to minimize the KL divergence which is mathematically equivalent to minimizing the average of negative log probabilities, as emulated by an NF, evaluated at many samples drawn from the actual target distribution. Furthermore, since the derivatives of the loss function with respect to the parameters of the base distribution as well as the hyperparameters of $g$ are all that is needed for optimizer techniques to perform a gradient descent, the constant term is trivial and is often dropped when defining a loss-function for NFs. Thus, the loss-function is simply defined to be 
\begin{align}
    L=-\frac{1}{m}\sum\limits_{i=1}^{m}{\ln \left( {{p}_{\text{em}}}\left( {{y}_{i}} \right) \right)}.
\end{align}

Informed by \citet{whyQ} and \citet{whyQ2}, our choice for $g$ is the `Autoregressive Coupling Rational Quadratic Spline' (ACRQS) \citep{Delbourgo1982RationalQS,durkan2019neuralsplineflows}. ACRQS is a single rational quadratic transformation whose parameters are given by a masked autoencoder autoregressive neural network (ANN). Explaining the inner workings of ANN is outside the scope of this work, and we refer the interested reader to \citet{made} for this purpose. However, in \S\ref{sec:ARQS}, we highlight other aspects of ACRQS that include the details of the construction and implementation of this mapping. Henceforth, when we use the term NF, we mean ACRQS NF.

\subsection{\label{sec:nf training}Training Procedure}

\begin{table*}\centering
\ra{1.3}
\begin{tabular}{@{}cccccccccccc@{}}\toprule
& \multicolumn{3}{c}{\textbf{ACRQS Hyperparameters}} & \phantom{abc}& \multicolumn{3}{c}{\textbf{Training Parameters}} &
\phantom{abc}\\
\cmidrule{2-4} \cmidrule{6-8} \cmidrule{10-12}
& bin-count & number of neurons & number of layers && learning rate & decay rate & batch size \\ \midrule
 & 8 & 50 & \cellcolor[HTML]{C0C0C0}$\bm{2}$ && $3\times 10^{-3}$ & \cellcolor[HTML]{C0C0C0}$\bm{0}$ & 100\\
 & \cellcolor[HTML]{C0C0C0} $\bm{16}$ & \cellcolor[HTML]{C0C0C0}$\bm{128}$ & 4 && \cellcolor[HTML]{C0C0C0}$\bm{1\times 10^{-4}}$ & 0.96 & \cellcolor[HTML]{C0C0C0}$\bm{1000}$\\
 & 32 &  & 8 && $1 \times 10^{-5}$ & & \\
\bottomrule
\end{tabular}
\caption{A table listing different sets of parameters used for training and constructing ACRQS NF. The combination of parameters highlighted with the color gray yields the best results as found empirically through many trials.}
\label{table1}
\end{table*}

\begin{figure}
\includegraphics[width=\linewidth]{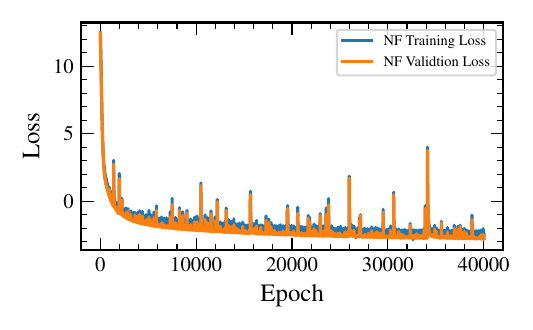}
    \caption{Training (blue) and validation (orange) loss comparisons for the NF used in this work. The training and validation-sets belong to the same instance of the holodeck’s phenomenological binary evolution library though they do not share samples in common. The validation-set contains much fewer samples ($10$ percent in size) as compared to the training-set.}
    \label{fig:nf loss}
\end{figure}

\begin{figure*}[!ht]
\centering
\subfloat[]{%
\includegraphics[width =0.49\linewidth]{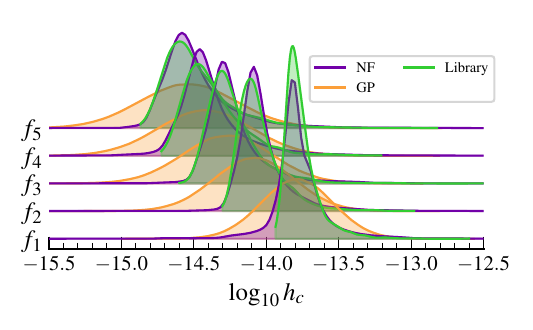}%
}\quad
\subfloat[]{%
\includegraphics[width =0.49\linewidth]{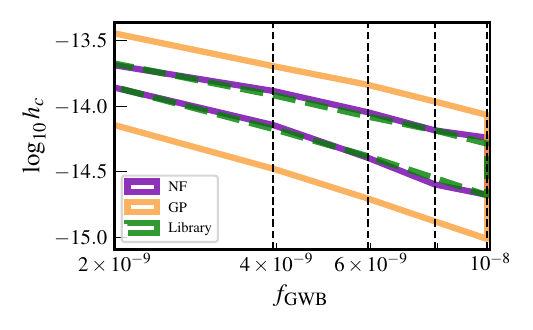}%
}
\caption{(a) A series of density plots comparing the characteristic-strain samples generated using NF (solid purple) and GP (solid orange) to those from the test-set (solid green). The samples form the probability distribution of the GWB characteristic-strain at each frequency bin for the set of binary evolution parameters $\psi_{0} = 7.6$, $m_{\psi_0} = -2.2$, $\mu = 11.6$,  $\epsilon_{\mu} = 8.4$, $\tau_{f} = 0.65$, and $\nu_{\rm{inner}} = -0.98$. Both NF and GP predict the median of the library's samples correctly; yet, NF emulates the entire distribution more successfully. (b) A plot featuring another visualization of the success of NF in emulating the GWB characteristic-strain distribution corresponding to the same set of binary evolution parameters as featured in (a). Similar to (a), the difference between NF and GP is in the spread of the generated samples with NF matching the library's samples more closely.}
\label{fig:hcsamples}
\end{figure*}

The training-set for the NF is \texttt{holodeck}’s phenomenological binary evolution library, as described in \S\ref{sec:library} and \S\ref{sec:gp}. Our goal is to learn the target probability distribution of the base-10 logarithm of the GWB characteristic-strain, $\log_{10}\bm{h_c}$, conditioned on the SMBH's binary evolution parameters $\bm{\theta} _{\text{evo}}$. We choose to use only five of the 30 frequency bins of the GWB spectrum available in the library; therefore, the parameters are $\{\log_{10}{h_{c;k}}\}$ where $k\in[1,5]$. For ease of reference, we refer to the collective as $\log_{10}\bm{h_c}$ from hereon in. Note that our choice to use fewer frequency bins is not due to an inherent limitation of NF but rather is made so that the trained NF is on equal footing with the trained GP used in \citet{15yr} for the fairness of comparisons made in \S\ref{sec:Results}. 

The training procedure for our NF-based astro-emulator goes as follows. First, we transform the library's content ($\bm{\theta}_{\text{evo}}$ as well as $\log_{10}\bm{h_c}$) into values bounded between arbitrary constants $-5$ and $5$. This step is required in order to use an ANN within our emulator (see \S\ref{sec:ARQS}).
Second, we set $g$ to be ACRQS in which $g^{-1}$ is tasked with transforming samples from a uniform base distribution, $U(\rm{lower} = -6, \rm{upper} = 6)$\footnote{The lower and upper bounds of the base distribution is lower and higher than the range of the normalized training data to avoid dealing with cases in which samples are very close to the edges of the base distribution.}, to the target distributions provided in the library for the five $\log_{10}\bm{h_{c}}$ parameters. More specifically, target distributions are treated as conditional probability distributions, $p\left( \left. {{\log }_{10}}\bm{h_{c}} \right|{{\bm{\theta} }_{\text{evo}}} \right)$, where a distribution of $\log_{10}\bm{h_{c}}$ requires, as context, a specific ${{\bm{\theta} }_{\text{evo}}}$ vector.
Third, an ANN is constructed to provide $g$ with the parameters it needs to perform the rational quadratic transformation of \autoref{rqs}. Last, the resulting pipeline (ANN plus $g$) is passed to an ADAM optimizer to iteratively minimize the loss-function and make the pipeline \emph{learn} the target distributions.

Let $m$ and $s$ denote batch sizes. In practice, we draw $m=1000$ unique samples from the training-set for $\bm{\theta}_\text{evo}$. Then, for every drawn sample of $\bm{\theta}_\text{evo}$, we draw $s=1000$ unique samples from the appropriate $\log_{10}\bm{h_{c}}$ distributions provided in the training-set. Based on \autoref{first loss}, the loss-function $L$ used for the optimization is calculated via
\begin{align}
L=-\frac{1}{\left( s+m \right)}\sum\limits_{j=1}^{s}{\sum\limits_{i=1}^{m}{\ln{{p}_{\text{em}}}\left( \left. {\left({\log_{10}\bm{h_{c}}}\right)^{i}} \right|\bm{\theta} _{\text{evo}}^{j} \right)}}.
\end{align}
Note that the NF considers the five $\log_{10}\bm{h_{c}}$ parameters belonging to a single multivariate distribution rather than five separate single-variate distributions. This retains frequency covariance information in NFs and is in contrast to how GPs are trained.

Moreover, we stop training once both validation and training loss are stable as shown in \autoref{fig:nf loss}. The validation-set is made prior to the training and contains random draws from the 10 percent of the library that is not included in the training-set. The validation-set contains $200$ random draws (out of the total of $2000$ samples from the library) for $\bm{\theta}_\text{evo}$ and $200$ random draws for $\log_{10}\bm{h_{c}}$ conditioned on the previously drawn $\bm{\theta}_\text{evo}$.

Nonetheless, a stable validation and training loss is not an adequate measure of the quality of an emulator. We test the performance of a trained NF through \emph{Hellinger distance} \citep{Hell} estimates. Hellinger distance is a measure of similarity between two probability distributions. The distance varies from $0$ to $1$ as the two distributions become more dissimilar. For two probability distributions $p$ and $q$ of discrete random variables, the Hellinger distance is defined as 
\begin{align}
    H=\frac{1}{\sqrt{2}} \sqrt{\sum\limits_{i}{{{\left( \sqrt{{{p}_{i}}}-\sqrt{{{q}_{i}}} \right)}^{2}}}} \label{helleq},
\end{align}
where $i$ ranges over the number of bins used in discretizing both $p$ and $q$. The Hellinger distance is estimated between the test-set's $\log_{10}\bm{h_{c}}$ distributions and those generated by NF or GP. 

For the test-set, we use an instance of \texttt{holodeck}'s library that is independent of the training-set. In practice, we first digitize a given $\log_{10}\bm{h_{c}}$ distribution from the test-set with 15 bins dividing the space of the GWB strain distribution. The resulting bins are then used to divide the emulated samples into the same 15 bins. Finally, the Hellinger distance is computed based on \autoref{helleq}. We report on the results of such comparisons in \S\ref{sec:Results}.

\autoref{table1} lists all the different parameters that one needs to tune in order to perform the NF training. We use the functionalities provided by \texttt{pyro} \citep{pyro} to implement our NF training. Empirically proven, the combination made by the highlighted cells gives the best results in terms of the quality of the NF. Note that the term `bin-count' is the same quantity as $K$ defined in \S\ref{sec:ARQS} which denotes the number of bins used for the spline rational quadratic transformation of \autoref{rqs} and is not related to the binning of the distributions discussed earlier. Lower values of this quantity result in smoother distributions emulated by NF while higher values of this quantity enable NF to capture more complicated features of the target distribution such as multi-modality. Furthermore, the number of neurons\footnote{The neurons of a neural network are functions that receive a sample and are tasked with performing a transformation on the received sample.} and the number layers are the two hyperparameters of ANN. Remarkably, increasing the number of layers of ANN leads to inferior emulation by NF. This is ideal as adding more layers to an ANN is computationally demanding. The training parameters of the ADAM optimizer significantly affect the emulation ability of NF. The ADAM optimizer's learning rate of $10^{-4}$ with no decaying while using a batch size of $1000$ yields the best results. Note that \autoref{table1} is provided for the purpose of reproducibility of the results in \S\ref{sec:Results} and is not necessarily extendable to other training-sets. \autoref{table1} acts as a general guideline rather than a strict recipe.

\section{\label{sec:Results} Comparing emulators: Flows versus Gaussian processes}

To showcase the relative strength of NFs over GPs in building an astro-emulator, we address two criteria: (i) how well the two techniques compare in predicting the GWB's characteristic-strain given a set of binary evolution parameters, and (ii) how well they both uncover the binary evolution parameters underlying a given distribution of a GWB's characteristic-strain. To answer these, we train each emulator technique on \texttt{holodeck}'s 15-year phenomenological binary evolution library, \S\ref{sec:library}, following the procedures outlined in \S\ref{sec:nf training} and \S\ref{sec:gp}. We then test their respective performance on the same test-set as discussed in \S\ref{sec:nf training}.

\subsection{\label{sec:sample-generation} Generating the ensemble distribution of characteristic-strain}

\begin{figure}
\includegraphics[width=\linewidth]{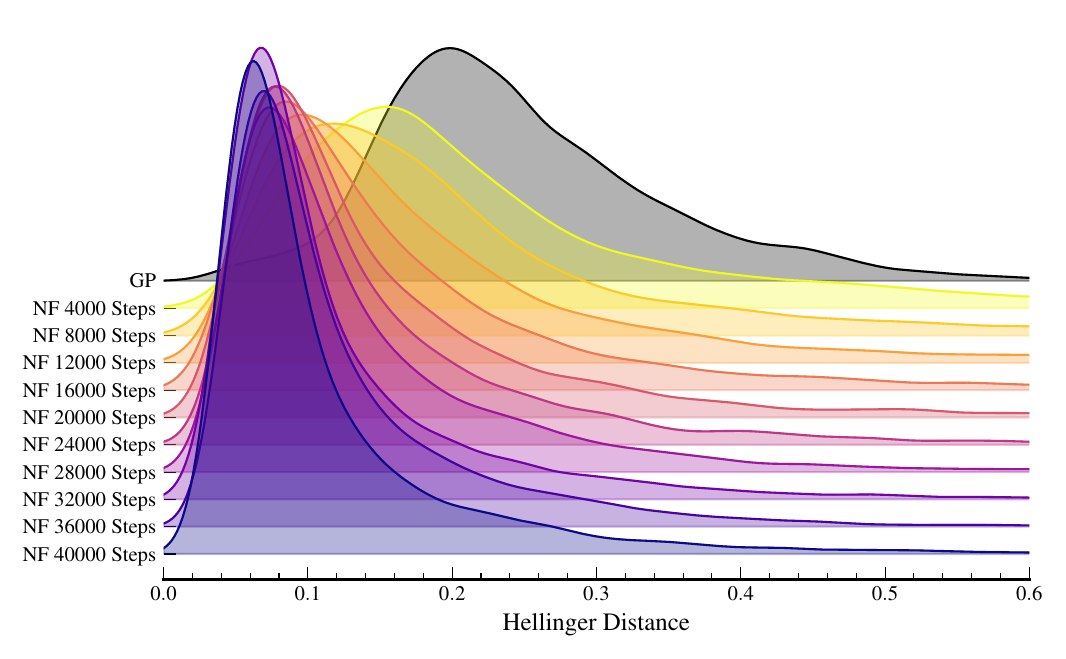}
    \caption{Distributions illustrating the Hellinger distance estimates. To quantify the degree of similarity between the generated characteristic-strain samples and those of the test-set library, Hellinger distance estimates are computed between each set of generated samples and the corresponding samples from the library. The colored distributions highlight how the distance estimates evolves as NF receives more training while the gray distribution showcases the distribution of Hellinger distances obtained using the best trained GP. As expected from prior figures, NF's samples match the library's distribution much more closely. The tail of each distribution corresponds to cases where both GP and NF fail in generating the correct samples. Such cases occur when one or more of the six parameters of $\bm{\theta}_{\rm{evo}}$ have insufficient support in the library; hence, adequate training cannot be provided to the emulator techniques for such $\bm{\theta}_{\rm{evo}}$ values. }
    \label{fig:hellcomp}
\end{figure}

\begin{figure}
\includegraphics[width=\linewidth]{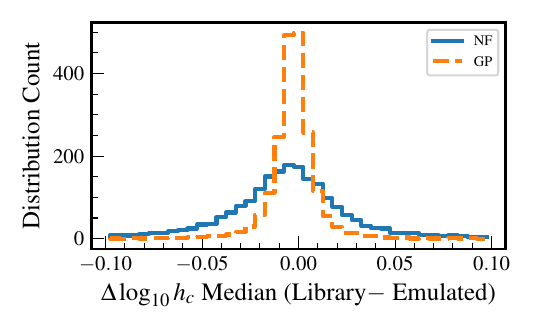}
    \caption{Two histograms showcasing the difference between NF (solid purple) and GP (dashed orange) in predicting the medians of the $\log_{10}\bm{h_c}$ distributions. The horizontal axis ranges over the differences in the median values of $\log_{10}\bm{h_c}$ of the library and those emulated by either NF or GP. The vertical axis ranges over the number of distributions considered for the comparison which sums to 2000. As evident by the figure, GP yields distributions with medians closer to the medians of the distributions in the library as compared to NF.}
    \label{fig:mediancomp}
\end{figure}

\begin{figure}
\includegraphics[width=\linewidth]{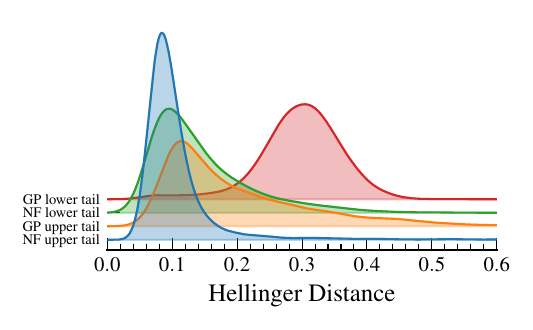}
    \caption{A series of distributions comparing the Hellinger distance estimates between tails of emulated distributions and tails of the distributions in the test-set library. The lower tail is the lowest 25 percentile of a distribution while the upper tail is the upper 25 percentile. To compute Hellinger distances, the library's characteristic-strain distributions are digitized with 15 bins per frequency-bin. The bins are then used to digitize the relevant emulated distributions from NF and GP. For both lower and upper tails, NF's emulation is more accurate evident by low Hellinger distance estimates.}
    \label{fig:tail}
\end{figure}

\autoref{fig:hcsamples} illustrates the differences between GP and NF in generating GWB characteristic-strain distributions conditioned on the following randomly selected set of SMBH-binary evolution parameters:
\begin{align}
\begin{split}
\phi_{0} &= 7.6,\\
m_{\phi,0} &= -2.2,\\
\mu &= 11.6, \\
\epsilon_{\mu} &= 8.4, \\
\tau_{f} &= 0.65,\\
\nu_{\rm{inner}} &= -0.98.
\end{split}
\label{randomdraw}
\end{align}
Refer to \S\ref{sec:library} for a definition of these parameters. Despite the consistency of median values of both the GP and NF samples with those from the library, NF samples are more accurately emulating the library's strain distribution at each frequency-bin. As evident by both panels of the figure, the difference between GP and NF is in the spread of the generated samples. The NF samples are significantly more constrained, mirroring this same behavior in the library.

We now perform a more systematic assessment of this relative performance across all astrophysical coordinates of the test-set. For every $\bm{\theta}_{\rm{evo}}$ in the test-set, we generate $10^{6}$ GWB characteristic-strain samples using both the trained GP and NF at each frequency-bin. The generated characteristic-strain samples for all $2000$ $\bm{\theta}_{\rm{evo}}$ are then compared to the relevant distributions from the test-set through Hellinger distances; \autoref{fig:hellcomp} shows the result of this comparison. We highlight several points shown in this figure. First, we are justified in stopping the NF's training after $40000$ steps as the improvement in NF's ability to generate samples consistent with the library has become marginal since $28000$ steps. Second, even the least trained NF is more capable than GP in generating samples consistent with those in the library. Last, the NF with the highest level of training shows remarkably low Hellinger distance range of $0.08^{+0.05}_{-0.02}$ (median and $50\%$ percentile) compared to GP's $0.20^{+0.10}_{-0.05}$. This is the main support for our claim about better suitability of NFs to GPs for building an astro-emulator.

Despite the high overall accuracy of NFs in emulating the true ensemble distribution of a GWB's characteristic-strain, GP performs better at estimating the median values, as  shown in \autoref{fig:mediancomp}.
Across all five frequency-bins, the medians of the ensemble distributions predicted by GPs more closely match the library values compared to NF. This is unsurprising, since GP is trained on the medians and the standard deviations of the distributions in the training-set library, whereas NF is trained on the entirety of the distributions. Thus, while GP has better performance in point statistics, NF achieves superior accuracy in emulating the overall distribution of the GWB's characteristic-strain over population realizations.  

To complete our evaluation of NF and GP's capability in emulating ensemble distributions of characteristic-strain, we focus on accuracy in the tails of the library's $\log_{10}{\bm{h_c}}$ distributions. \autoref{fig:tail} compares the tails of emulated strain distributions in terms of Hellinger distances. \emph{Upper (lower) tail} corresponds to the upper (lower) $25th$ percentile of the distribution. We can easily see from the figure that the performance of NF surpasses that of GP.

As a concluding note for this section, it is worth mentioning that NF does in fact produce long tails as seen in panel (a) of \autoref{fig:hcsamples}. The extent of the tails often go beyond what is supported in the library for a particular set of $\bm{\theta}_{\rm{evo}}$. Nevertheless, the tails never exceed the bounds set by the entire training-set (not just one $\bm{\theta}_{\rm{evo}}$) as the construction of ACRQS guarantees this outcome. There are two reasons for this: (1) the training-set itself contains distributions with long tails, which the NF tries to emulate, and (2) NF is trained on the entire samples available in the training-set; hence, the variation of $\log_{10}{\bm{h_c}}$ samples across all different $\bm{\theta}_{\rm{evo}}$ values manifest themselves in the form of long tails.

\subsection{\label{sec:estimating probabilities}Estimating probabilities}

\begin{figure}
\includegraphics[width=\linewidth]{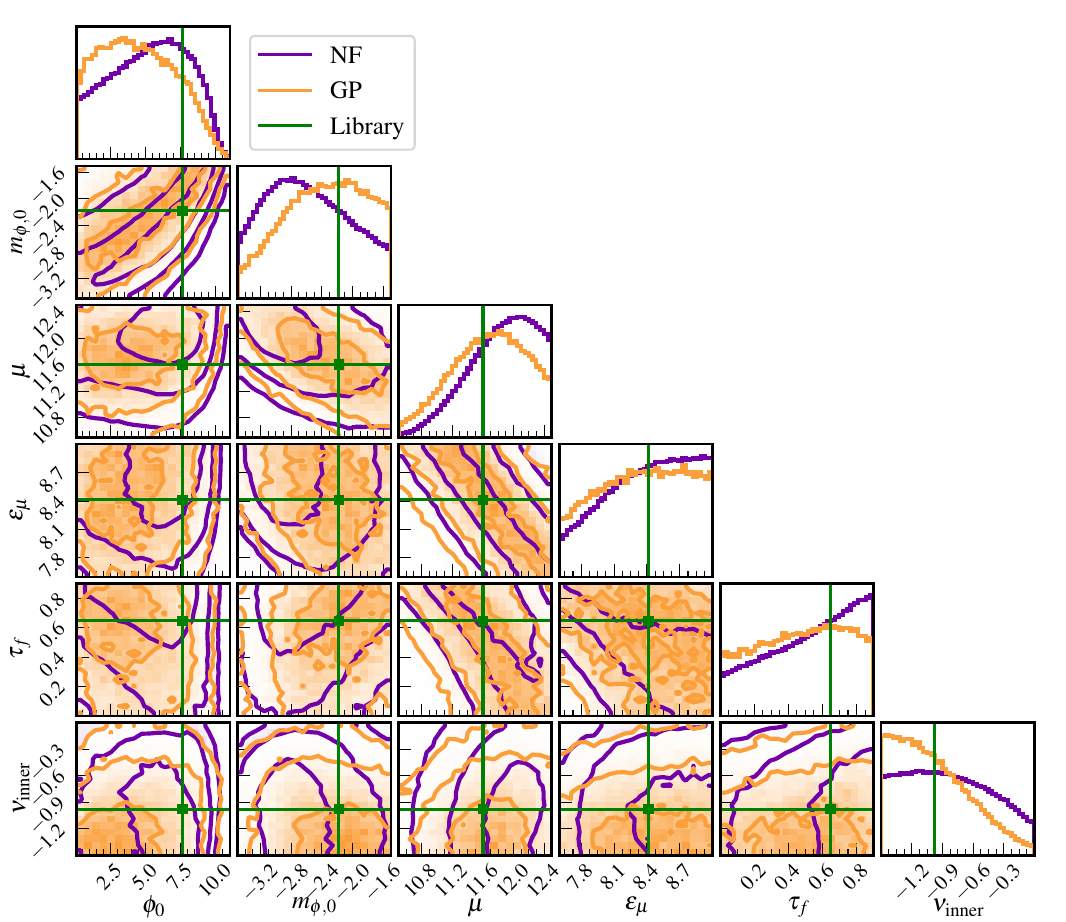}
    \caption{A series of histograms and contours comparing the ability of the best-trained NF and GP in predicting the underlying set of $\bm{\theta}_{\rm{evo}}$ through the use of a Monte Carlo simulation. To take advantage of GP, one-dimensional kernel-density estimators are used to learn the density of the GWB PSD distribution, per frequency-bin, corresponding to the same $\bm{\theta}_{\rm{evo}}$ stated in \autoref{randomdraw}. Subsequently, for GP, \autoref{GPLIK} is used as a likelihood within a Bayesian framework to infer the right $\bm{\theta}_{\rm{evo}}$ from the kernel-density estimates. For NF, the procedure outlined in steps 1 and 2 of \S\ref{sec:estimating probabilities} is used to perform the Bayesian inference. For both cases, the prior probability used for all six $\bm{\theta}_{\rm{evo}}$ is a uniform prior bounded by the minimum and maximum values in the holodeck's library. Surprisingly, NF does not outperform the GP in inferring the right $\bm{\theta}_{\rm{evo}}$.}
    \label{fig:envcomp}
\end{figure}

\begin{figure}
\includegraphics[width=\linewidth]{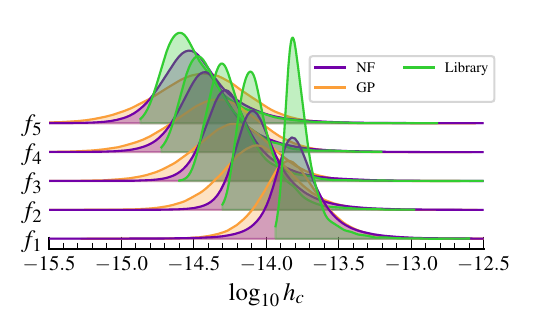}
    \caption{A series of density plots comparing the reconstructed posterior of GWB characteristic-strain parameters, per frequency-bin, obtained form NF (purple) and GP (orange). The true distribution from the library is shown in green. The reconstructed posterior for GWB is made by inputting the inferred $\bm{\theta}_{\rm{evo}}$, shown in \autoref{fig:envcomp}, into the emulators and generating many samples for each draw from the inferred $\bm{\theta}_{\rm{evo}}$ distributions. As seen before, NF is more capable of generating accurate samples compared to GP.}
    \label{fig:recon}
\end{figure}

\begin{figure}
\includegraphics[width=\linewidth]{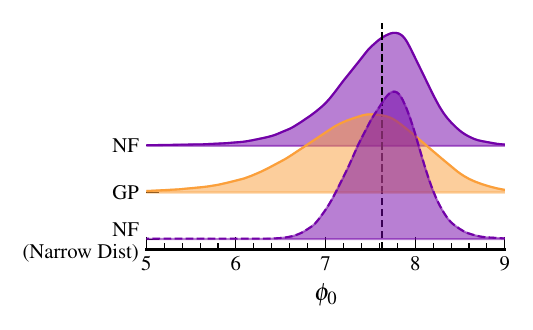}
    \caption{A series of density plots showcasing the ability of NF and GP in predicting the underlying set of $\bm{\theta}_{\rm{evo}}$ resulting from two variations of the analysis whose outcome is captured in \autoref{fig:envcomp}. For the first variation (top two plots), only $\phi_0$ is a variable while other $5$ $\bm{\theta}_{\rm{evo}}$ parameters are fixed to their true values. The dashed vertical line is the true value of $\phi_0$ which is recovered correctly by both GP and NF though NF's recovery is more constrained. For the second variation (bottom plot), the GWB PSD spectrum is artificially narrowed prior to the analysis while $\phi_0$ is still the only model parameter in the Bayesian inference setup. The recovered posterior for $\phi_0$ is narrower for this variation.}
    \label{fig:phi_0_comp}
\end{figure}

The other important aspect of an astrophysical emulator is its ability to assign probability to some measured $\log_{10}{\bm{h_c}}$ conditioned on a given $\bm{\theta}_{\rm{evo}}$. In \citet{15yrastro}, a GP trained on \texttt{holodeck}'s 15-year phenomenological binary evolution library was used within a Markov Chain Monte Carlo (MCMC) analysis to infer the posterior probability distribution of $\bm{\theta}_{\rm{evo}}$ given NANOGrav's 15 year GWB power-spectral-density (PSD) parameters. The PSD parameter $\rho$ (the rms timing deviation induced by the GWB) at a given frequency-bin $k$, is related to the characteristic-strain through the relation
\begin{align}
    \rho _{k}^{2}=\frac{h_{c;k}^{2}}{12{{\pi }^{2}}{{f}^{3}}{{T}_{\text{obs}}}},
\end{align}
where $T_\text{obs}$ is the observational baseline of the PTA. The likelihood used within the MCMC in \citet{15yrastro} evaluates
\begin{align}
\begin{split}
        p\left( \left. \bm{\delta t} \right|{{\bm{\theta} }_{\text{evo}}} \right)&=\prod\limits_{k=1}^{{{n}_{f}}}{\int_{{{\rho }_{\min }}}^{{{\rho }_{\max }}}{d{{\rho }_{k}}\left\{ p\left( \left. {{\rho }_{k}} \right|{{\bm{\theta} }_{\text{evo}}} \right)p\left( \left. \bm{\delta t} \right|{{\rho }_{k}} \right) \right\}}} \\
    &=\prod\limits_{k=1}^{{{n}_{f}}}{\int_{{{\rho }_{\min }}}^{{{\rho }_{\max }}}{d{{\rho }_{k}}\left\{ p\left( \left. {{\rho }_{k}} \right|{{\bm{\theta} }_{\text{evo}}} \right)\frac{p\left( \left. {{\rho }_{k}} \right|\bm{\delta t} \right)}{\pi \left( {{\rho }_{k}} \right)} \right\}}},
    \label{GPLIK}
    \end{split}
\end{align}
using an equally-distanced grid of $\rho_k$ values and kernel-density-estimates of posterior probability $p\left( \left. {{\rho }_{k}} \right| \bm{\delta t}  \right)$ \citep{fitting}. In the above, $n_f$ is the total number of frequency-bins used, $\bm{\delta t}$ is the timing residuals, and $\pi\left( \rho_k\right)$ is the prior used to obtain the $p\left( \left. {{\rho }_{k}} \right| 
\bm{\delta t}  \right)$ posterior probabilities\footnote{ \autoref{GPLIK} is used within a trapezoid rule to estimate the integral numerically.}. However, to perform the $\bm{\theta}_{\rm{evo}}$ inference using our NF-based emulator, \autoref{GPLIK} cannot be employed as we ask our NF to model the GWB spectrum as a multivariate distribution rather than $n_f$ independent single-variate distributions. Correlations between frequency-bins may contain important information that we would like NF to learn.  Hence, we either have to  rewrite \autoref{GPLIK} as a multivariate integral, which is computationally expensive to evaluate, or choose a different approach. 

The approach we take for using our NF to infer the posterior distribution on $\bm{\theta}_{\rm{evo}}$ is inspired by the treatment of pulsar white noise parameters in \citet{Laal:2023etp}. The procedure is as follows:
\begin{itemize}[leftmargin=*]
    \item[]\textbf{Step 1}: A value of the GWB PSD is drawn from each frequency-bin's PSD posterior distribution while respecting the covariances across frequency-bins. Let $\bm{\rho }_{0}$ denote the drawn vector of length $n_f$.
    
    \item[]\textbf{Step 2}: A short MCMC analysis is performed, consisting of $\sim30-50$ steps in the Metropolis-Hastings algorithm. At a given step $i$, $\bm{\theta}_\text{evo}$ is proposed and the likelihood $p\left( \left. {\bm{\rho }_{0}} \right|{{\bm{\theta} }_{\text{evo}}^{i}} \right)$ is evaluated by our NF-based emulator. This short MCMC aims to perform a quick Bayesian inference about ${\bm{\theta}_{\text{evo}}}$ given $\bm{\rho}_0$. 
\end{itemize}
The above two steps are repeated sequentially until the MCMC chain over the model parameters, $\bm{\theta}_{\rm{evo}}$, reaches a satisfactory convergence level \footnote{We used `effective sample size' to measure the convergence level \citep{arviz_2019}.}. 

To test the capability of GPs and NFs as emulators with which to recover posterior probability distributions on $\bm{\theta}_{\rm{evo}}$ given distributions of the GWB PSD, we treat the PSD distribution corresponding to $\bm{\theta}_{\rm{evo}}$ specified in \autoref{randomdraw} as our likelihood $p\left( \left.  \bm{\delta t}\right| {\bm{\rho }}   \right)$. This choice ensures that the result of the subsequent Bayesian inference is unaffected by the detection complications associated with estimating a GWB's PSD from the timing residuals of a finite number of pulsars observed for a finite period of time. In other words, we consider a perfect recovery of the GWB's PSD parameters which is the distribution formed by the GWB PSD samples in the library associated with $\bm{\theta}_{\rm{evo}}$ of \autoref{randomdraw}.
Lastly, to perform a Bayesian inference about $\bm{\theta}_{\rm{evo}}$, we use a uniform prior bounded by the lowest and the largest values in the library for all the six parameters of the vector $\bm{\theta}_{\rm{evo}}$.

Performing Bayesian inference about $\bm{\theta}_{\rm{evo}}$ is simple. To use GPs in the Bayesian pipeline, a MCMC simulation uses \autoref{GPLIK} as its likelihood where each $p\left( \left.  \bm{\delta t}\right| {{\rho_k }}   \right)$ is given by a kernel-density-estimator. Similarly, to use NF, a MCMC simulation uses the the same kernel-density-estimates while following steps highlighted above. The results of both inferences are shown in \autoref{fig:envcomp}. The orange posteriors belong to GPs while the purple posteriors belong to NFs. The straight lines are the true values of $\bm{\theta}_{\rm{evo}}$ specified in \autoref{randomdraw}. To help understand the results highlighted in \autoref{fig:envcomp}, the reconstructed characteristic-strain of GWB using the $\bm{\theta}_{\rm{evo}}$ posterior probabilities featured in \autoref{fig:envcomp} is shown for both NF and GP in \autoref{fig:recon}. To make \autoref{fig:recon}, samples that form the posterior probability distributions of \autoref{fig:envcomp} are given to their corresponding emulator in order to generate GWB characteristic-strain samples. The generated samples are then plotted and compared against the \emph{true} samples found within the library for $\bm{\theta}_{\rm{evo}}$ of \autoref{randomdraw}. As evident by the two figures, NF is not outperforming GP in recovering the underlying $\bm{\theta}_{\rm{evo}}$; however, NF's $\bm{\theta}_{\rm{evo}}$ can better reconstruct the right GWB characteristic-strain distributions.

Since an in-depth analysis of detectability of $\bm{\theta}_{\rm{evo}}$ parameters is out of the scope of this work, we hypothesize that both GP and NF's failure in recovering the right $\bm{\theta}_{\rm{evo}}$ is caused by the degeneracy of such parameters as well as the uniformity of the $\bm{\theta}_{\rm{evo}}$ distributions found in the library. This is the very reason why the training for both NF and GP is done in the conditional form of $p\left( \left. {{\log }_{10}}\bm{h_{c}} \right|{{\bm{\theta} }_{\text{evo}}} \right)$ rather than $p\left( \left.  {{\bm{\theta} }_{\text{evo}}}   \right| {{\ln }_{10}}\bm{h_{c}}\right)$ (i.e., reversing the conditional form so that one can generate ${{\bm{\theta} }_{\text{evo}}}$ samples given a estimates of the GWB characteristic-strain). For the latter to be possible, the training-set library has to be more informative with respect to $\bm{\theta}_{\rm{evo}}$ parameters meaning that the distributions of  $\bm{\theta}_{\rm{evo}}$ in the library must exhibit features and modes learnable by an emulator. Since the holodeck's $\bm{\theta}_{\rm{evo}}$ distributions is uniform, neither NF nor GP can learn the connection between $\bm{\theta}_{\rm{evo}}$ and the characteristic-strain parameters in the reverse direction of what has been shown in this paper. 

To test the stated hypothesis, we performed two variations of the Bayesian inference discussed above. In the first case, we simply fix all but the galaxy merger rate parameter, $\phi_0$, of $\bm{\theta}_{\rm{evo}}$ to their true values expressed in \autoref{randomdraw}. In the second case, we randomly choose a single realization out of the available 2000 realizations in the test-set, and set the PSD distribution to be a Gaussian distribution centered at the GWB PSD value that corresponds to the chosen realization and $\bm{\theta}_{\rm{evo}}$ of \autoref{randomdraw}. The variance of the Gaussian distribution is chosen to be vary small (less than $1$ percent of the PSD value). Additionally, similar to the first variation, we set all but $\phi_0$ to their true values. The result of performing an inference on these two cases is shown in \autoref{fig:phi_0_comp}. For the first variation, both GP and NF can recover the right value of $\phi_0$. We attribute this outcome to the breaking of the degeneracy between the binary parameters by only allowing $\phi_0$ to vary. For the second case where the distribution of the GWB spectrum is narrow, the obtained posterior for $\phi_0$ is narrower than the other cases as one might expect. We also performed a Bayesian inference with the narrow PSD distribution while varying all six components of $\bm{\theta}_{\rm{evo}}$; however, the resulting posterior distributions were not much different from those shown in \autoref{fig:envcomp}. Thus, we conclude that the features of the training-set including degeneracy between $\bm{\theta}_{\rm{evo}}$ components, the uniformity of $\bm{\theta}_{\rm{evo}}$ distribution in the training-set, and the training-set's GWB characteristic-strain ensemble distribution being broad are the causes for the uncertain posterior distributions shown in \autoref{fig:envcomp}.

\section{\label{sec: Conclusion}Conclusion}

In this paper, we demonstrated how NFs can outperform GPs in being an emulator of the ensemble distribution (over many Universes) of the GWB characteristic-strain. Through the use of holodeck's phenomenological binary evolution library, we were able to train an NF capable of emulating the connection between a GWB's characteristic-strain and the demographic and dynamical properties governing a population of SMBH-binaries. The connection is one-directional, meaning that our emulator is capable of generating and assign probability to GWB characteristic-strain ensemble distributions given specific values of binary evolution parameters. 

We compared the capability of our new emulator to the GP-based one in \citet{15yr}, concluding that the new NF-based emulator is far more capable, since the emulated ensemble distributions of characteristic-strain much more closely mimic those from underlying simulations. This was quantified through the estimation of Hellinger distances between ensemble distributions made by emulation and independent instances of the \texttt{holodeck} library that were not used during the training of either NF or GP. In particular, the best-trained NF-based emulator yields a Hellinger distance range of $0.08^{+0.05}_{-0.02}$ (median and $50\%$ percentile) compared to the GP emulator's range of $0.20^{+0.10}_{-0.05}$. The range of distances is over all 2000 available GWB characteristic-strain distributions in the \texttt{holodeck} library. 

The advantages of NF-based emulators do not end with more accurate sample generation from learned ensemble distributions. Due to the ingenious structure of the autoregressive coupling rational quadratic spline normalizing flow used within the emulator, the training can be performed quickly and efficiently at large dimensions. For the best-trained NF-based emulator used in this work, the training took 30 minutes on an NVIDIA RTX 3090 GPU and 80 minutes on a 13th Generation Intel Core i9-13900K CPU. Naturally, for training-sets with higher dimensions, the training becomes more expensive to perform. However, the type of NF used within our astro-emulator is still capable of handling very high dimensions as shown in \citet{whyQ}. For instance, if we increase the number of frequency-bins for the characteristic-strain parameters from 5 to 30, the same training takes about 9 hours on the same GPU. This is in contrast to training GPs which cannot handle 30 frequency-bins at all.

In the future, we will use our NF-based astro-emulator introduced in this work to perform more accurate astrophysical inference using GWB observations. NFs will allow us to explore the full, high-dimensional parameter spaces that more accurately characterize our uncertainties in supermassive black-hole binary formation and evolution.  As GWB measurements become more constraining, the techniques developed here will allow us to fully utilize those measurements in constraining the complete distributions of simulated spectra (e.g.~as opposed to only medians and standard deviations), and fully capturing their covariances (e.g.~instead of consider each frequency independently).  These tools will then also allow us to better predict how future measurements will constrain our models, and thus inform inform strategies for improving PTA observations.

\subsection{\label{sec:software}Software}
We took advantage of the functionalities provided by \texttt{holodeck} \citep{15yrastro}, \texttt{PTMCMC sampler} \citep{PTMCMC}, \texttt{pytorch} \citep{pytorch}, and \texttt{pyro} \citep{pyro}. Python package \texttt{matplotlib} \citep{plt} was used for generating the figures in this paper.
\begin{acknowledgments}
Our work was supported by the NANOGrav NSF Physics Frontier Center awards \#2020265 and \#1430284. S.R.T acknowledges support from NSF AST-2007993, and an NSF CAREER \#2146016. N.L is supported by the Vanderbilt Initiative in Data Intensive Astrophysics (VIDA) Fellowship. This work was conducted in part using the resources of the Advanced Computing Center for Research and Education (ACCRE) at Vanderbilt University, Nashville, TN.
\end{acknowledgments}

\appendix
\section{\label{sec:ARQS}Details of Construction of ARQS}
To use ACRQS, a batch of samples from the target distribution, denoted by $\bm{y}$, is randomly partitioned into two unequally-sized sets: $\bm{y}_{A}$ and $\bm{y}_{B}$ where $\bm{y}_{B}$ is fed into an ANN and is kept untransformed throughout. If not already, all samples from the target distribution need to be cast in the range between a lower, $-B$, and an upper, $B$, bound. The objective is to transform every sample in $\bm{y}_{A}$ separately and independently of every other sample in $\bm{y}_{A}$ by a spline interpolation rule where the location of the connecting points (i.e., knots) of the splines are determined by $\bm{y}_{B}$. Furthermore, the number of bins $K$ (the number of knots minus $1$), is a hyperparameter of $g$ and is always set prior to the construction of $g$ and ANN. 

The ANN outputs three quantities for every sample $y_{A;i}$ in $\bm{y}_{A}$ given the set $\bm{y}_{B}$ as its input. The three quantities are 
 (1) horizontal and 
 (2) vertical coordinates of the $K$ knots collected in matrices $H$ and $U$ and (3) the set of derivatives saved in matrix $\Delta$ whose elements are the slope of the tangent lines at the location of knots. Note that, $H$ and $U$ are two $n_A \times K$ matrices while $\Delta$ is a $n_A \times K-1$ matrix where $n_A$ is the number of samples in $\bm{y}_{A}$. For numerical stability reasons, $\Delta_{i;1}$ and $\Delta_{i;K+1}$ are always set to 1; hence, the ANN needs to output a $\Delta$ matrix of size $n_A \times K-1$. Furthermore, the location of the first knot is always set at $(-B, -B)$; thus, $H$ and $U$ should have the size $n_A \times K$ rather than $n_A \times K+1$. Since the output of ANN is unconstrained, all matrices go through a soft-type function in order to be regularized. Specifically, $H$ and $U$ go through a softmax function whereas $\Delta$ goes through a softplus function. Finally, after the spline knots are fully specified, each ${y}_{A;i\in k}$, that is each sample in the partition $A$ of $\bm{y}$ that belongs to the $k$th bin, is transformed according to a spline rational quadratic law \citep{Delbourgo1982RationalQS,durkan2019neuralsplineflows}:
     \begin{align}
g\left( {{y}_{A;i\in k}} \right)=\frac{\alpha y_{A;i\in k}^{2}+\beta {{y}_{A;i\in k}}+\gamma }{ay_{A;i\in k}^{2}+b{{y}_{A;i\in k}}+c} \label{rqs},
     \end{align}
where $\alpha$, $\beta$, $\gamma$, $a$, $b$, and $c$ are all functions of $\left\{ {{H}_{i;k}},{{U}_{i;k}}, {{\Delta }_{i;k}} \right\}$ and $\left\{ {{H}_{i;k+1}},{{U}_{i;k+1}}, {{\Delta }_{i;k+1}} \right\}$.

The provided details for the construction of $g$ might distract one from realizing the ingenuity of the outlined technique. By not transforming the samples in one partition of $\bm{y}$ while transforming the samples in the other partition individually and separately, one ensures that the Jacobian matrix of $g$ is lower-triangular; hence, the determinant of the Jacobian is simply the product of the diagonal elements of the Jacobian matrix. Furthermore, given the rational quadratic form of \autoref{rqs}, both ${\partial g}/{\partial y_{A;i}}$ and ${\partial g^{-1}}/{\partial y_{A;i}}$ can be analytically, and cheaply, determined. Lastly, the use of \autoref{rqs} ensures extreme flexibility of the transformation resulting in an state of the art NF technique capable of dealing with complicated target distributions at very high dimensions \citep{whyQ} which is ideal for our proposes in this work. In \S\ref{sec:Results} we put this NF technique in use to learn the connection between SMBHs' binary evolution parameters and their GWB characteristic-strain. 
\bibliography{main}

\providecommand{\noopsort}[1]{}\providecommand{\singleletter}[1]{#1}%
\begin{thebibliography}{52}%
\makeatletter
\providecommand \@ifxundefined [1]{%
 \@ifx{#1\undefined}
}%
\providecommand \@ifnum [1]{%
 \ifnum #1\expandafter \@firstoftwo
 \else \expandafter \@secondoftwo
 \fi
}%
\providecommand \@ifx [1]{%
 \ifx #1\expandafter \@firstoftwo
 \else \expandafter \@secondoftwo
 \fi
}%
\providecommand \natexlab [1]{#1}%
\providecommand \enquote  [1]{``#1''}%
\providecommand \bibnamefont  [1]{#1}%
\providecommand \bibfnamefont [1]{#1}%
\providecommand \citenamefont [1]{#1}%
\providecommand \href@noop [0]{\@secondoftwo}%
\providecommand \href [0]{\begingroup \@sanitize@url \@href}%
\providecommand \@href[1]{\@@startlink{#1}\@@href}%
\providecommand \@@href[1]{\endgroup#1\@@endlink}%
\providecommand \@sanitize@url [0]{\catcode `\\12\catcode `\$12\catcode `\&12\catcode `\#12\catcode `\^12\catcode `\_12\catcode `\%12\relax}%
\providecommand \@@startlink[1]{}%
\providecommand \@@endlink[0]{}%
\providecommand \url  [0]{\begingroup\@sanitize@url \@url }%
\providecommand \@url [1]{\endgroup\@href {#1}{\urlprefix }}%
\providecommand \urlprefix  [0]{URL }%
\providecommand \Eprint [0]{\href }%
\providecommand \doibase [0]{https://doi.org/}%
\providecommand \selectlanguage [0]{\@gobble}%
\providecommand \bibinfo  [0]{\@secondoftwo}%
\providecommand \bibfield  [0]{\@secondoftwo}%
\providecommand \translation [1]{[#1]}%
\providecommand \BibitemOpen [0]{}%
\providecommand \bibitemStop [0]{}%
\providecommand \bibitemNoStop [0]{.\EOS\space}%
\providecommand \EOS [0]{\spacefactor3000\relax}%
\providecommand \BibitemShut  [1]{\csname bibitem#1\endcsname}%
\let\auto@bib@innerbib\@empty
\bibitem [{\citenamefont {{Rajagopal}}\ and\ \citenamefont {{Romani}}(1995)}]{SMBHB_prime_1}%
  \BibitemOpen
  \bibfield  {author} {\bibinfo {author} {\bibfnamefont {M.}~\bibnamefont {{Rajagopal}}}\ and\ \bibinfo {author} {\bibfnamefont {R.~W.}\ \bibnamefont {{Romani}}},\ }\bibfield  {title} {\bibinfo {title} {{Ultra--Low-Frequency Gravitational Radiation from Massive Black Hole Binaries}},\ }\href {https://doi.org/10.1086/175813} {\bibfield  {journal} {\bibinfo  {journal} {\apj}\ }\textbf {\bibinfo {volume} {446}},\ \bibinfo {pages} {543} (\bibinfo {year} {1995})},\ \Eprint {https://arxiv.org/abs/astro-ph/9412038} {arXiv:astro-ph/9412038 [astro-ph]} \BibitemShut {NoStop}%
\bibitem [{\citenamefont {{Jaffe}}\ and\ \citenamefont {{Backer}}(2003)}]{SMBHB_prime_2}%
  \BibitemOpen
  \bibfield  {author} {\bibinfo {author} {\bibfnamefont {A.~H.}\ \bibnamefont {{Jaffe}}}\ and\ \bibinfo {author} {\bibfnamefont {D.~C.}\ \bibnamefont {{Backer}}},\ }\bibfield  {title} {\bibinfo {title} {{Gravitational Waves Probe the Coalescence Rate of Massive Black Hole Binaries}},\ }\href {https://doi.org/10.1086/345443} {\bibfield  {journal} {\bibinfo  {journal} {\apj}\ }\textbf {\bibinfo {volume} {583}},\ \bibinfo {pages} {616} (\bibinfo {year} {2003})},\ \Eprint {https://arxiv.org/abs/astro-ph/0210148} {arXiv:astro-ph/0210148 [astro-ph]} \BibitemShut {NoStop}%
\bibitem [{\citenamefont {{Wyithe}}\ and\ \citenamefont {{Loeb}}(2003)}]{SMBHB_prime_3}%
  \BibitemOpen
  \bibfield  {author} {\bibinfo {author} {\bibfnamefont {J.~S.~B.}\ \bibnamefont {{Wyithe}}}\ and\ \bibinfo {author} {\bibfnamefont {A.}~\bibnamefont {{Loeb}}},\ }\bibfield  {title} {\bibinfo {title} {{Low-Frequency Gravitational Waves from Massive Black Hole Binaries: Predictions for LISA and Pulsar Timing Arrays}},\ }\href {https://doi.org/10.1086/375187} {\bibfield  {journal} {\bibinfo  {journal} {\apj}\ }\textbf {\bibinfo {volume} {590}},\ \bibinfo {pages} {691} (\bibinfo {year} {2003})},\ \Eprint {https://arxiv.org/abs/astro-ph/0211556} {arXiv:astro-ph/0211556 [astro-ph]} \BibitemShut {NoStop}%
\bibitem [{\citenamefont {{Sesana}}\ \emph {et~al.}(2004)\citenamefont {{Sesana}}, \citenamefont {{Haardt}}, \citenamefont {{Madau}},\ and\ \citenamefont {{Volonteri}}}]{SMBHB_prime_4}%
  \BibitemOpen
  \bibfield  {author} {\bibinfo {author} {\bibfnamefont {A.}~\bibnamefont {{Sesana}}}, \bibinfo {author} {\bibfnamefont {F.}~\bibnamefont {{Haardt}}}, \bibinfo {author} {\bibfnamefont {P.}~\bibnamefont {{Madau}}},\ and\ \bibinfo {author} {\bibfnamefont {M.}~\bibnamefont {{Volonteri}}},\ }\bibfield  {title} {\bibinfo {title} {{Low-Frequency Gravitational Radiation from Coalescing Massive Black Hole Binaries in Hierarchical Cosmologies}},\ }\href {https://doi.org/10.1086/422185} {\bibfield  {journal} {\bibinfo  {journal} {\apj}\ }\textbf {\bibinfo {volume} {611}},\ \bibinfo {pages} {623} (\bibinfo {year} {2004})},\ \Eprint {https://arxiv.org/abs/astro-ph/0401543} {arXiv:astro-ph/0401543 [astro-ph]} \BibitemShut {NoStop}%
\bibitem [{\citenamefont {{Burke-Spolaor}}\ \emph {et~al.}(2019)\citenamefont {{Burke-Spolaor}}, \citenamefont {{Taylor}}, \citenamefont {{Charisi}}, \citenamefont {{Dolch}}, \citenamefont {{Hazboun}}, \citenamefont {{Holgado}}, \citenamefont {{Kelley}}, \citenamefont {{Lazio}}, \citenamefont {{Madison}}, \citenamefont {{McMann}}, \citenamefont {{Mingarelli}}, \citenamefont {{Rasskazov}}, \citenamefont {{Siemens}}, \citenamefont {{Simon}},\ and\ \citenamefont {{Smith}}}]{SMBHB_prime_5}%
  \BibitemOpen
  \bibfield  {author} {\bibinfo {author} {\bibfnamefont {S.}~\bibnamefont {{Burke-Spolaor}}}, \bibinfo {author} {\bibfnamefont {S.~R.}\ \bibnamefont {{Taylor}}}, \bibinfo {author} {\bibfnamefont {M.}~\bibnamefont {{Charisi}}}, \bibinfo {author} {\bibfnamefont {T.}~\bibnamefont {{Dolch}}}, \bibinfo {author} {\bibfnamefont {J.~S.}\ \bibnamefont {{Hazboun}}}, \bibinfo {author} {\bibfnamefont {A.~M.}\ \bibnamefont {{Holgado}}}, \bibinfo {author} {\bibfnamefont {L.~Z.}\ \bibnamefont {{Kelley}}}, \bibinfo {author} {\bibfnamefont {T.~J.~W.}\ \bibnamefont {{Lazio}}}, \bibinfo {author} {\bibfnamefont {D.~R.}\ \bibnamefont {{Madison}}}, \bibinfo {author} {\bibfnamefont {N.}~\bibnamefont {{McMann}}}, \bibinfo {author} {\bibfnamefont {C.~M.~F.}\ \bibnamefont {{Mingarelli}}}, \bibinfo {author} {\bibfnamefont {A.}~\bibnamefont {{Rasskazov}}}, \bibinfo {author} {\bibfnamefont {X.}~\bibnamefont {{Siemens}}}, \bibinfo {author} {\bibfnamefont {J.~J.}\ \bibnamefont {{Simon}}},\ and\ \bibinfo {author} {\bibfnamefont {T.~L.}\
  \bibnamefont {{Smith}}},\ }\bibfield  {title} {\bibinfo {title} {{The astrophysics of nanohertz gravitational waves}},\ }\href {https://doi.org/10.1007/s00159-019-0115-7} {\bibfield  {journal} {\bibinfo  {journal} {\aapr}\ }\textbf {\bibinfo {volume} {27}},\ \bibinfo {eid} {5} (\bibinfo {year} {2019})},\ \Eprint {https://arxiv.org/abs/1811.08826} {arXiv:1811.08826 [astro-ph.HE]} \BibitemShut {NoStop}%
\bibitem [{\citenamefont {{Sazhin}}(1978)}]{Shazin}%
  \BibitemOpen
  \bibfield  {author} {\bibinfo {author} {\bibfnamefont {M.~V.}\ \bibnamefont {{Sazhin}}},\ }\bibfield  {title} {\bibinfo {title} {{Opportunities for detecting ultralong gravitational waves}},\ }\href@noop {} {\bibfield  {journal} {\bibinfo  {journal} {Soviet Physics Journal}\ }\textbf {\bibinfo {volume} {22}},\ \bibinfo {pages} {36} (\bibinfo {year} {1978})}\BibitemShut {NoStop}%
\bibitem [{\citenamefont {{Detweiler}}(1979)}]{Detw}%
  \BibitemOpen
  \bibfield  {author} {\bibinfo {author} {\bibfnamefont {S.}~\bibnamefont {{Detweiler}}},\ }\bibfield  {title} {\bibinfo {title} {{Pulsar timing measurements and the search for gravitational waves}},\ }\href {https://doi.org/10.1086/157593} {\bibfield  {journal} {\bibinfo  {journal} {\apjl}\ }\textbf {\bibinfo {volume} {234}},\ \bibinfo {pages} {1100} (\bibinfo {year} {1979})}\BibitemShut {NoStop}%
\bibitem [{\citenamefont {{Foster}}\ and\ \citenamefont {{Backer}}(1990)}]{FosterBecker}%
  \BibitemOpen
  \bibfield  {author} {\bibinfo {author} {\bibfnamefont {R.~S.}\ \bibnamefont {{Foster}}}\ and\ \bibinfo {author} {\bibfnamefont {D.~C.}\ \bibnamefont {{Backer}}},\ }\bibfield  {title} {\bibinfo {title} {{Constructing a Pulsar Timing Array}},\ }\href {https://doi.org/10.1086/169195} {\bibfield  {journal} {\bibinfo  {journal} {\apj}\ }\textbf {\bibinfo {volume} {361}},\ \bibinfo {pages} {300} (\bibinfo {year} {1990})}\BibitemShut {NoStop}%
\bibitem [{\citenamefont {Taylor}(2021)}]{stevebook}%
  \BibitemOpen
  \bibfield  {author} {\bibinfo {author} {\bibfnamefont {S.~R.}\ \bibnamefont {Taylor}},\ }\href@noop {} {\emph {\bibinfo {title} {Nanohertz gravitational wave astronomy}}}\ (\bibinfo  {publisher} {CRC Press},\ \bibinfo {year} {2021})\BibitemShut {NoStop}%
\bibitem [{\citenamefont {{Agazie}}\ \emph {et~al.}(2023{\natexlab{a}})\citenamefont {{Agazie}}, \citenamefont {{Anumarlapudi}}, \citenamefont {{Archibald}}, \citenamefont {{Arzoumanian}}, \citenamefont {{Baker}}, \citenamefont {{B{\'e}csy}}, \citenamefont {{Blecha}}, \citenamefont {{Brazier}}, \citenamefont {{Brook}}, \citenamefont {{Burke-Spolaor}}, \citenamefont {{Burnette}}, \citenamefont {{Case}}, \citenamefont {{Charisi}}, \citenamefont {{Chatterjee}}, \citenamefont {{Chatziioannou}}, \citenamefont {{Cheeseboro}}, \citenamefont {{Chen}}, \citenamefont {{Cohen}}, \citenamefont {{Cordes}}, \citenamefont {{Cornish}}, \citenamefont {{Crawford}}, \citenamefont {{Cromartie}}, \citenamefont {{Crowter}}, \citenamefont {{Cutler}}, \citenamefont {{Decesar}}, \citenamefont {{Degan}}, \citenamefont {{Demorest}}, \citenamefont {{Deng}}, \citenamefont {{Dolch}}, \citenamefont {{Drachler}}, \citenamefont {{Ellis}}, \citenamefont {{Ferrara}}, \citenamefont {{Fiore}}, \citenamefont {{Fonseca}}, \citenamefont
  {{Freedman}}, \citenamefont {{Garver-Daniels}}, \citenamefont {{Gentile}}, \citenamefont {{Gersbach}}, \citenamefont {{Glaser}}, \citenamefont {{Good}}, \citenamefont {{G{\"u}ltekin}}, \citenamefont {{Hazboun}}, \citenamefont {{Hourihane}}, \citenamefont {{Islo}}, \citenamefont {{Jennings}}, \citenamefont {{Johnson}}, \citenamefont {{Jones}}, \citenamefont {{Kaiser}}, \citenamefont {{Kaplan}}, \citenamefont {{Kelley}}, \citenamefont {{Kerr}}, \citenamefont {{Key}}, \citenamefont {{Klein}}, \citenamefont {{Laal}}, \citenamefont {{Lam}}, \citenamefont {{Lamb}}, \citenamefont {{Lazio}}, \citenamefont {{Lewandowska}}, \citenamefont {{Littenberg}}, \citenamefont {{Liu}}, \citenamefont {{Lommen}}, \citenamefont {{Lorimer}}, \citenamefont {{Luo}}, \citenamefont {{Lynch}}, \citenamefont {{Ma}}, \citenamefont {{Madison}}, \citenamefont {{Mattson}}, \citenamefont {{McEwen}}, \citenamefont {{McKee}}, \citenamefont {{McLaughlin}}, \citenamefont {{McMann}}, \citenamefont {{Meyers}}, \citenamefont {{Meyers}},
  \citenamefont {{Mingarelli}}, \citenamefont {{Mitridate}}, \citenamefont {{Natarajan}}, \citenamefont {{Ng}}, \citenamefont {{Nice}}, \citenamefont {{Ocker}}, \citenamefont {{Olum}}, \citenamefont {{Pennucci}}, \citenamefont {{Perera}}, \citenamefont {{Petrov}}, \citenamefont {{Pol}}, \citenamefont {{Radovan}}, \citenamefont {{Ransom}}, \citenamefont {{Ray}}, \citenamefont {{Romano}}, \citenamefont {{Sardesai}}, \citenamefont {{Schmiedekamp}}, \citenamefont {{Schmiedekamp}}, \citenamefont {{Schmitz}}, \citenamefont {{Schult}}, \citenamefont {{Shapiro-Albert}}, \citenamefont {{Siemens}}, \citenamefont {{Simon}}, \citenamefont {{Siwek}}, \citenamefont {{Stairs}}, \citenamefont {{Stinebring}}, \citenamefont {{Stovall}}, \citenamefont {{Sun}}, \citenamefont {{Susobhanan}}, \citenamefont {{Swiggum}}, \citenamefont {{Taylor}}, \citenamefont {{Taylor}}, \citenamefont {{Turner}}, \citenamefont {{Unal}}, \citenamefont {{Vallisneri}}, \citenamefont {{van Haasteren}}, \citenamefont {{Vigeland}}, \citenamefont
  {{Wahl}}, \citenamefont {{Wang}}, \citenamefont {{Witt}}, \citenamefont {{Young}},\ and\ \citenamefont {{Nanograv Collaboration}}}]{15yr}%
  \BibitemOpen
  \bibfield  {author} {\bibinfo {author} {\bibfnamefont {G.}~\bibnamefont {{Agazie}}}, \bibinfo {author} {\bibfnamefont {A.}~\bibnamefont {{Anumarlapudi}}}, \bibinfo {author} {\bibfnamefont {A.~M.}\ \bibnamefont {{Archibald}}}, \bibinfo {author} {\bibfnamefont {Z.}~\bibnamefont {{Arzoumanian}}}, \bibinfo {author} {\bibfnamefont {P.~T.}\ \bibnamefont {{Baker}}}, \bibinfo {author} {\bibfnamefont {B.}~\bibnamefont {{B{\'e}csy}}}, \bibinfo {author} {\bibfnamefont {L.}~\bibnamefont {{Blecha}}}, \bibinfo {author} {\bibfnamefont {A.}~\bibnamefont {{Brazier}}}, \bibinfo {author} {\bibfnamefont {P.~R.}\ \bibnamefont {{Brook}}}, \bibinfo {author} {\bibfnamefont {S.}~\bibnamefont {{Burke-Spolaor}}}, \bibinfo {author} {\bibfnamefont {R.}~\bibnamefont {{Burnette}}}, \bibinfo {author} {\bibfnamefont {R.}~\bibnamefont {{Case}}}, \bibinfo {author} {\bibfnamefont {M.}~\bibnamefont {{Charisi}}}, \bibinfo {author} {\bibfnamefont {S.}~\bibnamefont {{Chatterjee}}}, \bibinfo {author} {\bibfnamefont {K.}~\bibnamefont
  {{Chatziioannou}}}, \bibinfo {author} {\bibfnamefont {B.~D.}\ \bibnamefont {{Cheeseboro}}}, \bibinfo {author} {\bibfnamefont {S.}~\bibnamefont {{Chen}}}, \bibinfo {author} {\bibfnamefont {T.}~\bibnamefont {{Cohen}}}, \bibinfo {author} {\bibfnamefont {J.~M.}\ \bibnamefont {{Cordes}}}, \bibinfo {author} {\bibfnamefont {N.~J.}\ \bibnamefont {{Cornish}}}, \bibinfo {author} {\bibfnamefont {F.}~\bibnamefont {{Crawford}}}, \bibinfo {author} {\bibfnamefont {H.~T.}\ \bibnamefont {{Cromartie}}}, \bibinfo {author} {\bibfnamefont {K.}~\bibnamefont {{Crowter}}}, \bibinfo {author} {\bibfnamefont {C.~J.}\ \bibnamefont {{Cutler}}}, \bibinfo {author} {\bibfnamefont {M.~E.}\ \bibnamefont {{Decesar}}}, \bibinfo {author} {\bibfnamefont {D.}~\bibnamefont {{Degan}}}, \bibinfo {author} {\bibfnamefont {P.~B.}\ \bibnamefont {{Demorest}}}, \bibinfo {author} {\bibfnamefont {H.}~\bibnamefont {{Deng}}}, \bibinfo {author} {\bibfnamefont {T.}~\bibnamefont {{Dolch}}}, \bibinfo {author} {\bibfnamefont {B.}~\bibnamefont {{Drachler}}},
  \bibinfo {author} {\bibfnamefont {J.~A.}\ \bibnamefont {{Ellis}}}, \bibinfo {author} {\bibfnamefont {E.~C.}\ \bibnamefont {{Ferrara}}}, \bibinfo {author} {\bibfnamefont {W.}~\bibnamefont {{Fiore}}}, \bibinfo {author} {\bibfnamefont {E.}~\bibnamefont {{Fonseca}}}, \bibinfo {author} {\bibfnamefont {G.~E.}\ \bibnamefont {{Freedman}}}, \bibinfo {author} {\bibfnamefont {N.}~\bibnamefont {{Garver-Daniels}}}, \bibinfo {author} {\bibfnamefont {P.~A.}\ \bibnamefont {{Gentile}}}, \bibinfo {author} {\bibfnamefont {K.~A.}\ \bibnamefont {{Gersbach}}}, \bibinfo {author} {\bibfnamefont {J.}~\bibnamefont {{Glaser}}}, \bibinfo {author} {\bibfnamefont {D.~C.}\ \bibnamefont {{Good}}}, \bibinfo {author} {\bibfnamefont {K.}~\bibnamefont {{G{\"u}ltekin}}}, \bibinfo {author} {\bibfnamefont {J.~S.}\ \bibnamefont {{Hazboun}}}, \bibinfo {author} {\bibfnamefont {S.}~\bibnamefont {{Hourihane}}}, \bibinfo {author} {\bibfnamefont {K.}~\bibnamefont {{Islo}}}, \bibinfo {author} {\bibfnamefont {R.~J.}\ \bibnamefont {{Jennings}}}, \bibinfo
  {author} {\bibfnamefont {A.~D.}\ \bibnamefont {{Johnson}}}, \bibinfo {author} {\bibfnamefont {M.~L.}\ \bibnamefont {{Jones}}}, \bibinfo {author} {\bibfnamefont {A.~R.}\ \bibnamefont {{Kaiser}}}, \bibinfo {author} {\bibfnamefont {D.~L.}\ \bibnamefont {{Kaplan}}}, \bibinfo {author} {\bibfnamefont {L.~Z.}\ \bibnamefont {{Kelley}}}, \bibinfo {author} {\bibfnamefont {M.}~\bibnamefont {{Kerr}}}, \bibinfo {author} {\bibfnamefont {J.~S.}\ \bibnamefont {{Key}}}, \bibinfo {author} {\bibfnamefont {T.~C.}\ \bibnamefont {{Klein}}}, \bibinfo {author} {\bibfnamefont {N.}~\bibnamefont {{Laal}}}, \bibinfo {author} {\bibfnamefont {M.~T.}\ \bibnamefont {{Lam}}}, \bibinfo {author} {\bibfnamefont {W.~G.}\ \bibnamefont {{Lamb}}}, \bibinfo {author} {\bibfnamefont {T.~J.~W.}\ \bibnamefont {{Lazio}}}, \bibinfo {author} {\bibfnamefont {N.}~\bibnamefont {{Lewandowska}}}, \bibinfo {author} {\bibfnamefont {T.~B.}\ \bibnamefont {{Littenberg}}}, \bibinfo {author} {\bibfnamefont {T.}~\bibnamefont {{Liu}}}, \bibinfo {author} {\bibfnamefont
  {A.}~\bibnamefont {{Lommen}}}, \bibinfo {author} {\bibfnamefont {D.~R.}\ \bibnamefont {{Lorimer}}}, \bibinfo {author} {\bibfnamefont {J.}~\bibnamefont {{Luo}}}, \bibinfo {author} {\bibfnamefont {R.~S.}\ \bibnamefont {{Lynch}}}, \bibinfo {author} {\bibfnamefont {C.-P.}\ \bibnamefont {{Ma}}}, \bibinfo {author} {\bibfnamefont {D.~R.}\ \bibnamefont {{Madison}}}, \bibinfo {author} {\bibfnamefont {M.~A.}\ \bibnamefont {{Mattson}}}, \bibinfo {author} {\bibfnamefont {A.}~\bibnamefont {{McEwen}}}, \bibinfo {author} {\bibfnamefont {J.~W.}\ \bibnamefont {{McKee}}}, \bibinfo {author} {\bibfnamefont {M.~A.}\ \bibnamefont {{McLaughlin}}}, \bibinfo {author} {\bibfnamefont {N.}~\bibnamefont {{McMann}}}, \bibinfo {author} {\bibfnamefont {B.~W.}\ \bibnamefont {{Meyers}}}, \bibinfo {author} {\bibfnamefont {P.~M.}\ \bibnamefont {{Meyers}}}, \bibinfo {author} {\bibfnamefont {C.~M.~F.}\ \bibnamefont {{Mingarelli}}}, \bibinfo {author} {\bibfnamefont {A.}~\bibnamefont {{Mitridate}}}, \bibinfo {author} {\bibfnamefont
  {P.}~\bibnamefont {{Natarajan}}}, \bibinfo {author} {\bibfnamefont {C.}~\bibnamefont {{Ng}}}, \bibinfo {author} {\bibfnamefont {D.~J.}\ \bibnamefont {{Nice}}}, \bibinfo {author} {\bibfnamefont {S.~K.}\ \bibnamefont {{Ocker}}}, \bibinfo {author} {\bibfnamefont {K.~D.}\ \bibnamefont {{Olum}}}, \bibinfo {author} {\bibfnamefont {T.~T.}\ \bibnamefont {{Pennucci}}}, \bibinfo {author} {\bibfnamefont {B.~B.~P.}\ \bibnamefont {{Perera}}}, \bibinfo {author} {\bibfnamefont {P.}~\bibnamefont {{Petrov}}}, \bibinfo {author} {\bibfnamefont {N.~S.}\ \bibnamefont {{Pol}}}, \bibinfo {author} {\bibfnamefont {H.~A.}\ \bibnamefont {{Radovan}}}, \bibinfo {author} {\bibfnamefont {S.~M.}\ \bibnamefont {{Ransom}}}, \bibinfo {author} {\bibfnamefont {P.~S.}\ \bibnamefont {{Ray}}}, \bibinfo {author} {\bibfnamefont {J.~D.}\ \bibnamefont {{Romano}}}, \bibinfo {author} {\bibfnamefont {S.~C.}\ \bibnamefont {{Sardesai}}}, \bibinfo {author} {\bibfnamefont {A.}~\bibnamefont {{Schmiedekamp}}}, \bibinfo {author} {\bibfnamefont
  {C.}~\bibnamefont {{Schmiedekamp}}}, \bibinfo {author} {\bibfnamefont {K.}~\bibnamefont {{Schmitz}}}, \bibinfo {author} {\bibfnamefont {L.}~\bibnamefont {{Schult}}}, \bibinfo {author} {\bibfnamefont {B.~J.}\ \bibnamefont {{Shapiro-Albert}}}, \bibinfo {author} {\bibfnamefont {X.}~\bibnamefont {{Siemens}}}, \bibinfo {author} {\bibfnamefont {J.}~\bibnamefont {{Simon}}}, \bibinfo {author} {\bibfnamefont {M.~S.}\ \bibnamefont {{Siwek}}}, \bibinfo {author} {\bibfnamefont {I.~H.}\ \bibnamefont {{Stairs}}}, \bibinfo {author} {\bibfnamefont {D.~R.}\ \bibnamefont {{Stinebring}}}, \bibinfo {author} {\bibfnamefont {K.}~\bibnamefont {{Stovall}}}, \bibinfo {author} {\bibfnamefont {J.~P.}\ \bibnamefont {{Sun}}}, \bibinfo {author} {\bibfnamefont {A.}~\bibnamefont {{Susobhanan}}}, \bibinfo {author} {\bibfnamefont {J.~K.}\ \bibnamefont {{Swiggum}}}, \bibinfo {author} {\bibfnamefont {J.}~\bibnamefont {{Taylor}}}, \bibinfo {author} {\bibfnamefont {S.~R.}\ \bibnamefont {{Taylor}}}, \bibinfo {author} {\bibfnamefont {J.~E.}\
  \bibnamefont {{Turner}}}, \bibinfo {author} {\bibfnamefont {C.}~\bibnamefont {{Unal}}}, \bibinfo {author} {\bibfnamefont {M.}~\bibnamefont {{Vallisneri}}}, \bibinfo {author} {\bibfnamefont {R.}~\bibnamefont {{van Haasteren}}}, \bibinfo {author} {\bibfnamefont {S.~J.}\ \bibnamefont {{Vigeland}}}, \bibinfo {author} {\bibfnamefont {H.~M.}\ \bibnamefont {{Wahl}}}, \bibinfo {author} {\bibfnamefont {Q.}~\bibnamefont {{Wang}}}, \bibinfo {author} {\bibfnamefont {C.~A.}\ \bibnamefont {{Witt}}}, \bibinfo {author} {\bibfnamefont {O.}~\bibnamefont {{Young}}},\ and\ \bibinfo {author} {\bibnamefont {{Nanograv Collaboration}}},\ }\bibfield  {title} {\bibinfo {title} {{The NANOGrav 15 yr Data Set: Evidence for a Gravitational-wave Background}},\ }\href {https://doi.org/10.3847/2041-8213/acdac6} {\bibfield  {journal} {\bibinfo  {journal} {\apjl}\ }\textbf {\bibinfo {volume} {951}},\ \bibinfo {eid} {L8} (\bibinfo {year} {2023}{\natexlab{a}})},\ \Eprint {https://arxiv.org/abs/2306.16213} {arXiv:2306.16213 [astro-ph.HE]}
  \BibitemShut {NoStop}%
\bibitem [{\citenamefont {{Antoniadis}}\ \emph {et~al.}(2023)\citenamefont {{Antoniadis}}, \citenamefont {{Arumugam}}, \citenamefont {{Arumugam}}, \citenamefont {{Babak}}, \citenamefont {{Bagchi}}, \citenamefont {{Bak Nielsen}}, \citenamefont {{Bassa}}, \citenamefont {{Bathula}}, \citenamefont {{Berthereau}}, \citenamefont {{Bonetti}}, \citenamefont {{Bortolas}}, \citenamefont {{Brook}}, \citenamefont {{Burgay}}, \citenamefont {{Caballero}}, \citenamefont {{Chalumeau}}, \citenamefont {{Champion}}, \citenamefont {{Chanlaridis}}, \citenamefont {{Chen}}, \citenamefont {{Cognard}}, \citenamefont {{Dandapat}}, \citenamefont {{Deb}}, \citenamefont {{Desai}}, \citenamefont {{Desvignes}}, \citenamefont {{Dhanda-Batra}}, \citenamefont {{Dwivedi}}, \citenamefont {{Falxa}}, \citenamefont {{Ferdman}}, \citenamefont {{Franchini}}, \citenamefont {{Gair}}, \citenamefont {{Goncharov}}, \citenamefont {{Gopakumar}}, \citenamefont {{Graikou}}, \citenamefont {{Grie{\ss}meier}}, \citenamefont {{Guillemot}}, \citenamefont {{Guo}},
  \citenamefont {{Gupta}}, \citenamefont {{Hisano}}, \citenamefont {{Hu}}, \citenamefont {{Iraci}}, \citenamefont {{Izquierdo-Villalba}}, \citenamefont {{Jang}}, \citenamefont {{Jawor}}, \citenamefont {{Janssen}}, \citenamefont {{Jessner}}, \citenamefont {{Joshi}}, \citenamefont {{Kareem}}, \citenamefont {{Karuppusamy}}, \citenamefont {{Keane}}, \citenamefont {{Keith}}, \citenamefont {{Kharbanda}}, \citenamefont {{Kikunaga}}, \citenamefont {{Kolhe}}, \citenamefont {{Kramer}}, \citenamefont {{Krishnakumar}}, \citenamefont {{Lackeos}}, \citenamefont {{Lee}}, \citenamefont {{Liu}}, \citenamefont {{Liu}}, \citenamefont {{Lyne}}, \citenamefont {{McKee}}, \citenamefont {{Maan}}, \citenamefont {{Main}}, \citenamefont {{Mickaliger}}, \citenamefont {{Nitu}}, \citenamefont {{Nobleson}}, \citenamefont {{Paladi}}, \citenamefont {{Parthasarathy}}, \citenamefont {{Perera}}, \citenamefont {{Perrodin}}, \citenamefont {{Petiteau}}, \citenamefont {{Porayko}}, \citenamefont {{Possenti}}, \citenamefont {{Prabu}}, \citenamefont
  {{Quelquejay Leclere}}, \citenamefont {{Rana}}, \citenamefont {{Samajdar}}, \citenamefont {{Sanidas}}, \citenamefont {{Sesana}}, \citenamefont {{Shaifullah}}, \citenamefont {{Singha}}, \citenamefont {{Speri}}, \citenamefont {{Spiewak}}, \citenamefont {{Srivastava}}, \citenamefont {{Stappers}}, \citenamefont {{Surnis}}, \citenamefont {{Susarla}}, \citenamefont {{Susobhanan}}, \citenamefont {{Takahashi}}, \citenamefont {{Tarafdar}}, \citenamefont {{Theureau}}, \citenamefont {{Tiburzi}}, \citenamefont {{van der Wateren}}, \citenamefont {{Vecchio}}, \citenamefont {{Venkatraman Krishnan}}, \citenamefont {{Verbiest}}, \citenamefont {{Wang}}, \citenamefont {{Wang}},\ and\ \citenamefont {{Wu}}}]{d2}%
  \BibitemOpen
  \bibfield  {author} {\bibinfo {author} {\bibfnamefont {J.}~\bibnamefont {{Antoniadis}}}, \bibinfo {author} {\bibfnamefont {P.}~\bibnamefont {{Arumugam}}}, \bibinfo {author} {\bibfnamefont {S.}~\bibnamefont {{Arumugam}}}, \bibinfo {author} {\bibfnamefont {S.}~\bibnamefont {{Babak}}}, \bibinfo {author} {\bibfnamefont {M.}~\bibnamefont {{Bagchi}}}, \bibinfo {author} {\bibfnamefont {A.~S.}\ \bibnamefont {{Bak Nielsen}}}, \bibinfo {author} {\bibfnamefont {C.~G.}\ \bibnamefont {{Bassa}}}, \bibinfo {author} {\bibfnamefont {A.}~\bibnamefont {{Bathula}}}, \bibinfo {author} {\bibfnamefont {A.}~\bibnamefont {{Berthereau}}}, \bibinfo {author} {\bibfnamefont {M.}~\bibnamefont {{Bonetti}}}, \bibinfo {author} {\bibfnamefont {E.}~\bibnamefont {{Bortolas}}}, \bibinfo {author} {\bibfnamefont {P.~R.}\ \bibnamefont {{Brook}}}, \bibinfo {author} {\bibfnamefont {M.}~\bibnamefont {{Burgay}}}, \bibinfo {author} {\bibfnamefont {R.~N.}\ \bibnamefont {{Caballero}}}, \bibinfo {author} {\bibfnamefont {A.}~\bibnamefont {{Chalumeau}}},
  \bibinfo {author} {\bibfnamefont {D.~J.}\ \bibnamefont {{Champion}}}, \bibinfo {author} {\bibfnamefont {S.}~\bibnamefont {{Chanlaridis}}}, \bibinfo {author} {\bibfnamefont {S.}~\bibnamefont {{Chen}}}, \bibinfo {author} {\bibfnamefont {I.}~\bibnamefont {{Cognard}}}, \bibinfo {author} {\bibfnamefont {S.}~\bibnamefont {{Dandapat}}}, \bibinfo {author} {\bibfnamefont {D.}~\bibnamefont {{Deb}}}, \bibinfo {author} {\bibfnamefont {S.}~\bibnamefont {{Desai}}}, \bibinfo {author} {\bibfnamefont {G.}~\bibnamefont {{Desvignes}}}, \bibinfo {author} {\bibfnamefont {N.}~\bibnamefont {{Dhanda-Batra}}}, \bibinfo {author} {\bibfnamefont {C.}~\bibnamefont {{Dwivedi}}}, \bibinfo {author} {\bibfnamefont {M.}~\bibnamefont {{Falxa}}}, \bibinfo {author} {\bibfnamefont {R.~D.}\ \bibnamefont {{Ferdman}}}, \bibinfo {author} {\bibfnamefont {A.}~\bibnamefont {{Franchini}}}, \bibinfo {author} {\bibfnamefont {J.~R.}\ \bibnamefont {{Gair}}}, \bibinfo {author} {\bibfnamefont {B.}~\bibnamefont {{Goncharov}}}, \bibinfo {author} {\bibfnamefont
  {A.}~\bibnamefont {{Gopakumar}}}, \bibinfo {author} {\bibfnamefont {E.}~\bibnamefont {{Graikou}}}, \bibinfo {author} {\bibfnamefont {J.~M.}\ \bibnamefont {{Grie{\ss}meier}}}, \bibinfo {author} {\bibfnamefont {L.}~\bibnamefont {{Guillemot}}}, \bibinfo {author} {\bibfnamefont {Y.~J.}\ \bibnamefont {{Guo}}}, \bibinfo {author} {\bibfnamefont {Y.}~\bibnamefont {{Gupta}}}, \bibinfo {author} {\bibfnamefont {S.}~\bibnamefont {{Hisano}}}, \bibinfo {author} {\bibfnamefont {H.}~\bibnamefont {{Hu}}}, \bibinfo {author} {\bibfnamefont {F.}~\bibnamefont {{Iraci}}}, \bibinfo {author} {\bibfnamefont {D.}~\bibnamefont {{Izquierdo-Villalba}}}, \bibinfo {author} {\bibfnamefont {J.}~\bibnamefont {{Jang}}}, \bibinfo {author} {\bibfnamefont {J.}~\bibnamefont {{Jawor}}}, \bibinfo {author} {\bibfnamefont {G.~H.}\ \bibnamefont {{Janssen}}}, \bibinfo {author} {\bibfnamefont {A.}~\bibnamefont {{Jessner}}}, \bibinfo {author} {\bibfnamefont {B.~C.}\ \bibnamefont {{Joshi}}}, \bibinfo {author} {\bibfnamefont {F.}~\bibnamefont {{Kareem}}},
  \bibinfo {author} {\bibfnamefont {R.}~\bibnamefont {{Karuppusamy}}}, \bibinfo {author} {\bibfnamefont {E.~F.}\ \bibnamefont {{Keane}}}, \bibinfo {author} {\bibfnamefont {M.~J.}\ \bibnamefont {{Keith}}}, \bibinfo {author} {\bibfnamefont {D.}~\bibnamefont {{Kharbanda}}}, \bibinfo {author} {\bibfnamefont {T.}~\bibnamefont {{Kikunaga}}}, \bibinfo {author} {\bibfnamefont {N.}~\bibnamefont {{Kolhe}}}, \bibinfo {author} {\bibfnamefont {M.}~\bibnamefont {{Kramer}}}, \bibinfo {author} {\bibfnamefont {M.~A.}\ \bibnamefont {{Krishnakumar}}}, \bibinfo {author} {\bibfnamefont {K.}~\bibnamefont {{Lackeos}}}, \bibinfo {author} {\bibfnamefont {K.~J.}\ \bibnamefont {{Lee}}}, \bibinfo {author} {\bibfnamefont {K.}~\bibnamefont {{Liu}}}, \bibinfo {author} {\bibfnamefont {Y.}~\bibnamefont {{Liu}}}, \bibinfo {author} {\bibfnamefont {A.~G.}\ \bibnamefont {{Lyne}}}, \bibinfo {author} {\bibfnamefont {J.~W.}\ \bibnamefont {{McKee}}}, \bibinfo {author} {\bibfnamefont {Y.}~\bibnamefont {{Maan}}}, \bibinfo {author} {\bibfnamefont
  {R.~A.}\ \bibnamefont {{Main}}}, \bibinfo {author} {\bibfnamefont {M.~B.}\ \bibnamefont {{Mickaliger}}}, \bibinfo {author} {\bibfnamefont {I.~C.}\ \bibnamefont {{Nitu}}}, \bibinfo {author} {\bibfnamefont {K.}~\bibnamefont {{Nobleson}}}, \bibinfo {author} {\bibfnamefont {A.~K.}\ \bibnamefont {{Paladi}}}, \bibinfo {author} {\bibfnamefont {A.}~\bibnamefont {{Parthasarathy}}}, \bibinfo {author} {\bibfnamefont {B.~B.~P.}\ \bibnamefont {{Perera}}}, \bibinfo {author} {\bibfnamefont {D.}~\bibnamefont {{Perrodin}}}, \bibinfo {author} {\bibfnamefont {A.}~\bibnamefont {{Petiteau}}}, \bibinfo {author} {\bibfnamefont {N.~K.}\ \bibnamefont {{Porayko}}}, \bibinfo {author} {\bibfnamefont {A.}~\bibnamefont {{Possenti}}}, \bibinfo {author} {\bibfnamefont {T.}~\bibnamefont {{Prabu}}}, \bibinfo {author} {\bibfnamefont {H.}~\bibnamefont {{Quelquejay Leclere}}}, \bibinfo {author} {\bibfnamefont {P.}~\bibnamefont {{Rana}}}, \bibinfo {author} {\bibfnamefont {A.}~\bibnamefont {{Samajdar}}}, \bibinfo {author} {\bibfnamefont {S.~A.}\
  \bibnamefont {{Sanidas}}}, \bibinfo {author} {\bibfnamefont {A.}~\bibnamefont {{Sesana}}}, \bibinfo {author} {\bibfnamefont {G.}~\bibnamefont {{Shaifullah}}}, \bibinfo {author} {\bibfnamefont {J.}~\bibnamefont {{Singha}}}, \bibinfo {author} {\bibfnamefont {L.}~\bibnamefont {{Speri}}}, \bibinfo {author} {\bibfnamefont {R.}~\bibnamefont {{Spiewak}}}, \bibinfo {author} {\bibfnamefont {A.}~\bibnamefont {{Srivastava}}}, \bibinfo {author} {\bibfnamefont {B.~W.}\ \bibnamefont {{Stappers}}}, \bibinfo {author} {\bibfnamefont {M.}~\bibnamefont {{Surnis}}}, \bibinfo {author} {\bibfnamefont {S.~C.}\ \bibnamefont {{Susarla}}}, \bibinfo {author} {\bibfnamefont {A.}~\bibnamefont {{Susobhanan}}}, \bibinfo {author} {\bibfnamefont {K.}~\bibnamefont {{Takahashi}}}, \bibinfo {author} {\bibfnamefont {P.}~\bibnamefont {{Tarafdar}}}, \bibinfo {author} {\bibfnamefont {G.}~\bibnamefont {{Theureau}}}, \bibinfo {author} {\bibfnamefont {C.}~\bibnamefont {{Tiburzi}}}, \bibinfo {author} {\bibfnamefont {E.}~\bibnamefont {{van der
  Wateren}}}, \bibinfo {author} {\bibfnamefont {A.}~\bibnamefont {{Vecchio}}}, \bibinfo {author} {\bibfnamefont {V.}~\bibnamefont {{Venkatraman Krishnan}}}, \bibinfo {author} {\bibfnamefont {J.~P.~W.}\ \bibnamefont {{Verbiest}}}, \bibinfo {author} {\bibfnamefont {J.}~\bibnamefont {{Wang}}}, \bibinfo {author} {\bibfnamefont {L.}~\bibnamefont {{Wang}}},\ and\ \bibinfo {author} {\bibfnamefont {Z.}~\bibnamefont {{Wu}}},\ }\bibfield  {title} {\bibinfo {title} {{The second data release from the European Pulsar Timing Array III. Search for gravitational wave signals}},\ }\href {https://doi.org/10.48550/arXiv.2306.16214} {\bibfield  {journal} {\bibinfo  {journal} {arXiv e-prints}\ ,\ \bibinfo {eid} {arXiv:2306.16214}} (\bibinfo {year} {2023})},\ \Eprint {https://arxiv.org/abs/2306.16214} {arXiv:2306.16214 [astro-ph.HE]} \BibitemShut {NoStop}%
\bibitem [{\citenamefont {{Reardon}}\ \emph {et~al.}(2023)\citenamefont {{Reardon}}, \citenamefont {{Zic}}, \citenamefont {{Shannon}}, \citenamefont {{Hobbs}}, \citenamefont {{Bailes}}, \citenamefont {{Di Marco}}, \citenamefont {{Kapur}}, \citenamefont {{Rogers}}, \citenamefont {{Thrane}}, \citenamefont {{Askew}}, \citenamefont {{Bhat}}, \citenamefont {{Cameron}}, \citenamefont {{Cury{\l}o}}, \citenamefont {{Coles}}, \citenamefont {{Dai}}, \citenamefont {{Goncharov}}, \citenamefont {{Kerr}}, \citenamefont {{Kulkarni}}, \citenamefont {{Levin}}, \citenamefont {{Lower}}, \citenamefont {{Manchester}}, \citenamefont {{Mandow}}, \citenamefont {{Miles}}, \citenamefont {{Nathan}}, \citenamefont {{Os{\l}owski}}, \citenamefont {{Russell}}, \citenamefont {{Spiewak}}, \citenamefont {{Zhang}},\ and\ \citenamefont {{Zhu}}}]{d3}%
  \BibitemOpen
  \bibfield  {author} {\bibinfo {author} {\bibfnamefont {D.~J.}\ \bibnamefont {{Reardon}}}, \bibinfo {author} {\bibfnamefont {A.}~\bibnamefont {{Zic}}}, \bibinfo {author} {\bibfnamefont {R.~M.}\ \bibnamefont {{Shannon}}}, \bibinfo {author} {\bibfnamefont {G.~B.}\ \bibnamefont {{Hobbs}}}, \bibinfo {author} {\bibfnamefont {M.}~\bibnamefont {{Bailes}}}, \bibinfo {author} {\bibfnamefont {V.}~\bibnamefont {{Di Marco}}}, \bibinfo {author} {\bibfnamefont {A.}~\bibnamefont {{Kapur}}}, \bibinfo {author} {\bibfnamefont {A.~F.}\ \bibnamefont {{Rogers}}}, \bibinfo {author} {\bibfnamefont {E.}~\bibnamefont {{Thrane}}}, \bibinfo {author} {\bibfnamefont {J.}~\bibnamefont {{Askew}}}, \bibinfo {author} {\bibfnamefont {N.~D.~R.}\ \bibnamefont {{Bhat}}}, \bibinfo {author} {\bibfnamefont {A.}~\bibnamefont {{Cameron}}}, \bibinfo {author} {\bibfnamefont {M.}~\bibnamefont {{Cury{\l}o}}}, \bibinfo {author} {\bibfnamefont {W.~A.}\ \bibnamefont {{Coles}}}, \bibinfo {author} {\bibfnamefont {S.}~\bibnamefont {{Dai}}}, \bibinfo {author}
  {\bibfnamefont {B.}~\bibnamefont {{Goncharov}}}, \bibinfo {author} {\bibfnamefont {M.}~\bibnamefont {{Kerr}}}, \bibinfo {author} {\bibfnamefont {A.}~\bibnamefont {{Kulkarni}}}, \bibinfo {author} {\bibfnamefont {Y.}~\bibnamefont {{Levin}}}, \bibinfo {author} {\bibfnamefont {M.~E.}\ \bibnamefont {{Lower}}}, \bibinfo {author} {\bibfnamefont {R.~N.}\ \bibnamefont {{Manchester}}}, \bibinfo {author} {\bibfnamefont {R.}~\bibnamefont {{Mandow}}}, \bibinfo {author} {\bibfnamefont {M.~T.}\ \bibnamefont {{Miles}}}, \bibinfo {author} {\bibfnamefont {R.~S.}\ \bibnamefont {{Nathan}}}, \bibinfo {author} {\bibfnamefont {S.}~\bibnamefont {{Os{\l}owski}}}, \bibinfo {author} {\bibfnamefont {C.~J.}\ \bibnamefont {{Russell}}}, \bibinfo {author} {\bibfnamefont {R.}~\bibnamefont {{Spiewak}}}, \bibinfo {author} {\bibfnamefont {S.}~\bibnamefont {{Zhang}}},\ and\ \bibinfo {author} {\bibfnamefont {X.-J.}\ \bibnamefont {{Zhu}}},\ }\bibfield  {title} {\bibinfo {title} {{Search for an Isotropic Gravitational-wave Background with the
  Parkes Pulsar Timing Array}},\ }\href {https://doi.org/10.3847/2041-8213/acdd02} {\bibfield  {journal} {\bibinfo  {journal} {\apjl}\ }\textbf {\bibinfo {volume} {951}},\ \bibinfo {eid} {L6} (\bibinfo {year} {2023})},\ \Eprint {https://arxiv.org/abs/2306.16215} {arXiv:2306.16215 [astro-ph.HE]} \BibitemShut {NoStop}%
\bibitem [{\citenamefont {{Hellings}}\ and\ \citenamefont {{Downs}}(1983)}]{hd83}%
  \BibitemOpen
  \bibfield  {author} {\bibinfo {author} {\bibfnamefont {R.~W.}\ \bibnamefont {{Hellings}}}\ and\ \bibinfo {author} {\bibfnamefont {G.~S.}\ \bibnamefont {{Downs}}},\ }\bibfield  {title} {\bibinfo {title} {{Upper limits on the isotropic gravitational radiation background from pulsar timing analysis}},\ }\href {https://doi.org/10.1086/183954} {\bibfield  {journal} {\bibinfo  {journal} {\apjl}\ }\textbf {\bibinfo {volume} {265}},\ \bibinfo {pages} {L39} (\bibinfo {year} {1983})}\BibitemShut {NoStop}%
\bibitem [{\citenamefont {{Bi}}\ \emph {et~al.}(2023)\citenamefont {{Bi}}, \citenamefont {{Wu}}, \citenamefont {{Chen}},\ and\ \citenamefont {{Huang}}}]{SMBHB-NANO-1}%
  \BibitemOpen
  \bibfield  {author} {\bibinfo {author} {\bibfnamefont {Y.-C.}\ \bibnamefont {{Bi}}}, \bibinfo {author} {\bibfnamefont {Y.-M.}\ \bibnamefont {{Wu}}}, \bibinfo {author} {\bibfnamefont {Z.-C.}\ \bibnamefont {{Chen}}},\ and\ \bibinfo {author} {\bibfnamefont {Q.-G.}\ \bibnamefont {{Huang}}},\ }\bibfield  {title} {\bibinfo {title} {{Implications for the supermassive black hole binaries from the NANOGrav 15-year data set}},\ }\href {https://doi.org/10.1007/s11433-023-2252-4} {\bibfield  {journal} {\bibinfo  {journal} {Science China Physics, Mechanics, and Astronomy}\ }\textbf {\bibinfo {volume} {66}},\ \bibinfo {eid} {120402} (\bibinfo {year} {2023})},\ \Eprint {https://arxiv.org/abs/2307.00722} {arXiv:2307.00722 [astro-ph.CO]} \BibitemShut {NoStop}%
\bibitem [{\citenamefont {{Ellis}}\ \emph {et~al.}(2024)\citenamefont {{Ellis}}, \citenamefont {{Fairbairn}}, \citenamefont {{H{\"u}tsi}}, \citenamefont {{Raidal}}, \citenamefont {{Urrutia}}, \citenamefont {{Vaskonen}},\ and\ \citenamefont {{Veerm{\"a}e}}}]{SMBHB-NANO-2}%
  \BibitemOpen
  \bibfield  {author} {\bibinfo {author} {\bibfnamefont {J.}~\bibnamefont {{Ellis}}}, \bibinfo {author} {\bibfnamefont {M.}~\bibnamefont {{Fairbairn}}}, \bibinfo {author} {\bibfnamefont {G.}~\bibnamefont {{H{\"u}tsi}}}, \bibinfo {author} {\bibfnamefont {J.}~\bibnamefont {{Raidal}}}, \bibinfo {author} {\bibfnamefont {J.}~\bibnamefont {{Urrutia}}}, \bibinfo {author} {\bibfnamefont {V.}~\bibnamefont {{Vaskonen}}},\ and\ \bibinfo {author} {\bibfnamefont {H.}~\bibnamefont {{Veerm{\"a}e}}},\ }\bibfield  {title} {\bibinfo {title} {{Gravitational waves from supermassive black hole binaries in light of the NANOGrav 15-year data}},\ }\href {https://doi.org/10.1103/PhysRevD.109.L021302} {\bibfield  {journal} {\bibinfo  {journal} {\prd}\ }\textbf {\bibinfo {volume} {109}},\ \bibinfo {eid} {L021302} (\bibinfo {year} {2024})},\ \Eprint {https://arxiv.org/abs/2306.17021} {arXiv:2306.17021 [astro-ph.CO]} \BibitemShut {NoStop}%
\bibitem [{\citenamefont {{Liepold}}\ and\ \citenamefont {{Ma}}(2024)}]{SMBHB-NANO-3}%
  \BibitemOpen
  \bibfield  {author} {\bibinfo {author} {\bibfnamefont {E.~R.}\ \bibnamefont {{Liepold}}}\ and\ \bibinfo {author} {\bibfnamefont {C.-P.}\ \bibnamefont {{Ma}}},\ }\bibfield  {title} {\bibinfo {title} {{Big Galaxies and Big Black Holes: The Massive Ends of the Local Stellar and Black Hole Mass Functions and the Implications for Nanohertz Gravitational Waves}},\ }\href {https://doi.org/10.3847/2041-8213/ad66b8} {\bibfield  {journal} {\bibinfo  {journal} {\apjl}\ }\textbf {\bibinfo {volume} {971}},\ \bibinfo {eid} {L29} (\bibinfo {year} {2024})},\ \Eprint {https://arxiv.org/abs/2407.14595} {arXiv:2407.14595 [astro-ph.GA]} \BibitemShut {NoStop}%
\bibitem [{\citenamefont {{Sato-Polito}}\ \emph {et~al.}(2024)\citenamefont {{Sato-Polito}}, \citenamefont {{Zaldarriaga}},\ and\ \citenamefont {{Quataert}}}]{SMBHB-NANO-4}%
  \BibitemOpen
  \bibfield  {author} {\bibinfo {author} {\bibfnamefont {G.}~\bibnamefont {{Sato-Polito}}}, \bibinfo {author} {\bibfnamefont {M.}~\bibnamefont {{Zaldarriaga}}},\ and\ \bibinfo {author} {\bibfnamefont {E.}~\bibnamefont {{Quataert}}},\ }\bibfield  {title} {\bibinfo {title} {{Where are the supermassive black holes measured by PTAs?}},\ }\href {https://doi.org/10.1103/PhysRevD.110.063020} {\bibfield  {journal} {\bibinfo  {journal} {\prd}\ }\textbf {\bibinfo {volume} {110}},\ \bibinfo {eid} {063020} (\bibinfo {year} {2024})}\BibitemShut {NoStop}%
\bibitem [{\citenamefont {{Goncharov}}\ \emph {et~al.}(2024)\citenamefont {{Goncharov}}, \citenamefont {{Sardana}}, \citenamefont {{Sesana}}, \citenamefont {{Antoniadis}}, \citenamefont {{Chalumeau}}, \citenamefont {{Champion}}, \citenamefont {{Chen}}, \citenamefont {{Keane}}, \citenamefont {{Shaifullah}},\ and\ \citenamefont {{Speri}}}]{SMBHB-NANO-5}%
  \BibitemOpen
  \bibfield  {author} {\bibinfo {author} {\bibfnamefont {B.}~\bibnamefont {{Goncharov}}}, \bibinfo {author} {\bibfnamefont {S.}~\bibnamefont {{Sardana}}}, \bibinfo {author} {\bibfnamefont {A.}~\bibnamefont {{Sesana}}}, \bibinfo {author} {\bibfnamefont {J.}~\bibnamefont {{Antoniadis}}}, \bibinfo {author} {\bibfnamefont {A.}~\bibnamefont {{Chalumeau}}}, \bibinfo {author} {\bibfnamefont {D.}~\bibnamefont {{Champion}}}, \bibinfo {author} {\bibfnamefont {S.}~\bibnamefont {{Chen}}}, \bibinfo {author} {\bibfnamefont {E.~F.}\ \bibnamefont {{Keane}}}, \bibinfo {author} {\bibfnamefont {G.}~\bibnamefont {{Shaifullah}}},\ and\ \bibinfo {author} {\bibfnamefont {L.}~\bibnamefont {{Speri}}},\ }\bibfield  {title} {\bibinfo {title} {{Fewer supermassive binary black holes in pulsar timing array observations}},\ }\href {https://doi.org/10.48550/arXiv.2409.03627} {\bibfield  {journal} {\bibinfo  {journal} {arXiv e-prints}\ ,\ \bibinfo {eid} {arXiv:2409.03627}} (\bibinfo {year} {2024})},\ \Eprint
  {https://arxiv.org/abs/2409.03627} {arXiv:2409.03627 [astro-ph.HE]} \BibitemShut {NoStop}%
\bibitem [{\citenamefont {{Agazie}}\ \emph {et~al.}(2023{\natexlab{b}})\citenamefont {{Agazie}}, \citenamefont {{Anumarlapudi}}, \citenamefont {{Archibald}}, \citenamefont {{Baker}}, \citenamefont {{B{\'e}csy}}, \citenamefont {{Blecha}}, \citenamefont {{Bonilla}}, \citenamefont {{Brazier}}, \citenamefont {{Brook}}, \citenamefont {{Burke-Spolaor}}, \citenamefont {{Burnette}}, \citenamefont {{Case}}, \citenamefont {{Casey-Clyde}}, \citenamefont {{Charisi}}, \citenamefont {{Chatterjee}}, \citenamefont {{Chatziioannou}}, \citenamefont {{Cheeseboro}}, \citenamefont {{Chen}}, \citenamefont {{Cohen}}, \citenamefont {{Cordes}}, \citenamefont {{Cornish}}, \citenamefont {{Crawford}}, \citenamefont {{Cromartie}}, \citenamefont {{Crowter}}, \citenamefont {{Cutler}}, \citenamefont {{D'Orazio}}, \citenamefont {{Decesar}}, \citenamefont {{Degan}}, \citenamefont {{Demorest}}, \citenamefont {{Deng}}, \citenamefont {{Dolch}}, \citenamefont {{Drachler}}, \citenamefont {{Ferrara}}, \citenamefont {{Fiore}}, \citenamefont
  {{Fonseca}}, \citenamefont {{Freedman}}, \citenamefont {{Gardiner}}, \citenamefont {{Garver-Daniels}}, \citenamefont {{Gentile}}, \citenamefont {{Gersbach}}, \citenamefont {{Glaser}}, \citenamefont {{Good}}, \citenamefont {{G{\"u}ltekin}}, \citenamefont {{Hazboun}}, \citenamefont {{Hourihane}}, \citenamefont {{Islo}}, \citenamefont {{Jennings}}, \citenamefont {{Johnson}}, \citenamefont {{Jones}}, \citenamefont {{Kaiser}}, \citenamefont {{Kaplan}}, \citenamefont {{Kelley}}, \citenamefont {{Kerr}}, \citenamefont {{Key}}, \citenamefont {{Laal}}, \citenamefont {{Lam}}, \citenamefont {{Lamb}}, \citenamefont {{Lazio}}, \citenamefont {{Lewandowska}}, \citenamefont {{Littenberg}}, \citenamefont {{Liu}}, \citenamefont {{Luo}}, \citenamefont {{Lynch}}, \citenamefont {{Ma}}, \citenamefont {{Madison}}, \citenamefont {{McEwen}}, \citenamefont {{McKee}}, \citenamefont {{McLaughlin}}, \citenamefont {{McMann}}, \citenamefont {{Meyers}}, \citenamefont {{Meyers}}, \citenamefont {{Mingarelli}}, \citenamefont {{Mitridate}},
  \citenamefont {{Natarajan}}, \citenamefont {{Ng}}, \citenamefont {{Nice}}, \citenamefont {{Ocker}}, \citenamefont {{Olum}}, \citenamefont {{Pennucci}}, \citenamefont {{Perera}}, \citenamefont {{Petrov}}, \citenamefont {{Pol}}, \citenamefont {{Radovan}}, \citenamefont {{Ransom}}, \citenamefont {{Ray}}, \citenamefont {{Romano}}, \citenamefont {{Runnoe}}, \citenamefont {{Sardesai}}, \citenamefont {{Schmiedekamp}}, \citenamefont {{Schmiedekamp}}, \citenamefont {{Schmitz}}, \citenamefont {{Schult}}, \citenamefont {{Shapiro-Albert}}, \citenamefont {{Siemens}}, \citenamefont {{Simon}}, \citenamefont {{Siwek}}, \citenamefont {{Stairs}}, \citenamefont {{Stinebring}}, \citenamefont {{Stovall}}, \citenamefont {{Sun}}, \citenamefont {{Susobhanan}}, \citenamefont {{Swiggum}}, \citenamefont {{Taylor}}, \citenamefont {{Taylor}}, \citenamefont {{Turner}}, \citenamefont {{Unal}}, \citenamefont {{Vallisneri}}, \citenamefont {{Vigeland}}, \citenamefont {{Wachter}}, \citenamefont {{Wahl}}, \citenamefont {{Wang}}, \citenamefont
  {{Witt}}, \citenamefont {{Wright}}, \citenamefont {{Young}},\ and\ \citenamefont {{Nanograv Collaboration}}}]{15yrastro}%
  \BibitemOpen
  \bibfield  {author} {\bibinfo {author} {\bibfnamefont {G.}~\bibnamefont {{Agazie}}}, \bibinfo {author} {\bibfnamefont {A.}~\bibnamefont {{Anumarlapudi}}}, \bibinfo {author} {\bibfnamefont {A.~M.}\ \bibnamefont {{Archibald}}}, \bibinfo {author} {\bibfnamefont {P.~T.}\ \bibnamefont {{Baker}}}, \bibinfo {author} {\bibfnamefont {B.}~\bibnamefont {{B{\'e}csy}}}, \bibinfo {author} {\bibfnamefont {L.}~\bibnamefont {{Blecha}}}, \bibinfo {author} {\bibfnamefont {A.}~\bibnamefont {{Bonilla}}}, \bibinfo {author} {\bibfnamefont {A.}~\bibnamefont {{Brazier}}}, \bibinfo {author} {\bibfnamefont {P.~R.}\ \bibnamefont {{Brook}}}, \bibinfo {author} {\bibfnamefont {S.}~\bibnamefont {{Burke-Spolaor}}}, \bibinfo {author} {\bibfnamefont {R.}~\bibnamefont {{Burnette}}}, \bibinfo {author} {\bibfnamefont {R.}~\bibnamefont {{Case}}}, \bibinfo {author} {\bibfnamefont {J.~A.}\ \bibnamefont {{Casey-Clyde}}}, \bibinfo {author} {\bibfnamefont {M.}~\bibnamefont {{Charisi}}}, \bibinfo {author} {\bibfnamefont {S.}~\bibnamefont
  {{Chatterjee}}}, \bibinfo {author} {\bibfnamefont {K.}~\bibnamefont {{Chatziioannou}}}, \bibinfo {author} {\bibfnamefont {B.~D.}\ \bibnamefont {{Cheeseboro}}}, \bibinfo {author} {\bibfnamefont {S.}~\bibnamefont {{Chen}}}, \bibinfo {author} {\bibfnamefont {T.}~\bibnamefont {{Cohen}}}, \bibinfo {author} {\bibfnamefont {J.~M.}\ \bibnamefont {{Cordes}}}, \bibinfo {author} {\bibfnamefont {N.~J.}\ \bibnamefont {{Cornish}}}, \bibinfo {author} {\bibfnamefont {F.}~\bibnamefont {{Crawford}}}, \bibinfo {author} {\bibfnamefont {H.~T.}\ \bibnamefont {{Cromartie}}}, \bibinfo {author} {\bibfnamefont {K.}~\bibnamefont {{Crowter}}}, \bibinfo {author} {\bibfnamefont {C.~J.}\ \bibnamefont {{Cutler}}}, \bibinfo {author} {\bibfnamefont {D.~J.}\ \bibnamefont {{D'Orazio}}}, \bibinfo {author} {\bibfnamefont {M.~E.}\ \bibnamefont {{Decesar}}}, \bibinfo {author} {\bibfnamefont {D.}~\bibnamefont {{Degan}}}, \bibinfo {author} {\bibfnamefont {P.~B.}\ \bibnamefont {{Demorest}}}, \bibinfo {author} {\bibfnamefont {H.}~\bibnamefont
  {{Deng}}}, \bibinfo {author} {\bibfnamefont {T.}~\bibnamefont {{Dolch}}}, \bibinfo {author} {\bibfnamefont {B.}~\bibnamefont {{Drachler}}}, \bibinfo {author} {\bibfnamefont {E.~C.}\ \bibnamefont {{Ferrara}}}, \bibinfo {author} {\bibfnamefont {W.}~\bibnamefont {{Fiore}}}, \bibinfo {author} {\bibfnamefont {E.}~\bibnamefont {{Fonseca}}}, \bibinfo {author} {\bibfnamefont {G.~E.}\ \bibnamefont {{Freedman}}}, \bibinfo {author} {\bibfnamefont {E.}~\bibnamefont {{Gardiner}}}, \bibinfo {author} {\bibfnamefont {N.}~\bibnamefont {{Garver-Daniels}}}, \bibinfo {author} {\bibfnamefont {P.~A.}\ \bibnamefont {{Gentile}}}, \bibinfo {author} {\bibfnamefont {K.~A.}\ \bibnamefont {{Gersbach}}}, \bibinfo {author} {\bibfnamefont {J.}~\bibnamefont {{Glaser}}}, \bibinfo {author} {\bibfnamefont {D.~C.}\ \bibnamefont {{Good}}}, \bibinfo {author} {\bibfnamefont {K.}~\bibnamefont {{G{\"u}ltekin}}}, \bibinfo {author} {\bibfnamefont {J.~S.}\ \bibnamefont {{Hazboun}}}, \bibinfo {author} {\bibfnamefont {S.}~\bibnamefont {{Hourihane}}},
  \bibinfo {author} {\bibfnamefont {K.}~\bibnamefont {{Islo}}}, \bibinfo {author} {\bibfnamefont {R.~J.}\ \bibnamefont {{Jennings}}}, \bibinfo {author} {\bibfnamefont {A.}~\bibnamefont {{Johnson}}}, \bibinfo {author} {\bibfnamefont {M.~L.}\ \bibnamefont {{Jones}}}, \bibinfo {author} {\bibfnamefont {A.~R.}\ \bibnamefont {{Kaiser}}}, \bibinfo {author} {\bibfnamefont {D.~L.}\ \bibnamefont {{Kaplan}}}, \bibinfo {author} {\bibfnamefont {L.~Z.}\ \bibnamefont {{Kelley}}}, \bibinfo {author} {\bibfnamefont {M.}~\bibnamefont {{Kerr}}}, \bibinfo {author} {\bibfnamefont {J.~S.}\ \bibnamefont {{Key}}}, \bibinfo {author} {\bibfnamefont {N.}~\bibnamefont {{Laal}}}, \bibinfo {author} {\bibfnamefont {M.~T.}\ \bibnamefont {{Lam}}}, \bibinfo {author} {\bibfnamefont {W.~G.}\ \bibnamefont {{Lamb}}}, \bibinfo {author} {\bibfnamefont {T.~J.~W.}\ \bibnamefont {{Lazio}}}, \bibinfo {author} {\bibfnamefont {N.}~\bibnamefont {{Lewandowska}}}, \bibinfo {author} {\bibfnamefont {T.~B.}\ \bibnamefont {{Littenberg}}}, \bibinfo {author}
  {\bibfnamefont {T.}~\bibnamefont {{Liu}}}, \bibinfo {author} {\bibfnamefont {J.}~\bibnamefont {{Luo}}}, \bibinfo {author} {\bibfnamefont {R.~S.}\ \bibnamefont {{Lynch}}}, \bibinfo {author} {\bibfnamefont {C.-P.}\ \bibnamefont {{Ma}}}, \bibinfo {author} {\bibfnamefont {D.~R.}\ \bibnamefont {{Madison}}}, \bibinfo {author} {\bibfnamefont {A.}~\bibnamefont {{McEwen}}}, \bibinfo {author} {\bibfnamefont {J.~W.}\ \bibnamefont {{McKee}}}, \bibinfo {author} {\bibfnamefont {M.~A.}\ \bibnamefont {{McLaughlin}}}, \bibinfo {author} {\bibfnamefont {N.}~\bibnamefont {{McMann}}}, \bibinfo {author} {\bibfnamefont {B.~W.}\ \bibnamefont {{Meyers}}}, \bibinfo {author} {\bibfnamefont {P.~M.}\ \bibnamefont {{Meyers}}}, \bibinfo {author} {\bibfnamefont {C.~M.~F.}\ \bibnamefont {{Mingarelli}}}, \bibinfo {author} {\bibfnamefont {A.}~\bibnamefont {{Mitridate}}}, \bibinfo {author} {\bibfnamefont {P.}~\bibnamefont {{Natarajan}}}, \bibinfo {author} {\bibfnamefont {C.}~\bibnamefont {{Ng}}}, \bibinfo {author} {\bibfnamefont {D.~J.}\
  \bibnamefont {{Nice}}}, \bibinfo {author} {\bibfnamefont {S.~K.}\ \bibnamefont {{Ocker}}}, \bibinfo {author} {\bibfnamefont {K.~D.}\ \bibnamefont {{Olum}}}, \bibinfo {author} {\bibfnamefont {T.~T.}\ \bibnamefont {{Pennucci}}}, \bibinfo {author} {\bibfnamefont {B.~B.~P.}\ \bibnamefont {{Perera}}}, \bibinfo {author} {\bibfnamefont {P.}~\bibnamefont {{Petrov}}}, \bibinfo {author} {\bibfnamefont {N.~S.}\ \bibnamefont {{Pol}}}, \bibinfo {author} {\bibfnamefont {H.~A.}\ \bibnamefont {{Radovan}}}, \bibinfo {author} {\bibfnamefont {S.~M.}\ \bibnamefont {{Ransom}}}, \bibinfo {author} {\bibfnamefont {P.~S.}\ \bibnamefont {{Ray}}}, \bibinfo {author} {\bibfnamefont {J.~D.}\ \bibnamefont {{Romano}}}, \bibinfo {author} {\bibfnamefont {J.~C.}\ \bibnamefont {{Runnoe}}}, \bibinfo {author} {\bibfnamefont {S.~C.}\ \bibnamefont {{Sardesai}}}, \bibinfo {author} {\bibfnamefont {A.}~\bibnamefont {{Schmiedekamp}}}, \bibinfo {author} {\bibfnamefont {C.}~\bibnamefont {{Schmiedekamp}}}, \bibinfo {author} {\bibfnamefont
  {K.}~\bibnamefont {{Schmitz}}}, \bibinfo {author} {\bibfnamefont {L.}~\bibnamefont {{Schult}}}, \bibinfo {author} {\bibfnamefont {B.~J.}\ \bibnamefont {{Shapiro-Albert}}}, \bibinfo {author} {\bibfnamefont {X.}~\bibnamefont {{Siemens}}}, \bibinfo {author} {\bibfnamefont {J.}~\bibnamefont {{Simon}}}, \bibinfo {author} {\bibfnamefont {M.~S.}\ \bibnamefont {{Siwek}}}, \bibinfo {author} {\bibfnamefont {I.~H.}\ \bibnamefont {{Stairs}}}, \bibinfo {author} {\bibfnamefont {D.~R.}\ \bibnamefont {{Stinebring}}}, \bibinfo {author} {\bibfnamefont {K.}~\bibnamefont {{Stovall}}}, \bibinfo {author} {\bibfnamefont {J.~P.}\ \bibnamefont {{Sun}}}, \bibinfo {author} {\bibfnamefont {A.}~\bibnamefont {{Susobhanan}}}, \bibinfo {author} {\bibfnamefont {J.~K.}\ \bibnamefont {{Swiggum}}}, \bibinfo {author} {\bibfnamefont {J.}~\bibnamefont {{Taylor}}}, \bibinfo {author} {\bibfnamefont {S.~R.}\ \bibnamefont {{Taylor}}}, \bibinfo {author} {\bibfnamefont {J.~E.}\ \bibnamefont {{Turner}}}, \bibinfo {author} {\bibfnamefont
  {C.}~\bibnamefont {{Unal}}}, \bibinfo {author} {\bibfnamefont {M.}~\bibnamefont {{Vallisneri}}}, \bibinfo {author} {\bibfnamefont {S.~J.}\ \bibnamefont {{Vigeland}}}, \bibinfo {author} {\bibfnamefont {J.~M.}\ \bibnamefont {{Wachter}}}, \bibinfo {author} {\bibfnamefont {H.~M.}\ \bibnamefont {{Wahl}}}, \bibinfo {author} {\bibfnamefont {Q.}~\bibnamefont {{Wang}}}, \bibinfo {author} {\bibfnamefont {C.~A.}\ \bibnamefont {{Witt}}}, \bibinfo {author} {\bibfnamefont {D.}~\bibnamefont {{Wright}}}, \bibinfo {author} {\bibfnamefont {O.}~\bibnamefont {{Young}}},\ and\ \bibinfo {author} {\bibnamefont {{Nanograv Collaboration}}},\ }\bibfield  {title} {\bibinfo {title} {{The NANOGrav 15 yr Data Set: Constraints on Supermassive Black Hole Binaries from the Gravitational-wave Background}},\ }\href {https://doi.org/10.3847/2041-8213/ace18b} {\bibfield  {journal} {\bibinfo  {journal} {\apjl}\ }\textbf {\bibinfo {volume} {952}},\ \bibinfo {eid} {L37} (\bibinfo {year} {2023}{\natexlab{b}})},\ \Eprint
  {https://arxiv.org/abs/2306.16220} {arXiv:2306.16220 [astro-ph.HE]} \BibitemShut {NoStop}%
\bibitem [{\citenamefont {Afzal}\ \emph {et~al.}(2023)\citenamefont {Afzal}, \citenamefont {Agazie}, \citenamefont {Anumarlapudi}, \citenamefont {Archibald}, \citenamefont {Arzoumanian}, \citenamefont {Baker}, \citenamefont {Bécsy}, \citenamefont {Blanco-Pillado}, \citenamefont {Blecha}, \citenamefont {Boddy}, \citenamefont {Brazier}, \citenamefont {Brook}, \citenamefont {Burke-Spolaor}, \citenamefont {Burnette}, \citenamefont {Case}, \citenamefont {Charisi}, \citenamefont {Chatterjee}, \citenamefont {Chatziioannou}, \citenamefont {Cheeseboro}, \citenamefont {Chen}, \citenamefont {Cohen}, \citenamefont {Cordes}, \citenamefont {Cornish}, \citenamefont {Crawford}, \citenamefont {Cromartie}, \citenamefont {Crowter}, \citenamefont {Cutler}, \citenamefont {DeCesar}, \citenamefont {DeGan}, \citenamefont {Demorest}, \citenamefont {Deng}, \citenamefont {Dolch}, \citenamefont {Drachler}, \citenamefont {von Eckardstein}, \citenamefont {Ferrara}, \citenamefont {Fiore}, \citenamefont {Fonseca}, \citenamefont {Freedman},
  \citenamefont {Garver-Daniels}, \citenamefont {Gentile}, \citenamefont {Gersbach}, \citenamefont {Glaser}, \citenamefont {Good}, \citenamefont {Guertin}, \citenamefont {Gültekin}, \citenamefont {Hazboun}, \citenamefont {Hourihane}, \citenamefont {Islo}, \citenamefont {Jennings}, \citenamefont {Johnson}, \citenamefont {Jones}, \citenamefont {Kaiser}, \citenamefont {Kaplan}, \citenamefont {Kelley}, \citenamefont {Kerr}, \citenamefont {Key}, \citenamefont {Laal}, \citenamefont {Lam}, \citenamefont {Lamb}, \citenamefont {Lazio}, \citenamefont {Lee}, \citenamefont {Lewandowska}, \citenamefont {dos Santos}, \citenamefont {Littenberg}, \citenamefont {Liu}, \citenamefont {Lorimer}, \citenamefont {Luo}, \citenamefont {Lynch}, \citenamefont {Ma}, \citenamefont {Madison}, \citenamefont {McEwen}, \citenamefont {McKee}, \citenamefont {McLaughlin}, \citenamefont {McMann}, \citenamefont {Meyers}, \citenamefont {Meyers}, \citenamefont {Mingarelli}, \citenamefont {Mitridate}, \citenamefont {Nay}, \citenamefont {Natarajan},
  \citenamefont {Ng}, \citenamefont {Nice}, \citenamefont {Ocker}, \citenamefont {Olum}, \citenamefont {Pennucci}, \citenamefont {Perera}, \citenamefont {Petrov}, \citenamefont {Pol}, \citenamefont {Radovan}, \citenamefont {Ransom}, \citenamefont {Ray}, \citenamefont {Romano}, \citenamefont {Sardesai}, \citenamefont {Schmiedekamp}, \citenamefont {Schmiedekamp}, \citenamefont {Schmitz}, \citenamefont {Schröder}, \citenamefont {Schult}, \citenamefont {Shapiro-Albert}, \citenamefont {Siemens}, \citenamefont {Simon}, \citenamefont {Siwek}, \citenamefont {Stairs}, \citenamefont {Stinebring}, \citenamefont {Stovall}, \citenamefont {Stratmann}, \citenamefont {Sun}, \citenamefont {Susobhanan}, \citenamefont {Swiggum}, \citenamefont {Taylor}, \citenamefont {Taylor}, \citenamefont {Trickle}, \citenamefont {Turner}, \citenamefont {Unal}, \citenamefont {Vallisneri}, \citenamefont {Verma}, \citenamefont {Vigeland}, \citenamefont {Wahl}, \citenamefont {Wang}, \citenamefont {Witt}, \citenamefont {Wright}, \citenamefont
  {Young}, \citenamefont {Zurek},\ and\ \citenamefont {Collaboration}}]{15yrnewphysics}%
  \BibitemOpen
  \bibfield  {author} {\bibinfo {author} {\bibfnamefont {A.}~\bibnamefont {Afzal}}, \bibinfo {author} {\bibfnamefont {G.}~\bibnamefont {Agazie}}, \bibinfo {author} {\bibfnamefont {A.}~\bibnamefont {Anumarlapudi}}, \bibinfo {author} {\bibfnamefont {A.~M.}\ \bibnamefont {Archibald}}, \bibinfo {author} {\bibfnamefont {Z.}~\bibnamefont {Arzoumanian}}, \bibinfo {author} {\bibfnamefont {P.~T.}\ \bibnamefont {Baker}}, \bibinfo {author} {\bibfnamefont {B.}~\bibnamefont {Bécsy}}, \bibinfo {author} {\bibfnamefont {J.~J.}\ \bibnamefont {Blanco-Pillado}}, \bibinfo {author} {\bibfnamefont {L.}~\bibnamefont {Blecha}}, \bibinfo {author} {\bibfnamefont {K.~K.}\ \bibnamefont {Boddy}}, \bibinfo {author} {\bibfnamefont {A.}~\bibnamefont {Brazier}}, \bibinfo {author} {\bibfnamefont {P.~R.}\ \bibnamefont {Brook}}, \bibinfo {author} {\bibfnamefont {S.}~\bibnamefont {Burke-Spolaor}}, \bibinfo {author} {\bibfnamefont {R.}~\bibnamefont {Burnette}}, \bibinfo {author} {\bibfnamefont {R.}~\bibnamefont {Case}}, \bibinfo {author}
  {\bibfnamefont {M.}~\bibnamefont {Charisi}}, \bibinfo {author} {\bibfnamefont {S.}~\bibnamefont {Chatterjee}}, \bibinfo {author} {\bibfnamefont {K.}~\bibnamefont {Chatziioannou}}, \bibinfo {author} {\bibfnamefont {B.~D.}\ \bibnamefont {Cheeseboro}}, \bibinfo {author} {\bibfnamefont {S.}~\bibnamefont {Chen}}, \bibinfo {author} {\bibfnamefont {T.}~\bibnamefont {Cohen}}, \bibinfo {author} {\bibfnamefont {J.~M.}\ \bibnamefont {Cordes}}, \bibinfo {author} {\bibfnamefont {N.~J.}\ \bibnamefont {Cornish}}, \bibinfo {author} {\bibfnamefont {F.}~\bibnamefont {Crawford}}, \bibinfo {author} {\bibfnamefont {H.~T.}\ \bibnamefont {Cromartie}}, \bibinfo {author} {\bibfnamefont {K.}~\bibnamefont {Crowter}}, \bibinfo {author} {\bibfnamefont {C.~J.}\ \bibnamefont {Cutler}}, \bibinfo {author} {\bibfnamefont {M.~E.}\ \bibnamefont {DeCesar}}, \bibinfo {author} {\bibfnamefont {D.}~\bibnamefont {DeGan}}, \bibinfo {author} {\bibfnamefont {P.~B.}\ \bibnamefont {Demorest}}, \bibinfo {author} {\bibfnamefont {H.}~\bibnamefont {Deng}},
  \bibinfo {author} {\bibfnamefont {T.}~\bibnamefont {Dolch}}, \bibinfo {author} {\bibfnamefont {B.}~\bibnamefont {Drachler}}, \bibinfo {author} {\bibfnamefont {R.}~\bibnamefont {von Eckardstein}}, \bibinfo {author} {\bibfnamefont {E.~C.}\ \bibnamefont {Ferrara}}, \bibinfo {author} {\bibfnamefont {W.}~\bibnamefont {Fiore}}, \bibinfo {author} {\bibfnamefont {E.}~\bibnamefont {Fonseca}}, \bibinfo {author} {\bibfnamefont {G.~E.}\ \bibnamefont {Freedman}}, \bibinfo {author} {\bibfnamefont {N.}~\bibnamefont {Garver-Daniels}}, \bibinfo {author} {\bibfnamefont {P.~A.}\ \bibnamefont {Gentile}}, \bibinfo {author} {\bibfnamefont {K.~A.}\ \bibnamefont {Gersbach}}, \bibinfo {author} {\bibfnamefont {J.}~\bibnamefont {Glaser}}, \bibinfo {author} {\bibfnamefont {D.~C.}\ \bibnamefont {Good}}, \bibinfo {author} {\bibfnamefont {L.}~\bibnamefont {Guertin}}, \bibinfo {author} {\bibfnamefont {K.}~\bibnamefont {Gültekin}}, \bibinfo {author} {\bibfnamefont {J.~S.}\ \bibnamefont {Hazboun}}, \bibinfo {author} {\bibfnamefont
  {S.}~\bibnamefont {Hourihane}}, \bibinfo {author} {\bibfnamefont {K.}~\bibnamefont {Islo}}, \bibinfo {author} {\bibfnamefont {R.~J.}\ \bibnamefont {Jennings}}, \bibinfo {author} {\bibfnamefont {A.~D.}\ \bibnamefont {Johnson}}, \bibinfo {author} {\bibfnamefont {M.~L.}\ \bibnamefont {Jones}}, \bibinfo {author} {\bibfnamefont {A.~R.}\ \bibnamefont {Kaiser}}, \bibinfo {author} {\bibfnamefont {D.~L.}\ \bibnamefont {Kaplan}}, \bibinfo {author} {\bibfnamefont {L.~Z.}\ \bibnamefont {Kelley}}, \bibinfo {author} {\bibfnamefont {M.}~\bibnamefont {Kerr}}, \bibinfo {author} {\bibfnamefont {J.~S.}\ \bibnamefont {Key}}, \bibinfo {author} {\bibfnamefont {N.}~\bibnamefont {Laal}}, \bibinfo {author} {\bibfnamefont {M.~T.}\ \bibnamefont {Lam}}, \bibinfo {author} {\bibfnamefont {W.~G.}\ \bibnamefont {Lamb}}, \bibinfo {author} {\bibfnamefont {T.~J.~W.}\ \bibnamefont {Lazio}}, \bibinfo {author} {\bibfnamefont {V.~S.~H.}\ \bibnamefont {Lee}}, \bibinfo {author} {\bibfnamefont {N.}~\bibnamefont {Lewandowska}}, \bibinfo {author}
  {\bibfnamefont {R.~R.~L.}\ \bibnamefont {dos Santos}}, \bibinfo {author} {\bibfnamefont {T.~B.}\ \bibnamefont {Littenberg}}, \bibinfo {author} {\bibfnamefont {T.}~\bibnamefont {Liu}}, \bibinfo {author} {\bibfnamefont {D.~R.}\ \bibnamefont {Lorimer}}, \bibinfo {author} {\bibfnamefont {J.}~\bibnamefont {Luo}}, \bibinfo {author} {\bibfnamefont {R.~S.}\ \bibnamefont {Lynch}}, \bibinfo {author} {\bibfnamefont {C.-P.}\ \bibnamefont {Ma}}, \bibinfo {author} {\bibfnamefont {D.~R.}\ \bibnamefont {Madison}}, \bibinfo {author} {\bibfnamefont {A.}~\bibnamefont {McEwen}}, \bibinfo {author} {\bibfnamefont {J.~W.}\ \bibnamefont {McKee}}, \bibinfo {author} {\bibfnamefont {M.~A.}\ \bibnamefont {McLaughlin}}, \bibinfo {author} {\bibfnamefont {N.}~\bibnamefont {McMann}}, \bibinfo {author} {\bibfnamefont {B.~W.}\ \bibnamefont {Meyers}}, \bibinfo {author} {\bibfnamefont {P.~M.}\ \bibnamefont {Meyers}}, \bibinfo {author} {\bibfnamefont {C.~M.~F.}\ \bibnamefont {Mingarelli}}, \bibinfo {author} {\bibfnamefont {A.}~\bibnamefont
  {Mitridate}}, \bibinfo {author} {\bibfnamefont {J.}~\bibnamefont {Nay}}, \bibinfo {author} {\bibfnamefont {P.}~\bibnamefont {Natarajan}}, \bibinfo {author} {\bibfnamefont {C.}~\bibnamefont {Ng}}, \bibinfo {author} {\bibfnamefont {D.~J.}\ \bibnamefont {Nice}}, \bibinfo {author} {\bibfnamefont {S.~K.}\ \bibnamefont {Ocker}}, \bibinfo {author} {\bibfnamefont {K.~D.}\ \bibnamefont {Olum}}, \bibinfo {author} {\bibfnamefont {T.~T.}\ \bibnamefont {Pennucci}}, \bibinfo {author} {\bibfnamefont {B.~B.~P.}\ \bibnamefont {Perera}}, \bibinfo {author} {\bibfnamefont {P.}~\bibnamefont {Petrov}}, \bibinfo {author} {\bibfnamefont {N.~S.}\ \bibnamefont {Pol}}, \bibinfo {author} {\bibfnamefont {H.~A.}\ \bibnamefont {Radovan}}, \bibinfo {author} {\bibfnamefont {S.~M.}\ \bibnamefont {Ransom}}, \bibinfo {author} {\bibfnamefont {P.~S.}\ \bibnamefont {Ray}}, \bibinfo {author} {\bibfnamefont {J.~D.}\ \bibnamefont {Romano}}, \bibinfo {author} {\bibfnamefont {S.~C.}\ \bibnamefont {Sardesai}}, \bibinfo {author} {\bibfnamefont
  {A.}~\bibnamefont {Schmiedekamp}}, \bibinfo {author} {\bibfnamefont {C.}~\bibnamefont {Schmiedekamp}}, \bibinfo {author} {\bibfnamefont {K.}~\bibnamefont {Schmitz}}, \bibinfo {author} {\bibfnamefont {T.}~\bibnamefont {Schröder}}, \bibinfo {author} {\bibfnamefont {L.}~\bibnamefont {Schult}}, \bibinfo {author} {\bibfnamefont {B.~J.}\ \bibnamefont {Shapiro-Albert}}, \bibinfo {author} {\bibfnamefont {X.}~\bibnamefont {Siemens}}, \bibinfo {author} {\bibfnamefont {J.}~\bibnamefont {Simon}}, \bibinfo {author} {\bibfnamefont {M.~S.}\ \bibnamefont {Siwek}}, \bibinfo {author} {\bibfnamefont {I.~H.}\ \bibnamefont {Stairs}}, \bibinfo {author} {\bibfnamefont {D.~R.}\ \bibnamefont {Stinebring}}, \bibinfo {author} {\bibfnamefont {K.}~\bibnamefont {Stovall}}, \bibinfo {author} {\bibfnamefont {P.}~\bibnamefont {Stratmann}}, \bibinfo {author} {\bibfnamefont {J.~P.}\ \bibnamefont {Sun}}, \bibinfo {author} {\bibfnamefont {A.}~\bibnamefont {Susobhanan}}, \bibinfo {author} {\bibfnamefont {J.~K.}\ \bibnamefont {Swiggum}},
  \bibinfo {author} {\bibfnamefont {J.}~\bibnamefont {Taylor}}, \bibinfo {author} {\bibfnamefont {S.~R.}\ \bibnamefont {Taylor}}, \bibinfo {author} {\bibfnamefont {T.}~\bibnamefont {Trickle}}, \bibinfo {author} {\bibfnamefont {J.~E.}\ \bibnamefont {Turner}}, \bibinfo {author} {\bibfnamefont {C.}~\bibnamefont {Unal}}, \bibinfo {author} {\bibfnamefont {M.}~\bibnamefont {Vallisneri}}, \bibinfo {author} {\bibfnamefont {S.}~\bibnamefont {Verma}}, \bibinfo {author} {\bibfnamefont {S.~J.}\ \bibnamefont {Vigeland}}, \bibinfo {author} {\bibfnamefont {H.~M.}\ \bibnamefont {Wahl}}, \bibinfo {author} {\bibfnamefont {Q.}~\bibnamefont {Wang}}, \bibinfo {author} {\bibfnamefont {C.~A.}\ \bibnamefont {Witt}}, \bibinfo {author} {\bibfnamefont {D.}~\bibnamefont {Wright}}, \bibinfo {author} {\bibfnamefont {O.}~\bibnamefont {Young}}, \bibinfo {author} {\bibfnamefont {K.~M.}\ \bibnamefont {Zurek}},\ and\ \bibinfo {author} {\bibfnamefont {T.~N.}\ \bibnamefont {Collaboration}},\ }\bibfield  {title} {\bibinfo {title} {The nanograv 15
  yr data set: Search for signals from new physics},\ }\href {https://doi.org/10.3847/2041-8213/acdc91} {\bibfield  {journal} {\bibinfo  {journal} {The Astrophysical Journal Letters}\ }\textbf {\bibinfo {volume} {951}},\ \bibinfo {pages} {L11} (\bibinfo {year} {2023})}\BibitemShut {NoStop}%
\bibitem [{\citenamefont {{Taylor}}\ \emph {et~al.}(2017)\citenamefont {{Taylor}}, \citenamefont {{Simon}},\ and\ \citenamefont {{Sampson}}}]{tss17}%
  \BibitemOpen
  \bibfield  {author} {\bibinfo {author} {\bibfnamefont {S.~R.}\ \bibnamefont {{Taylor}}}, \bibinfo {author} {\bibfnamefont {J.}~\bibnamefont {{Simon}}},\ and\ \bibinfo {author} {\bibfnamefont {L.}~\bibnamefont {{Sampson}}},\ }\bibfield  {title} {\bibinfo {title} {{Constraints on the Dynamical Environments of Supermassive Black-Hole Binaries Using Pulsar-Timing Arrays}},\ }\href {https://doi.org/10.1103/PhysRevLett.118.181102} {\bibfield  {journal} {\bibinfo  {journal} {\prl}\ }\textbf {\bibinfo {volume} {118}},\ \bibinfo {eid} {181102} (\bibinfo {year} {2017})},\ \Eprint {https://arxiv.org/abs/1612.02817} {arXiv:1612.02817 [astro-ph.GA]} \BibitemShut {NoStop}%
\bibitem [{\citenamefont {{Wong}}\ and\ \citenamefont {{Gerosa}}(2019)}]{GP-limit}%
  \BibitemOpen
  \bibfield  {author} {\bibinfo {author} {\bibfnamefont {K.~W.~K.}\ \bibnamefont {{Wong}}}\ and\ \bibinfo {author} {\bibfnamefont {D.}~\bibnamefont {{Gerosa}}},\ }\bibfield  {title} {\bibinfo {title} {{Machine-learning interpolation of population-synthesis simulations to interpret gravitational-wave observations: A case study}},\ }\href {https://doi.org/10.1103/PhysRevD.100.083015} {\bibfield  {journal} {\bibinfo  {journal} {\prd}\ }\textbf {\bibinfo {volume} {100}},\ \bibinfo {eid} {083015} (\bibinfo {year} {2019})},\ \Eprint {https://arxiv.org/abs/1909.06373} {arXiv:1909.06373 [astro-ph.HE]} \BibitemShut {NoStop}%
\bibitem [{\citenamefont {{Coccaro}}\ \emph {et~al.}(2023)\citenamefont {{Coccaro}}, \citenamefont {{Letizia}}, \citenamefont {{Reyes-Gonzalez}},\ and\ \citenamefont {{Torre}}}]{whyQ}%
  \BibitemOpen
  \bibfield  {author} {\bibinfo {author} {\bibfnamefont {A.}~\bibnamefont {{Coccaro}}}, \bibinfo {author} {\bibfnamefont {M.}~\bibnamefont {{Letizia}}}, \bibinfo {author} {\bibfnamefont {H.}~\bibnamefont {{Reyes-Gonzalez}}},\ and\ \bibinfo {author} {\bibfnamefont {R.}~\bibnamefont {{Torre}}},\ }\bibfield  {title} {\bibinfo {title} {{Comparative Study of Coupling and Autoregressive Flows through Robust Statistical Tests}},\ }\href {https://doi.org/10.48550/arXiv.2302.12024} {\bibfield  {journal} {\bibinfo  {journal} {arXiv e-prints}\ ,\ \bibinfo {eid} {arXiv:2302.12024}} (\bibinfo {year} {2023})},\ \Eprint {https://arxiv.org/abs/2302.12024} {arXiv:2302.12024 [stat.ML]} \BibitemShut {NoStop}%
\bibitem [{\citenamefont {{Crenshaw}}\ \emph {et~al.}(2024)\citenamefont {{Crenshaw}}, \citenamefont {{Bryce Kalmbach}}, \citenamefont {{Gagliano}}, \citenamefont {{Yan}}, \citenamefont {{Connolly}}, \citenamefont {{Malz}}, \citenamefont {{Schmidt}},\ and\ \citenamefont {{The LSST Dark Energy Science Collaboration}}}]{whyQ2}%
  \BibitemOpen
  \bibfield  {author} {\bibinfo {author} {\bibfnamefont {J.~F.}\ \bibnamefont {{Crenshaw}}}, \bibinfo {author} {\bibfnamefont {J.}~\bibnamefont {{Bryce Kalmbach}}}, \bibinfo {author} {\bibfnamefont {A.}~\bibnamefont {{Gagliano}}}, \bibinfo {author} {\bibfnamefont {Z.}~\bibnamefont {{Yan}}}, \bibinfo {author} {\bibfnamefont {A.~J.}\ \bibnamefont {{Connolly}}}, \bibinfo {author} {\bibfnamefont {A.~I.}\ \bibnamefont {{Malz}}}, \bibinfo {author} {\bibfnamefont {S.~J.}\ \bibnamefont {{Schmidt}}},\ and\ \bibinfo {author} {\bibnamefont {{The LSST Dark Energy Science Collaboration}}},\ }\bibfield  {title} {\bibinfo {title} {{Probabilistic Forward Modeling of Galaxy Catalogs with Normalizing Flows}},\ }\href {https://doi.org/10.48550/arXiv.2405.04740} {\bibfield  {journal} {\bibinfo  {journal} {arXiv e-prints}\ ,\ \bibinfo {eid} {arXiv:2405.04740}} (\bibinfo {year} {2024})},\ \Eprint {https://arxiv.org/abs/2405.04740} {arXiv:2405.04740 [astro-ph.IM]} \BibitemShut {NoStop}%
\bibitem [{\citenamefont {{Wong}}\ \emph {et~al.}(2020)\citenamefont {{Wong}}, \citenamefont {{Contardo}},\ and\ \citenamefont {{Ho}}}]{EarlierNF}%
  \BibitemOpen
  \bibfield  {author} {\bibinfo {author} {\bibfnamefont {K.~W.~K.}\ \bibnamefont {{Wong}}}, \bibinfo {author} {\bibfnamefont {G.}~\bibnamefont {{Contardo}}},\ and\ \bibinfo {author} {\bibfnamefont {S.}~\bibnamefont {{Ho}}},\ }\bibfield  {title} {\bibinfo {title} {{Gravitational-wave population inference with deep flow-based generative network}},\ }\href {https://doi.org/10.1103/PhysRevD.101.123005} {\bibfield  {journal} {\bibinfo  {journal} {\prd}\ }\textbf {\bibinfo {volume} {101}},\ \bibinfo {eid} {123005} (\bibinfo {year} {2020})},\ \Eprint {https://arxiv.org/abs/2002.09491} {arXiv:2002.09491 [astro-ph.IM]} \BibitemShut {NoStop}%
\bibitem [{\citenamefont {{Bonetti}}\ \emph {et~al.}(2024)\citenamefont {{Bonetti}}, \citenamefont {{Franchini}}, \citenamefont {{Galuzzi}},\ and\ \citenamefont {{Sesana}}}]{NN}%
  \BibitemOpen
  \bibfield  {author} {\bibinfo {author} {\bibfnamefont {M.}~\bibnamefont {{Bonetti}}}, \bibinfo {author} {\bibfnamefont {A.}~\bibnamefont {{Franchini}}}, \bibinfo {author} {\bibfnamefont {B.~G.}\ \bibnamefont {{Galuzzi}}},\ and\ \bibinfo {author} {\bibfnamefont {A.}~\bibnamefont {{Sesana}}},\ }\bibfield  {title} {\bibinfo {title} {{Neural networks unveiling the properties of gravitational wave background from supermassive black hole binaries}},\ }\href {https://doi.org/10.1051/0004-6361/202348433} {\bibfield  {journal} {\bibinfo  {journal} {\aap}\ }\textbf {\bibinfo {volume} {687}},\ \bibinfo {eid} {A42} (\bibinfo {year} {2024})},\ \Eprint {https://arxiv.org/abs/2311.04276} {arXiv:2311.04276 [astro-ph.HE]} \BibitemShut {NoStop}%
\bibitem [{\citenamefont {Delbourgo}\ and\ \citenamefont {Gregory}(1982)}]{Delbourgo1982RationalQS}%
  \BibitemOpen
  \bibfield  {author} {\bibinfo {author} {\bibfnamefont {R.}~\bibnamefont {Delbourgo}}\ and\ \bibinfo {author} {\bibfnamefont {J.~A.}\ \bibnamefont {Gregory}},\ }\bibfield  {title} {\bibinfo {title} {Rational quadratic spline interpolation to monotonic data.}\ }(\bibinfo {year} {1982})\BibitemShut {NoStop}%
\bibitem [{\citenamefont {Durkan}\ \emph {et~al.}(2019)\citenamefont {Durkan}, \citenamefont {Bekasov}, \citenamefont {Murray},\ and\ \citenamefont {Papamakarios}}]{durkan2019neuralsplineflows}%
  \BibitemOpen
  \bibfield  {author} {\bibinfo {author} {\bibfnamefont {C.}~\bibnamefont {Durkan}}, \bibinfo {author} {\bibfnamefont {A.}~\bibnamefont {Bekasov}}, \bibinfo {author} {\bibfnamefont {I.}~\bibnamefont {Murray}},\ and\ \bibinfo {author} {\bibfnamefont {G.}~\bibnamefont {Papamakarios}},\ }\href {https://arxiv.org/abs/1906.04032} {\bibinfo {title} {Neural spline flows}} (\bibinfo {year} {2019}),\ \Eprint {https://arxiv.org/abs/1906.04032} {arXiv:1906.04032 [stat.ML]} \BibitemShut {NoStop}%
\bibitem [{\citenamefont {{Phinney}}(2001)}]{Phinney-2001}%
  \BibitemOpen
  \bibfield  {author} {\bibinfo {author} {\bibfnamefont {E.~S.}\ \bibnamefont {{Phinney}}},\ }\bibfield  {title} {\bibinfo {title} {{A Practical Theorem on Gravitational Wave Backgrounds}},\ }\href@noop {} {\bibfield  {journal} {\bibinfo  {journal} {ArXiv Astrophysics e-prints}\ } (\bibinfo {year} {2001})},\ \Eprint {https://arxiv.org/abs/astro-ph/0108028} {astro-ph/0108028} \BibitemShut {NoStop}%
\bibitem [{\citenamefont {{Finn}}\ and\ \citenamefont {{Thorne}}(2000)}]{Finn+Thorne-2000}%
  \BibitemOpen
  \bibfield  {author} {\bibinfo {author} {\bibfnamefont {L.~S.}\ \bibnamefont {{Finn}}}\ and\ \bibinfo {author} {\bibfnamefont {K.~S.}\ \bibnamefont {{Thorne}}},\ }\bibfield  {title} {\bibinfo {title} {{Gravitational waves from a compact star in a circular, inspiral orbit, in the equatorial plane of a massive, spinning black hole, as observed by LISA}},\ }\href {https://doi.org/10.1103/PhysRevD.62.124021} {\bibfield  {journal} {\bibinfo  {journal} {\prd}\ }\textbf {\bibinfo {volume} {62}},\ \bibinfo {eid} {124021} (\bibinfo {year} {2000})},\ \Eprint {https://arxiv.org/abs/gr-qc/0007074} {arXiv:gr-qc/0007074 [gr-qc]} \BibitemShut {NoStop}%
\bibitem [{\citenamefont {{Chen}}\ \emph {et~al.}(2019)\citenamefont {{Chen}}, \citenamefont {{Sesana}},\ and\ \citenamefont {{Conselice}}}]{Chen2019}%
  \BibitemOpen
  \bibfield  {author} {\bibinfo {author} {\bibfnamefont {S.}~\bibnamefont {{Chen}}}, \bibinfo {author} {\bibfnamefont {A.}~\bibnamefont {{Sesana}}},\ and\ \bibinfo {author} {\bibfnamefont {C.~J.}\ \bibnamefont {{Conselice}}},\ }\bibfield  {title} {\bibinfo {title} {{Constraining astrophysical observables of galaxy and supermassive black hole binary mergers using pulsar timing arrays}},\ }\href {https://doi.org/10.1093/mnras/stz1722} {\bibfield  {journal} {\bibinfo  {journal} {\mnras}\ }\textbf {\bibinfo {volume} {488}},\ \bibinfo {pages} {401} (\bibinfo {year} {2019})},\ \Eprint {https://arxiv.org/abs/1810.04184} {arXiv:1810.04184 [astro-ph.GA]} \BibitemShut {NoStop}%
\bibitem [{\citenamefont {{Kormendy}}\ and\ \citenamefont {{Ho}}(2013)}]{Kormendy+Ho2013}%
  \BibitemOpen
  \bibfield  {author} {\bibinfo {author} {\bibfnamefont {J.}~\bibnamefont {{Kormendy}}}\ and\ \bibinfo {author} {\bibfnamefont {L.~C.}\ \bibnamefont {{Ho}}},\ }\bibfield  {title} {\bibinfo {title} {{Coevolution (Or Not) of Supermassive Black Holes and Host Galaxies}},\ }\href {https://doi.org/10.1146/annurev-astro-082708-101811} {\bibfield  {journal} {\bibinfo  {journal} {\araa}\ }\textbf {\bibinfo {volume} {51}},\ \bibinfo {pages} {511} (\bibinfo {year} {2013})},\ \Eprint {https://arxiv.org/abs/1304.7762} {arXiv:1304.7762 [astro-ph.CO]} \BibitemShut {NoStop}%
\bibitem [{\citenamefont {{Hinshaw}}\ \emph {et~al.}(2013)\citenamefont {{Hinshaw}}, \citenamefont {{Larson}}, \citenamefont {{Komatsu}}, \citenamefont {{Spergel}}, \citenamefont {{Bennett}}, \citenamefont {{Dunkley}}, \citenamefont {{Nolta}}, \citenamefont {{Halpern}}, \citenamefont {{Hill}}, \citenamefont {{Odegard}}, \citenamefont {{Page}}, \citenamefont {{Smith}}, \citenamefont {{Weiland}}, \citenamefont {{Gold}}, \citenamefont {{Jarosik}}, \citenamefont {{Kogut}}, \citenamefont {{Limon}}, \citenamefont {{Meyer}}, \citenamefont {{Tucker}}, \citenamefont {{Wollack}},\ and\ \citenamefont {{Wright}}}]{wmap9}%
  \BibitemOpen
  \bibfield  {author} {\bibinfo {author} {\bibfnamefont {G.}~\bibnamefont {{Hinshaw}}}, \bibinfo {author} {\bibfnamefont {D.}~\bibnamefont {{Larson}}}, \bibinfo {author} {\bibfnamefont {E.}~\bibnamefont {{Komatsu}}}, \bibinfo {author} {\bibfnamefont {D.~N.}\ \bibnamefont {{Spergel}}}, \bibinfo {author} {\bibfnamefont {C.~L.}\ \bibnamefont {{Bennett}}}, \bibinfo {author} {\bibfnamefont {J.}~\bibnamefont {{Dunkley}}}, \bibinfo {author} {\bibfnamefont {M.~R.}\ \bibnamefont {{Nolta}}}, \bibinfo {author} {\bibfnamefont {M.}~\bibnamefont {{Halpern}}}, \bibinfo {author} {\bibfnamefont {R.~S.}\ \bibnamefont {{Hill}}}, \bibinfo {author} {\bibfnamefont {N.}~\bibnamefont {{Odegard}}}, \bibinfo {author} {\bibfnamefont {L.}~\bibnamefont {{Page}}}, \bibinfo {author} {\bibfnamefont {K.~M.}\ \bibnamefont {{Smith}}}, \bibinfo {author} {\bibfnamefont {J.~L.}\ \bibnamefont {{Weiland}}}, \bibinfo {author} {\bibfnamefont {B.}~\bibnamefont {{Gold}}}, \bibinfo {author} {\bibfnamefont {N.}~\bibnamefont {{Jarosik}}}, \bibinfo
  {author} {\bibfnamefont {A.}~\bibnamefont {{Kogut}}}, \bibinfo {author} {\bibfnamefont {M.}~\bibnamefont {{Limon}}}, \bibinfo {author} {\bibfnamefont {S.~S.}\ \bibnamefont {{Meyer}}}, \bibinfo {author} {\bibfnamefont {G.~S.}\ \bibnamefont {{Tucker}}}, \bibinfo {author} {\bibfnamefont {E.}~\bibnamefont {{Wollack}}},\ and\ \bibinfo {author} {\bibfnamefont {E.~L.}\ \bibnamefont {{Wright}}},\ }\bibfield  {title} {\bibinfo {title} {{Nine-year Wilkinson Microwave Anisotropy Probe (WMAP) Observations: Cosmological Parameter Results}},\ }\href {https://doi.org/10.1088/0067-0049/208/2/19} {\bibfield  {journal} {\bibinfo  {journal} {\apjs}\ }\textbf {\bibinfo {volume} {208}},\ \bibinfo {eid} {19} (\bibinfo {year} {2013})},\ \Eprint {https://arxiv.org/abs/1212.5226} {arXiv:1212.5226 [astro-ph.CO]} \BibitemShut {NoStop}%
\bibitem [{\citenamefont {{Aigrain}}\ and\ \citenamefont {{Foreman-Mackey}}(2023)}]{GPsReview}%
  \BibitemOpen
  \bibfield  {author} {\bibinfo {author} {\bibfnamefont {S.}~\bibnamefont {{Aigrain}}}\ and\ \bibinfo {author} {\bibfnamefont {D.}~\bibnamefont {{Foreman-Mackey}}},\ }\bibfield  {title} {\bibinfo {title} {{Gaussian Process Regression for Astronomical Time Series}},\ }\href {https://doi.org/10.1146/annurev-astro-052920-103508} {\bibfield  {journal} {\bibinfo  {journal} {\araa}\ }\textbf {\bibinfo {volume} {61}},\ \bibinfo {pages} {329} (\bibinfo {year} {2023})},\ \Eprint {https://arxiv.org/abs/2209.08940} {arXiv:2209.08940 [astro-ph.IM]} \BibitemShut {NoStop}%
\bibitem [{\citenamefont {{Lamb}}\ and\ \citenamefont {{Taylor}}(2024)}]{MVSK}%
  \BibitemOpen
  \bibfield  {author} {\bibinfo {author} {\bibfnamefont {W.~G.}\ \bibnamefont {{Lamb}}}\ and\ \bibinfo {author} {\bibfnamefont {S.~R.}\ \bibnamefont {{Taylor}}},\ }\bibfield  {title} {\bibinfo {title} {{Spectral Variance in a Stochastic Gravitational-wave Background from a Binary Population}},\ }\href {https://doi.org/10.3847/2041-8213/ad654a} {\bibfield  {journal} {\bibinfo  {journal} {\apjl}\ }\textbf {\bibinfo {volume} {971}},\ \bibinfo {eid} {L10} (\bibinfo {year} {2024})},\ \Eprint {https://arxiv.org/abs/2407.06270} {arXiv:2407.06270 [gr-qc]} \BibitemShut {NoStop}%
\bibitem [{\citenamefont {{Hazboun}}\ \emph {et~al.}(2020)\citenamefont {{Hazboun}}, \citenamefont {{Simon}}, \citenamefont {{Taylor}}, \citenamefont {{Lam}}, \citenamefont {{Vigeland}}, \citenamefont {{Islo}}, \citenamefont {{Key}}, \citenamefont {{Arzoumanian}}, \citenamefont {{Baker}}, \citenamefont {{Brazier}}, \citenamefont {{Brook}}, \citenamefont {{Burke-Spolaor}}, \citenamefont {{Chatterjee}}, \citenamefont {{Cordes}}, \citenamefont {{Cornish}}, \citenamefont {{Crawford}}, \citenamefont {{Crowter}}, \citenamefont {{Cromartie}}, \citenamefont {{DeCesar}}, \citenamefont {{Demorest}}, \citenamefont {{Dolch}}, \citenamefont {{Ellis}}, \citenamefont {{Ferdman}}, \citenamefont {{Ferrara}}, \citenamefont {{Fonseca}}, \citenamefont {{Garver-Daniels}}, \citenamefont {{Gentile}}, \citenamefont {{Good}}, \citenamefont {{Holgado}}, \citenamefont {{Huerta}}, \citenamefont {{Jennings}}, \citenamefont {{Jones}}, \citenamefont {{Jones}}, \citenamefont {{Kaiser}}, \citenamefont {{Kaplan}}, \citenamefont {{Kelley}},
  \citenamefont {{Lazio}}, \citenamefont {{Levin}}, \citenamefont {{Lommen}}, \citenamefont {{Lorimer}}, \citenamefont {{Luo}}, \citenamefont {{Lynch}}, \citenamefont {{Madison}}, \citenamefont {{McLaughlin}}, \citenamefont {{McWilliams}}, \citenamefont {{Mingarelli}}, \citenamefont {{Ng}}, \citenamefont {{Nice}}, \citenamefont {{Pennucci}}, \citenamefont {{Pol}}, \citenamefont {{Ransom}}, \citenamefont {{Ray}}, \citenamefont {{Siemens}}, \citenamefont {{Spiewak}}, \citenamefont {{Stairs}}, \citenamefont {{Stinebring}}, \citenamefont {{Stovall}}, \citenamefont {{Swiggum}}, \citenamefont {{Turner}}, \citenamefont {{Vallisneri}}, \citenamefont {{van Haasteren}}, \citenamefont {{Witt}},\ and\ \citenamefont {{Zhu}}}]{11yr}%
  \BibitemOpen
  \bibfield  {author} {\bibinfo {author} {\bibfnamefont {J.~S.}\ \bibnamefont {{Hazboun}}}, \bibinfo {author} {\bibfnamefont {J.}~\bibnamefont {{Simon}}}, \bibinfo {author} {\bibfnamefont {S.~R.}\ \bibnamefont {{Taylor}}}, \bibinfo {author} {\bibfnamefont {M.~T.}\ \bibnamefont {{Lam}}}, \bibinfo {author} {\bibfnamefont {S.~J.}\ \bibnamefont {{Vigeland}}}, \bibinfo {author} {\bibfnamefont {K.}~\bibnamefont {{Islo}}}, \bibinfo {author} {\bibfnamefont {J.~S.}\ \bibnamefont {{Key}}}, \bibinfo {author} {\bibfnamefont {Z.}~\bibnamefont {{Arzoumanian}}}, \bibinfo {author} {\bibfnamefont {P.~T.}\ \bibnamefont {{Baker}}}, \bibinfo {author} {\bibfnamefont {A.}~\bibnamefont {{Brazier}}}, \bibinfo {author} {\bibfnamefont {P.~R.}\ \bibnamefont {{Brook}}}, \bibinfo {author} {\bibfnamefont {S.}~\bibnamefont {{Burke-Spolaor}}}, \bibinfo {author} {\bibfnamefont {S.}~\bibnamefont {{Chatterjee}}}, \bibinfo {author} {\bibfnamefont {J.~M.}\ \bibnamefont {{Cordes}}}, \bibinfo {author} {\bibfnamefont {N.~J.}\ \bibnamefont
  {{Cornish}}}, \bibinfo {author} {\bibfnamefont {F.}~\bibnamefont {{Crawford}}}, \bibinfo {author} {\bibfnamefont {K.}~\bibnamefont {{Crowter}}}, \bibinfo {author} {\bibfnamefont {H.~T.}\ \bibnamefont {{Cromartie}}}, \bibinfo {author} {\bibfnamefont {M.}~\bibnamefont {{DeCesar}}}, \bibinfo {author} {\bibfnamefont {P.~B.}\ \bibnamefont {{Demorest}}}, \bibinfo {author} {\bibfnamefont {T.}~\bibnamefont {{Dolch}}}, \bibinfo {author} {\bibfnamefont {J.~A.}\ \bibnamefont {{Ellis}}}, \bibinfo {author} {\bibfnamefont {R.~D.}\ \bibnamefont {{Ferdman}}}, \bibinfo {author} {\bibfnamefont {E.}~\bibnamefont {{Ferrara}}}, \bibinfo {author} {\bibfnamefont {E.}~\bibnamefont {{Fonseca}}}, \bibinfo {author} {\bibfnamefont {N.}~\bibnamefont {{Garver-Daniels}}}, \bibinfo {author} {\bibfnamefont {P.}~\bibnamefont {{Gentile}}}, \bibinfo {author} {\bibfnamefont {D.}~\bibnamefont {{Good}}}, \bibinfo {author} {\bibfnamefont {A.~M.}\ \bibnamefont {{Holgado}}}, \bibinfo {author} {\bibfnamefont {E.~A.}\ \bibnamefont {{Huerta}}},
  \bibinfo {author} {\bibfnamefont {R.}~\bibnamefont {{Jennings}}}, \bibinfo {author} {\bibfnamefont {G.}~\bibnamefont {{Jones}}}, \bibinfo {author} {\bibfnamefont {M.~L.}\ \bibnamefont {{Jones}}}, \bibinfo {author} {\bibfnamefont {A.~R.}\ \bibnamefont {{Kaiser}}}, \bibinfo {author} {\bibfnamefont {D.~L.}\ \bibnamefont {{Kaplan}}}, \bibinfo {author} {\bibfnamefont {L.~Z.}\ \bibnamefont {{Kelley}}}, \bibinfo {author} {\bibfnamefont {T.~J.~W.}\ \bibnamefont {{Lazio}}}, \bibinfo {author} {\bibfnamefont {L.}~\bibnamefont {{Levin}}}, \bibinfo {author} {\bibfnamefont {A.~N.}\ \bibnamefont {{Lommen}}}, \bibinfo {author} {\bibfnamefont {D.~R.}\ \bibnamefont {{Lorimer}}}, \bibinfo {author} {\bibfnamefont {J.}~\bibnamefont {{Luo}}}, \bibinfo {author} {\bibfnamefont {R.~S.}\ \bibnamefont {{Lynch}}}, \bibinfo {author} {\bibfnamefont {D.~R.}\ \bibnamefont {{Madison}}}, \bibinfo {author} {\bibfnamefont {M.~A.}\ \bibnamefont {{McLaughlin}}}, \bibinfo {author} {\bibfnamefont {S.~T.}\ \bibnamefont {{McWilliams}}}, \bibinfo
  {author} {\bibfnamefont {C.~M.~F.}\ \bibnamefont {{Mingarelli}}}, \bibinfo {author} {\bibfnamefont {C.}~\bibnamefont {{Ng}}}, \bibinfo {author} {\bibfnamefont {D.~J.}\ \bibnamefont {{Nice}}}, \bibinfo {author} {\bibfnamefont {T.~T.}\ \bibnamefont {{Pennucci}}}, \bibinfo {author} {\bibfnamefont {N.~S.}\ \bibnamefont {{Pol}}}, \bibinfo {author} {\bibfnamefont {S.~M.}\ \bibnamefont {{Ransom}}}, \bibinfo {author} {\bibfnamefont {P.~S.}\ \bibnamefont {{Ray}}}, \bibinfo {author} {\bibfnamefont {X.}~\bibnamefont {{Siemens}}}, \bibinfo {author} {\bibfnamefont {R.}~\bibnamefont {{Spiewak}}}, \bibinfo {author} {\bibfnamefont {I.~H.}\ \bibnamefont {{Stairs}}}, \bibinfo {author} {\bibfnamefont {D.~R.}\ \bibnamefont {{Stinebring}}}, \bibinfo {author} {\bibfnamefont {K.}~\bibnamefont {{Stovall}}}, \bibinfo {author} {\bibfnamefont {J.}~\bibnamefont {{Swiggum}}}, \bibinfo {author} {\bibfnamefont {J.~E.}\ \bibnamefont {{Turner}}}, \bibinfo {author} {\bibfnamefont {M.}~\bibnamefont {{Vallisneri}}}, \bibinfo {author}
  {\bibfnamefont {R.}~\bibnamefont {{van Haasteren}}}, \bibinfo {author} {\bibfnamefont {C.~A.}\ \bibnamefont {{Witt}}},\ and\ \bibinfo {author} {\bibfnamefont {W.~W.}\ \bibnamefont {{Zhu}}},\ }\bibfield  {title} {\bibinfo {title} {{The NANOGrav 11 yr Data Set: Evolution of Gravitational-wave Background Statistics}},\ }\href {https://doi.org/10.3847/1538-4357/ab68db} {\bibfield  {journal} {\bibinfo  {journal} {\apj}\ }\textbf {\bibinfo {volume} {890}},\ \bibinfo {eid} {108} (\bibinfo {year} {2020})},\ \Eprint {https://arxiv.org/abs/1909.08644} {arXiv:1909.08644 [astro-ph.HE]} \BibitemShut {NoStop}%
\bibitem [{\citenamefont {Ambikasaran}\ \emph {et~al.}(2016)\citenamefont {Ambikasaran}, \citenamefont {Foreman-Mackey}, \citenamefont {Greengard}, \citenamefont {Hogg},\ and\ \citenamefont {O’Neil}}]{George}%
  \BibitemOpen
  \bibfield  {author} {\bibinfo {author} {\bibfnamefont {S.}~\bibnamefont {Ambikasaran}}, \bibinfo {author} {\bibfnamefont {D.}~\bibnamefont {Foreman-Mackey}}, \bibinfo {author} {\bibfnamefont {L.}~\bibnamefont {Greengard}}, \bibinfo {author} {\bibfnamefont {D.~W.}\ \bibnamefont {Hogg}},\ and\ \bibinfo {author} {\bibfnamefont {M.}~\bibnamefont {O’Neil}},\ }\bibfield  {title} {\bibinfo {title} {Fast direct methods for gaussian processes},\ }\href {https://doi.org/10.1109/TPAMI.2015.2448083} {\bibfield  {journal} {\bibinfo  {journal} {IEEE Transactions on Pattern Analysis and Machine Intelligence}\ }\textbf {\bibinfo {volume} {38}},\ \bibinfo {pages} {252} (\bibinfo {year} {2016})}\BibitemShut {NoStop}%
\bibitem [{\citenamefont {{Taylor}}\ and\ \citenamefont {{Gerosa}}(2018)}]{LHC}%
  \BibitemOpen
  \bibfield  {author} {\bibinfo {author} {\bibfnamefont {S.~R.}\ \bibnamefont {{Taylor}}}\ and\ \bibinfo {author} {\bibfnamefont {D.}~\bibnamefont {{Gerosa}}},\ }\bibfield  {title} {\bibinfo {title} {{Mining gravitational-wave catalogs to understand binary stellar evolution: A new hierarchical Bayesian framework}},\ }\href {https://doi.org/10.1103/PhysRevD.98.083017} {\bibfield  {journal} {\bibinfo  {journal} {\prd}\ }\textbf {\bibinfo {volume} {98}},\ \bibinfo {eid} {083017} (\bibinfo {year} {2018})},\ \Eprint {https://arxiv.org/abs/1806.08365} {arXiv:1806.08365 [astro-ph.HE]} \BibitemShut {NoStop}%
\bibitem [{\citenamefont {Agnelli}\ \emph {et~al.}(2010)\citenamefont {Agnelli}, \citenamefont {Cadeiras}, \citenamefont {Tabak}, \citenamefont {Turner},\ and\ \citenamefont {Vanden-Eijnden}}]{nf0}%
  \BibitemOpen
  \bibfield  {author} {\bibinfo {author} {\bibfnamefont {J.~P.}\ \bibnamefont {Agnelli}}, \bibinfo {author} {\bibfnamefont {M.}~\bibnamefont {Cadeiras}}, \bibinfo {author} {\bibfnamefont {E.~G.}\ \bibnamefont {Tabak}}, \bibinfo {author} {\bibfnamefont {C.~V.}\ \bibnamefont {Turner}},\ and\ \bibinfo {author} {\bibfnamefont {E.}~\bibnamefont {Vanden-Eijnden}},\ }\bibfield  {title} {\bibinfo {title} {Clustering and classification through normalizing flows in feature space},\ }\href {https://doi.org/10.1137/100783522} {\bibfield  {journal} {\bibinfo  {journal} {Multiscale Modeling \& Simulation}\ }\textbf {\bibinfo {volume} {8}},\ \bibinfo {pages} {1784} (\bibinfo {year} {2010})},\ \Eprint {https://arxiv.org/abs/https://doi.org/10.1137/100783522} {https://doi.org/10.1137/100783522} \BibitemShut {NoStop}%
\bibitem [{\citenamefont {Tabak}\ and\ \citenamefont {Turner}(2013)}]{nf1}%
  \BibitemOpen
  \bibfield  {author} {\bibinfo {author} {\bibfnamefont {E.~G.}\ \bibnamefont {Tabak}}\ and\ \bibinfo {author} {\bibfnamefont {C.~V.}\ \bibnamefont {Turner}},\ }\bibfield  {title} {\bibinfo {title} {A family of nonparametric density estimation algorithms},\ }\href {https://doi.org/https://doi.org/10.1002/cpa.21423} {\bibfield  {journal} {\bibinfo  {journal} {Communications on Pure and Applied Mathematics}\ }\textbf {\bibinfo {volume} {66}},\ \bibinfo {pages} {145} (\bibinfo {year} {2013})},\ \Eprint {https://arxiv.org/abs/https://onlinelibrary.wiley.com/doi/pdf/10.1002/cpa.21423} {https://onlinelibrary.wiley.com/doi/pdf/10.1002/cpa.21423} \BibitemShut {NoStop}%
\bibitem [{\citenamefont {Kobyzev}\ \emph {et~al.}(2021)\citenamefont {Kobyzev}, \citenamefont {Prince},\ and\ \citenamefont {Brubaker}}]{NFreview}%
  \BibitemOpen
  \bibfield  {author} {\bibinfo {author} {\bibfnamefont {I.}~\bibnamefont {Kobyzev}}, \bibinfo {author} {\bibfnamefont {S.~J.}\ \bibnamefont {Prince}},\ and\ \bibinfo {author} {\bibfnamefont {M.~A.}\ \bibnamefont {Brubaker}},\ }\bibfield  {title} {\bibinfo {title} {Normalizing flows: An introduction and review of current methods},\ }\href {https://doi.org/10.1109/TPAMI.2020.2992934} {\bibfield  {journal} {\bibinfo  {journal} {IEEE Transactions on Pattern Analysis and Machine Intelligence}\ }\textbf {\bibinfo {volume} {43}},\ \bibinfo {pages} {3964} (\bibinfo {year} {2021})}\BibitemShut {NoStop}%
\bibitem [{\citenamefont {Kingma}\ and\ \citenamefont {Ba}(2014)}]{ADAM}%
  \BibitemOpen
  \bibfield  {author} {\bibinfo {author} {\bibfnamefont {D.~P.}\ \bibnamefont {Kingma}}\ and\ \bibinfo {author} {\bibfnamefont {J.}~\bibnamefont {Ba}},\ }\bibfield  {title} {\bibinfo {title} {Adam: A method for stochastic optimization},\ }\href {https://api.semanticscholar.org/CorpusID:6628106} {\bibfield  {journal} {\bibinfo  {journal} {CoRR}\ }\textbf {\bibinfo {volume} {abs/1412.6980}} (\bibinfo {year} {2014})}\BibitemShut {NoStop}%
\bibitem [{\citenamefont {Kullback}\ and\ \citenamefont {Leibler}(1951)}]{KL-divergenece}%
  \BibitemOpen
  \bibfield  {author} {\bibinfo {author} {\bibfnamefont {S.}~\bibnamefont {Kullback}}\ and\ \bibinfo {author} {\bibfnamefont {R.~A.}\ \bibnamefont {Leibler}},\ }\bibfield  {title} {\bibinfo {title} {On information and sufficiency},\ }\href {http://www.jstor.org/stable/2236703} {\bibfield  {journal} {\bibinfo  {journal} {The Annals of Mathematical Statistics}\ }\textbf {\bibinfo {volume} {22}},\ \bibinfo {pages} {79} (\bibinfo {year} {1951})}\BibitemShut {NoStop}%
\bibitem [{\citenamefont {Germain}\ \emph {et~al.}(2015)\citenamefont {Germain}, \citenamefont {Gregor}, \citenamefont {Murray},\ and\ \citenamefont {Larochelle}}]{made}%
  \BibitemOpen
  \bibfield  {author} {\bibinfo {author} {\bibfnamefont {M.}~\bibnamefont {Germain}}, \bibinfo {author} {\bibfnamefont {K.}~\bibnamefont {Gregor}}, \bibinfo {author} {\bibfnamefont {I.}~\bibnamefont {Murray}},\ and\ \bibinfo {author} {\bibfnamefont {H.}~\bibnamefont {Larochelle}},\ }\bibfield  {title} {\bibinfo {title} {Made: Masked autoencoder for distribution estimation},\ }in\ \href {https://proceedings.mlr.press/v37/germain15.html} {\emph {\bibinfo {booktitle} {Proceedings of the 32nd International Conference on Machine Learning}}},\ \bibinfo {series} {Proceedings of Machine Learning Research}, Vol.~\bibinfo {volume} {37},\ \bibinfo {editor} {edited by\ \bibinfo {editor} {\bibfnamefont {F.}~\bibnamefont {Bach}}\ and\ \bibinfo {editor} {\bibfnamefont {D.}~\bibnamefont {Blei}}}\ (\bibinfo  {publisher} {PMLR},\ \bibinfo {address} {Lille, France},\ \bibinfo {year} {2015})\ pp.\ \bibinfo {pages} {881--889}\BibitemShut {NoStop}%
\bibitem [{\citenamefont {Hellinger}(1909)}]{Hell}%
  \BibitemOpen
  \bibfield  {author} {\bibinfo {author} {\bibfnamefont {E.}~\bibnamefont {Hellinger}},\ }\bibfield  {title} {\bibinfo {title} {Neue begründung der theorie quadratischer formen von unendlichvielen veränderlichen.},\ }\href {https://doi.org/doi:10.1515/crll.1909.136.210} {\bibfield  {journal} {\bibinfo  {journal} {Journal für die reine und angewandte Mathematik}\ }\textbf {\bibinfo {volume} {1909}},\ \bibinfo {pages} {210} (\bibinfo {year} {1909})}\BibitemShut {NoStop}%
\bibitem [{\citenamefont {Bingham}\ \emph {et~al.}(2019)\citenamefont {Bingham}, \citenamefont {Chen}, \citenamefont {Jankowiak}, \citenamefont {Obermeyer}, \citenamefont {Pradhan}, \citenamefont {Karaletsos}, \citenamefont {Singh}, \citenamefont {Szerlip}, \citenamefont {Horsfall},\ and\ \citenamefont {Goodman}}]{pyro}%
  \BibitemOpen
  \bibfield  {author} {\bibinfo {author} {\bibfnamefont {E.}~\bibnamefont {Bingham}}, \bibinfo {author} {\bibfnamefont {J.~P.}\ \bibnamefont {Chen}}, \bibinfo {author} {\bibfnamefont {M.}~\bibnamefont {Jankowiak}}, \bibinfo {author} {\bibfnamefont {F.}~\bibnamefont {Obermeyer}}, \bibinfo {author} {\bibfnamefont {N.}~\bibnamefont {Pradhan}}, \bibinfo {author} {\bibfnamefont {T.}~\bibnamefont {Karaletsos}}, \bibinfo {author} {\bibfnamefont {R.}~\bibnamefont {Singh}}, \bibinfo {author} {\bibfnamefont {P.~A.}\ \bibnamefont {Szerlip}}, \bibinfo {author} {\bibfnamefont {P.}~\bibnamefont {Horsfall}},\ and\ \bibinfo {author} {\bibfnamefont {N.~D.}\ \bibnamefont {Goodman}},\ }\bibfield  {title} {\bibinfo {title} {Pyro: Deep universal probabilistic programming},\ }\href {http://jmlr.org/papers/v20/18-403.html} {\bibfield  {journal} {\bibinfo  {journal} {J. Mach. Learn. Res.}\ }\textbf {\bibinfo {volume} {20}},\ \bibinfo {pages} {28:1} (\bibinfo {year} {2019})}\BibitemShut {NoStop}%
\bibitem [{\citenamefont {{Lamb}}\ \emph {et~al.}(2023)\citenamefont {{Lamb}}, \citenamefont {{Taylor}},\ and\ \citenamefont {{van Haasteren}}}]{fitting}%
  \BibitemOpen
  \bibfield  {author} {\bibinfo {author} {\bibfnamefont {W.~G.}\ \bibnamefont {{Lamb}}}, \bibinfo {author} {\bibfnamefont {S.~R.}\ \bibnamefont {{Taylor}}},\ and\ \bibinfo {author} {\bibfnamefont {R.}~\bibnamefont {{van Haasteren}}},\ }\bibfield  {title} {\bibinfo {title} {{The Need For Speed: Rapid Refitting Techniques for Bayesian Spectral Characterization of the Gravitational Wave Background Using PTAs}},\ }\href {https://doi.org/10.48550/arXiv.2303.15442} {\bibfield  {journal} {\bibinfo  {journal} {arXiv e-prints}\ ,\ \bibinfo {eid} {arXiv:2303.15442}} (\bibinfo {year} {2023})},\ \Eprint {https://arxiv.org/abs/2303.15442} {arXiv:2303.15442 [astro-ph.HE]} \BibitemShut {NoStop}%
\bibitem [{\citenamefont {Laal}\ \emph {et~al.}(2023)\citenamefont {Laal}, \citenamefont {Lamb}, \citenamefont {Romano}, \citenamefont {Siemens}, \citenamefont {Taylor},\ and\ \citenamefont {van Haasteren}}]{Laal:2023etp}%
  \BibitemOpen
  \bibfield  {author} {\bibinfo {author} {\bibfnamefont {N.}~\bibnamefont {Laal}}, \bibinfo {author} {\bibfnamefont {W.~G.}\ \bibnamefont {Lamb}}, \bibinfo {author} {\bibfnamefont {J.~D.}\ \bibnamefont {Romano}}, \bibinfo {author} {\bibfnamefont {X.}~\bibnamefont {Siemens}}, \bibinfo {author} {\bibfnamefont {S.~R.}\ \bibnamefont {Taylor}},\ and\ \bibinfo {author} {\bibfnamefont {R.}~\bibnamefont {van Haasteren}},\ }\bibfield  {title} {\bibinfo {title} {{Exploring the capabilities of Gibbs sampling in pulsar timing arrays}},\ }\href {https://doi.org/10.1103/PhysRevD.108.063008} {\bibfield  {journal} {\bibinfo  {journal} {Phys. Rev. D}\ }\textbf {\bibinfo {volume} {108}},\ \bibinfo {pages} {063008} (\bibinfo {year} {2023})},\ \Eprint {https://arxiv.org/abs/2305.12285} {arXiv:2305.12285 [astro-ph.IM]} \BibitemShut {NoStop}%
\bibitem [{\citenamefont {Kumar}\ \emph {et~al.}(2019)\citenamefont {Kumar}, \citenamefont {Carroll}, \citenamefont {Hartikainen},\ and\ \citenamefont {Martin}}]{arviz_2019}%
  \BibitemOpen
  \bibfield  {author} {\bibinfo {author} {\bibfnamefont {R.}~\bibnamefont {Kumar}}, \bibinfo {author} {\bibfnamefont {C.}~\bibnamefont {Carroll}}, \bibinfo {author} {\bibfnamefont {A.}~\bibnamefont {Hartikainen}},\ and\ \bibinfo {author} {\bibfnamefont {O.}~\bibnamefont {Martin}},\ }\bibfield  {title} {\bibinfo {title} {Arviz a unified library for exploratory analysis of bayesian models in python},\ }\href {https://doi.org/10.21105/joss.01143} {\bibfield  {journal} {\bibinfo  {journal} {Journal of Open Source Software}\ }\textbf {\bibinfo {volume} {4}},\ \bibinfo {pages} {1143} (\bibinfo {year} {2019})}\BibitemShut {NoStop}%
\bibitem [{\citenamefont {Ellis}\ and\ \citenamefont {van Haasteren}(2017)}]{PTMCMC}%
  \BibitemOpen
  \bibfield  {author} {\bibinfo {author} {\bibfnamefont {J.}~\bibnamefont {Ellis}}\ and\ \bibinfo {author} {\bibfnamefont {R.}~\bibnamefont {van Haasteren}},\ }\href {https://doi.org/10.5281/zenodo.1037579} {\bibinfo {title} {jellis18/ptmcmcsampler: Official release}} (\bibinfo {year} {2017})\BibitemShut {NoStop}%
\bibitem [{\citenamefont {Ansel}\ \emph {et~al.}(2024)\citenamefont {Ansel}, \citenamefont {Yang}, \citenamefont {He}, \citenamefont {Gimelshein}, \citenamefont {Jain}, \citenamefont {Voznesensky}, \citenamefont {Bao}, \citenamefont {Bell}, \citenamefont {Berard}, \citenamefont {Burovski}, \citenamefont {Chauhan}, \citenamefont {Chourdia}, \citenamefont {Constable}, \citenamefont {Desmaison}, \citenamefont {DeVito}, \citenamefont {Ellison}, \citenamefont {Feng}, \citenamefont {Gong}, \citenamefont {Gschwind}, \citenamefont {Hirsh}, \citenamefont {Huang}, \citenamefont {Kalambarkar}, \citenamefont {Kirsch}, \citenamefont {Lazos}, \citenamefont {Lezcano}, \citenamefont {Liang}, \citenamefont {Liang}, \citenamefont {Lu}, \citenamefont {Luk}, \citenamefont {Maher}, \citenamefont {Pan}, \citenamefont {Puhrsch}, \citenamefont {Reso}, \citenamefont {Saroufim}, \citenamefont {Siraichi}, \citenamefont {Suk}, \citenamefont {Suo}, \citenamefont {Tillet}, \citenamefont {Wang}, \citenamefont {Wang}, \citenamefont {Wen},
  \citenamefont {Zhang}, \citenamefont {Zhao}, \citenamefont {Zhou}, \citenamefont {Zou}, \citenamefont {Mathews}, \citenamefont {Chanan}, \citenamefont {Wu},\ and\ \citenamefont {Chintala}}]{pytorch}%
  \BibitemOpen
  \bibfield  {author} {\bibinfo {author} {\bibfnamefont {J.}~\bibnamefont {Ansel}}, \bibinfo {author} {\bibfnamefont {E.}~\bibnamefont {Yang}}, \bibinfo {author} {\bibfnamefont {H.}~\bibnamefont {He}}, \bibinfo {author} {\bibfnamefont {N.}~\bibnamefont {Gimelshein}}, \bibinfo {author} {\bibfnamefont {A.}~\bibnamefont {Jain}}, \bibinfo {author} {\bibfnamefont {M.}~\bibnamefont {Voznesensky}}, \bibinfo {author} {\bibfnamefont {B.}~\bibnamefont {Bao}}, \bibinfo {author} {\bibfnamefont {P.}~\bibnamefont {Bell}}, \bibinfo {author} {\bibfnamefont {D.}~\bibnamefont {Berard}}, \bibinfo {author} {\bibfnamefont {E.}~\bibnamefont {Burovski}}, \bibinfo {author} {\bibfnamefont {G.}~\bibnamefont {Chauhan}}, \bibinfo {author} {\bibfnamefont {A.}~\bibnamefont {Chourdia}}, \bibinfo {author} {\bibfnamefont {W.}~\bibnamefont {Constable}}, \bibinfo {author} {\bibfnamefont {A.}~\bibnamefont {Desmaison}}, \bibinfo {author} {\bibfnamefont {Z.}~\bibnamefont {DeVito}}, \bibinfo {author} {\bibfnamefont {E.}~\bibnamefont {Ellison}},
  \bibinfo {author} {\bibfnamefont {W.}~\bibnamefont {Feng}}, \bibinfo {author} {\bibfnamefont {J.}~\bibnamefont {Gong}}, \bibinfo {author} {\bibfnamefont {M.}~\bibnamefont {Gschwind}}, \bibinfo {author} {\bibfnamefont {B.}~\bibnamefont {Hirsh}}, \bibinfo {author} {\bibfnamefont {S.}~\bibnamefont {Huang}}, \bibinfo {author} {\bibfnamefont {K.}~\bibnamefont {Kalambarkar}}, \bibinfo {author} {\bibfnamefont {L.}~\bibnamefont {Kirsch}}, \bibinfo {author} {\bibfnamefont {M.}~\bibnamefont {Lazos}}, \bibinfo {author} {\bibfnamefont {M.}~\bibnamefont {Lezcano}}, \bibinfo {author} {\bibfnamefont {Y.}~\bibnamefont {Liang}}, \bibinfo {author} {\bibfnamefont {J.}~\bibnamefont {Liang}}, \bibinfo {author} {\bibfnamefont {Y.}~\bibnamefont {Lu}}, \bibinfo {author} {\bibfnamefont {C.}~\bibnamefont {Luk}}, \bibinfo {author} {\bibfnamefont {B.}~\bibnamefont {Maher}}, \bibinfo {author} {\bibfnamefont {Y.}~\bibnamefont {Pan}}, \bibinfo {author} {\bibfnamefont {C.}~\bibnamefont {Puhrsch}}, \bibinfo {author} {\bibfnamefont
  {M.}~\bibnamefont {Reso}}, \bibinfo {author} {\bibfnamefont {M.}~\bibnamefont {Saroufim}}, \bibinfo {author} {\bibfnamefont {M.~Y.}\ \bibnamefont {Siraichi}}, \bibinfo {author} {\bibfnamefont {H.}~\bibnamefont {Suk}}, \bibinfo {author} {\bibfnamefont {M.}~\bibnamefont {Suo}}, \bibinfo {author} {\bibfnamefont {P.}~\bibnamefont {Tillet}}, \bibinfo {author} {\bibfnamefont {E.}~\bibnamefont {Wang}}, \bibinfo {author} {\bibfnamefont {X.}~\bibnamefont {Wang}}, \bibinfo {author} {\bibfnamefont {W.}~\bibnamefont {Wen}}, \bibinfo {author} {\bibfnamefont {S.}~\bibnamefont {Zhang}}, \bibinfo {author} {\bibfnamefont {X.}~\bibnamefont {Zhao}}, \bibinfo {author} {\bibfnamefont {K.}~\bibnamefont {Zhou}}, \bibinfo {author} {\bibfnamefont {R.}~\bibnamefont {Zou}}, \bibinfo {author} {\bibfnamefont {A.}~\bibnamefont {Mathews}}, \bibinfo {author} {\bibfnamefont {G.}~\bibnamefont {Chanan}}, \bibinfo {author} {\bibfnamefont {P.}~\bibnamefont {Wu}},\ and\ \bibinfo {author} {\bibfnamefont {S.}~\bibnamefont {Chintala}},\ }\bibfield
   {title} {\bibinfo {title} {{PyTorch 2: Faster Machine Learning Through Dynamic Python Bytecode Transformation and Graph Compilation}},\ }in\ \href {https://doi.org/10.1145/3620665.3640366} {\emph {\bibinfo {booktitle} {29th ACM International Conference on Architectural Support for Programming Languages and Operating Systems, Volume 2 (ASPLOS '24)}}}\ (\bibinfo  {publisher} {ACM},\ \bibinfo {year} {2024})\BibitemShut {NoStop}%
\bibitem [{\citenamefont {Hunter}(2007)}]{plt}%
  \BibitemOpen
  \bibfield  {author} {\bibinfo {author} {\bibfnamefont {J.~D.}\ \bibnamefont {Hunter}},\ }\bibfield  {title} {\bibinfo {title} {Matplotlib: A 2d graphics environment},\ }\href {https://doi.org/10.1109/MCSE.2007.55} {\bibfield  {journal} {\bibinfo  {journal} {Computing in Science \& Engineering}\ }\textbf {\bibinfo {volume} {9}},\ \bibinfo {pages} {90} (\bibinfo {year} {2007})}\BibitemShut {NoStop}%
\end{thebibliography}%
\end{document}